\documentclass[aip, pof, graphicx, 12pt, floatfix, preprint]{revtex4-1}

\usepackage{kotex}
\usepackage{subcaption}
\usepackage{graphicx}
\usepackage{dcolumn}
\usepackage{multirow}
\usepackage{ctable}
\usepackage{booktabs}
\usepackage{amsmath}
\usepackage{amssymb}
\usepackage{xcolor}
\usepackage{svg}
\usepackage{makecell}
\usepackage{comment}

\definecolor{g}{rgb}{0., 0.4, 0.1}

\begin{document}

\title{A second-order particle Fokker-Planck-Master method for diatomic gas flows}



\author{Joonbeom Kim}
\affiliation{Department of Aerospace Engineering, Korea Advanced Institute of Science and Technology, Daejeon 34141, South Korea}

\author{Eunji Jun}
\email{eunji.jun@kaist.ac.kr}
\affiliation{Department of Aerospace Engineering, Korea Advanced Institute of Science and Technology, Daejeon 34141, South Korea}

\date{\today}

\begin{abstract}

The direct simulation Monte Carlo (DSMC) method is widely used to describe rarefied gas flows. The DSMC method accounts for the transport and collisions of computational particles, resulting in higher computational costs in the continuum regime. The Fokker-Planck (FP) model approximates particle collisions as Brownian motion to reduce computational cost. Advanced FP models have been developed to enhance physical fidelity, ensuring the correct Prandtl number and the H-theorem. The FP model has further been extended to handle diatomic gases, such as the Fokker–Planck–Master (FPM) model. Alongside these developments in modeling, computational efficiency has also been improved by achieving second-order spatial and temporal accuracy, as demonstrated in the unified stochastic particle FP (USP-FP) method. However, these accuracy improvements have not yet been extended to diatomic gases, which are essential for engineering applications such as atmospheric reentry. This study proposes a unified stochastic particle Fokker-Planck-Master (USP-FPM) method for diatomic gases that achieves second-order accuracy in both time and space. Temporal accuracy is enhanced by reproducing second-order energy, viscous stress, and heat flux relaxations. Spatial accuracy is improved by employing a first-order polynomial reconstruction method. Three test cases are investigated: homogeneous relaxation, Poiseuille flow, and hypersonic flow around a cylinder. The results show that the USP-FPM method provides accurate solutions even with coarser cell sizes and larger time steps compared to the DSMC and FPM methods. In particular, for the hypersonic flow around a cylinder, the USP-FPM method achieves a speed-up factor of 28 compared to the DSMC method, while maintaining accuracy.

\end{abstract}


\maketitle 

\section{\label{sec:intro}Introduction}


With the rise of reusable spacecraft and suborbital missions, accurate prediction of reentry flows is essential for mission safety and efficiency. During atmospheric reentry, hypersonic vehicles encounter strong shock waves and thermal nonequilibrium. Due to the breakdown of the continuum assumption, the Navier–Stokes–Fourier equations fail to describe these nonequilibrium flows. Instead, the Boltzmann equation serves as the governing equation for these non-continuum flows \cite{Bird_94_DSMC-book}. The direct simulation Monte Carlo (DSMC) method is widely used to numerically solve the Boltzmann equation. By representing molecular motion and collisions through computational particles, the DSMC method covers a broad range of flow regimes, from rarefied to continuum flows. However, the DSMC method becomes inefficient in the near-continuum regime due to more frequent particle collisions.


To reduce the computational cost of the DSMC method, particle-based alternatives have been extensively investigated \cite{Jun11_particle, Jun13_particle, Jun18_LD-DSMC, Jun18_cublcFP-DSMC-cylinder, Kim24_FP-DSMC-hypersonic}. Among them, kinetic models such as the Fokker-Planck (FP) and Bhatnagar-Gross-Krook (BGK) models approximate the collision operator to improve computational efficiency \cite{Kim23_Compare_CFD-Kinetic, Jenny10_linearFP, Gorji11_cubicFP, Mathiaud16_ESFP, Gorji21_Quad-EFP, Kim22_FP-nonlinear, Gorji12_cubicFP-mixture, Hepp20_cubicFP-mixture_HS, Hepp20_cubicFP-mixture_VHS, Kim25_FP_mixture, Kim25_FPM-mixture, Jun19_Compare_FP, Fei20_compare, Pfeiffer18_compare_BGK, Yao23_SBGK, Park24_BGK_PDF}. The FP model represents the particle collision process as a continuous drift-diffusion process in velocity space. Jenny et al. pioneered the stochastic particle FP model to describe rarefied gas flows \cite{Jenny10_linearFP}. The linear FP model yields an incorrect Prandtl number $(\mathrm{Pr})$ for monatomic gases \cite{Jenny10_linearFP}. Since then, advanced FP models have been developed to recover the correct $\mathrm{Pr}$. Gorji et al. proposed the cubic-FP model by introducing a cubic term to achieve the correct $\mathrm{Pr}$ \cite{Gorji11_cubicFP}. However, the cubic-FP model fails to satisfy the H-theorem consistently. To address this limitation, Gorji et al. developed the quadratic entropic FP (Quad-EFP) model that satisfies the H-theorem \cite{Gorji21_Quad-EFP}. Mathiaud et al. proposed the ellipsoidal statistical FP (ES-FP) model, which satisfies both the correct $\mathrm{Pr}$ and the H-theorem \cite{Mathiaud16_ESFP}. One limitation of the ES-FP model is that it fails to maintain the correct $\mathrm{Pr}$ under highly anisotropic flows. Beyond monatomic gases, the FP model has been extended to diatomic gases with internal energy modes. Gorji et al. extended the cubic-FP model to account for diatomic gases by employing rotational and vibrational velocities \cite{Gorji13_cubicFP-diatomic}. Mathiaud et al. extended the ES-FP model to account for diatomic gases by introducing continuous internal energy \cite{Mathiaud17_ESFP-polyatomic}. Nevertheless, neither FP model accounts for the discrete nature of the vibrational energy. Hepp et al. introduced the master equation into the cubic-FP model to explain the discrete vibrational energy \cite{Hepp20_cubicFPM-diatomic}. Most recently, Kim et al. proposed the Fokker–Planck–Master (FPM) model by incorporating the master equation into the ES-FP model to describe diatomic gases with discrete vibrational energy states \cite{Kim24-FPM}.


Although the FP model has advanced in terms of physical fidelity and applicability, its spatio-temporal accuracy remains limited. The FP model estimates macroscopic properties for particle updates by averaging moments of particles within each cell. In most FP models \cite{Gorji14_efficient-algorithm-FP, Jun18_cubicFP_slit, Jun19_cubicFP_slit, Pfeiffer19_compare_particle, Kim24_assessment_FP-conf}, macroscopic properties are assumed constant during each time step, resulting in first-order temporal accuracy \cite{Kim23_assessment-FP}. These models also directly use cell-averaged macroscopic properties without accounting for their spatial gradients, leading to first-order spatial accuracy \cite{Kim24_FP_poly}. This first-order accuracy in time and space imposes strict constraints on the allowable time step and cell size, thereby limiting the computational efficiency of the FP model. Despite its importance, only a few studies have investigated the spatio-temporal accuracy of the FP model. Kim et al. proposed the unified stochastic particle FP (USP-FP) method, which achieves second-order temporal accuracy by incorporating second-order relaxation terms for viscous stress and heat flux \cite{Kim24_USP-FP}. Cui et al. developed the multiscale stochastic particle (MSP) method, which obtains second-order temporal accuracy through exact time integration and operator-splitting correction \cite{Cui25_MSP}. For second-order spatial accuracy, the USP-FP method incorporates the random interpolation scheme \cite{Fei21_USP, Kim24_USP-FP}, while the MSP method uses linear interpolation. Another effort to improve spatial accuracy is found in the polynomial reconstruction method proposed by Kim et al., which achieves second-order spatial accuracy and supports higher-order extensions while conserving cell-averaged values \cite{Kim24_FP_poly}. Nevertheless, these studies have been confined to monatomic gases, where only translational motion is considered. The FP models for monatomic gases are inadequate for practical applications including atmospheric reentry, where diatomic species such as nitrogen and oxygen are prevalent. To extend second-order FP models to diatomic gases, internal energy modes and their temporal evolution must also be considered.


This paper develops a unified stochastic particle Fokker-Planck-Master (USP-FPM) method for diatomic gases to achieve second-order spatio-temporal accuracy. The second-order temporal accuracy is attained by adopting second-order moment relaxations for energy and heat flux in the translational, rotational, and vibrational modes, as well as for viscous stress. To attain second-order spatial accuracy, the polynomial reconstruction method is employed to resolve spatial gradients of macroscopic properties \cite{Kim24_FP_poly}. The remainder of the paper is organized as follows. Section \ref{sec:review} reviews the FPM model for diatomic gases. Section \ref{sec:usp-fpm} details the USP-FPM method in terms of its temporal and spatial accuracy. Section \ref{sec:implementation} describes numerical implementations and the algorithm of the USP-FPM method. Section \ref{sec:results} presents results of three test cases, including homogeneous relaxation with two sub-cases, Poiseuille flow, and hypersonic flow around a cylinder. Section \ref{sec:conclusions} concludes the paper.

\section{\label{sec:review}Review of stochastic particle Fokker–Planck–Master model}

\subsection{\label{subsec:kinetic} Kinetic theory of diatomic gases}


From a microscopic perspective, the state of a diatomic molecule is characterized by time $t$, position $\boldsymbol{x}$, velocity $\boldsymbol{c}$, rotational energy $\varepsilon_{rot}$, and vibrational energy $\varepsilon_{vib}$ \cite{Bird_94_DSMC-book}. Above room temperature, the rotational energy can be treated as continuous, whereas the vibrational energy should be considered as discrete energy levels for an accurate description \cite{Boyd17_Neq-book}. According to the harmonic oscillator model, the vibrational energy is given by $\varepsilon_{vib} = IR\Theta_{vib}$, where $I$ is the vibrational energy level, $R$ is the specific gas constant, and $\Theta_{vib}$ is the characteristic vibrational temperature. In kinetic theory, the probability density function (PDF), denoted by $f(t, \boldsymbol{x}, \boldsymbol{c}, \varepsilon_{rot}, \varepsilon_{vib})$, represents the statistical behavior of diatomic gases. Macroscopic properties are obtained by evaluating moments of the mass density function (MDF), denoted as $\mathcal{F} = \rho f$, as follows:
\begin{equation} 
\begin{aligned}
    \rho = \langle \mathcal{F} \rangle , &&
    \rho U_i = \langle c_i \,\mathcal{F} \rangle ,
\end{aligned}
\end{equation}
\begin{equation} 
\begin{aligned}
    p_{ij} = \rho \Pi_{ij} = \langle C_i C_j\, \mathcal{F} \rangle , &&
    \sigma_{ij} = \langle (C_i C_j - \frac{1}{3}C^2 \delta_{ij} ) \,\mathcal{F} \rangle ,
\end{aligned}
\end{equation}
\begin{equation} 
\begin{aligned}
    E_{tr} = \rho e_{tr} = \frac{1}{2} \langle C^2\, \mathcal{F} \rangle ,
    &&
    E_{rot} = \rho e_{rot} = \langle \varepsilon_{rot}\, \mathcal{F} \rangle ,
    &&
    E_{vib} = \rho e_{vib} = \langle \varepsilon_{vib}\, \mathcal{F} \rangle ,
\end{aligned}
\end{equation}
\begin{equation} 
\begin{aligned}
    q_{tr,\, i} = \frac{1}{2} \langle C^2 C_i\, \mathcal{F} \rangle,
    &&
    q_{rot,\, i} = \langle C_i \varepsilon_{rot}\, \mathcal{F} \rangle,
    &&
    q_{vib,\, i} = \langle C_i \varepsilon_{vib}\, \mathcal{F} \rangle,
\end{aligned}
\end{equation}
where $\rho$ is the density, $U_i$ is the bulk velocity, $p_{ij}$ is the pressure tensor, $\Pi_{ij}$ is the temperature tensor, $C_i = c_i - U_i$ is the thermal velocity, $\sigma_{ij}$ is the viscous stress, $\delta_{ij}$ is the Kronecker delta, $E$ is the mass-weight energy, $e$ is the specific energy, $q_i$ is the heat flux. The Einstein notation is adopted throughout this paper. The subscripts $tr$, $rot$, and $vib$ denote the translational, rotational, and vibrational modes, respectively. The angular bracket $\langle \phi \mathcal{F} \rangle$ for any quantity $\phi$ denotes the ensemble average over velocity and internal energy spaces:
\begin{equation}
    \langle \phi \mathcal{F} \rangle 
    = \sum_{I=0}^{\infty} \int_{\mathbb{R}^+} \int_{\mathbb{R}^3}
    \phi  \mathcal{F} \,
    d\boldsymbol{c} \, d\varepsilon_{rot} .
\end{equation}
The specific energies for each mode can be expressed as functions of temperature $T$:
\begin{equation} 
\begin{aligned}
    e_{tr} (T) = \frac{3}{2} R T ,
    &&
    e_{rot} (T) = R T ,
    &&
    e_{vib} (T) = \frac{R \Theta_{vib}}{\exp{(\Theta_{vib}/T)} - 1} .
\end{aligned}
\end{equation}
The specific heat capacities for each mode, defined as $c_v=\partial e / \partial T$, are obtained from their corresponding specific energies: 
\begin{equation} 
\begin{aligned}
    c_{v,\,tr} = \frac{3}{2} R ,
    &&
    c_{v,\,rot} = R ,
    &&
    c_{v,\,vib} (T) = \frac{R \, (\Theta_{vib} / 2T)^2}{\sinh{(\Theta_{vib}/2T)}} .
\end{aligned}
\end{equation}
The energy relaxations in homogeneous flows are determined by the Landau-Teller equation \cite{Boyd17_Neq-book}:
\begin{equation} 
    \frac{d \, e_{tr}(T_{tr})}{dt}
    = \frac{e_{rot}(T_{rot}) - e_{rot}(T_{tr}) }{\tau_{rot}}
    + \frac{e_{vib}(T_{vib}) - e_{vib}(T_{tr}) }{\tau_{vib}} ,
\end{equation}
\begin{equation} 
    \frac{d \, e_{rot}(T_{rot})}{dt} 
    = \frac{e_{rot}(T_{tr}) - e_{rot}(T_{rot}) }{\tau_{rot}} ,
\end{equation}
\begin{equation} 
    \frac{d \, e_{vib}(T_{vib})}{dt}
    = \frac{e_{vib}(T_{tr}) - e_{vib}(T_{vib}) }{\tau_{vib}} ,
\label{eq:LT-evib}
\end{equation}
where $\tau_{rot}=Z_{rot}\tau_c$ is the rotational relaxation time, $\tau_{vib}=Z_{vib}\tau_c$ is the vibrational relaxation time, $\tau_c$ is the mean collision time, $Z_{rot}$ is the rotational collision number, and $Z_{vib}$ is the vibrational collision number.

\subsection{\label{subsec:FPM} FPM model for diatomic gases}


The time evolution of the PDF is driven by particle motion and collisions, as governed by the Boltzmann equation \cite{Bird_94_DSMC-book}. For diatomic gases, the Boltzmann equation can be approximated by the FPM model, which describes particle collisions as a drift-diffusion-jump process \cite{Kim24-FPM}. In the FPM model, the FP equation describes the drift-diffusion process for continuous particle velocity and rotational energy, while the master equation governs the jump process for the discrete vibrational energy \cite{Tabar19_stochastic-book}. The FPM model is defined as follows \cite{Kim24-FPM}:
\begin{equation}
\begin{aligned}
    \frac{\partial \mathcal{F}}{\partial t}
    + 
    c_i \frac{\partial \mathcal{F}}{\partial c_i}
    =
    & - \frac{\partial}{\partial c_i} (A_{tr,\, i} \, \mathcal{F})
    + \frac{\partial^2}{\partial c_i \partial c_j} (D_{tr,\, ij} \, \mathcal{F}) \\
    & - \frac{\partial}{\partial \varepsilon_{rot}} (A_{rot} \, \mathcal{F})
    + \frac{\partial^2}{\partial \varepsilon_{rot}^2} (D_{rot} \, \mathcal{F})
    + \sum_{J=0}^{\infty} (\omega_{J,I} \, \mathcal{F}_J - \omega_{I,J} \, \mathcal{F}_I) 
    ,
\end{aligned}
\end{equation}
where $A_{tr,\, i}$ and $A_{rot}$ are the drift coefficients, $D_{tr,\, ij}$ and $D_{rot}$ are the diffusion coefficients, $\mathcal{F}_I$ is a MDF for particles in the $I$th vibrational level, and $\omega_{I,J}$ is the rate coefficient for transitions from the $I$th to the $J$th vibrational level. The coefficients are defined as follows:
\begin{equation}
    A_{tr,\, i} = -\frac{C_i}{\tau} ,
\end{equation}
\begin{equation}
    D_{tr,\, ij} = \frac{RT_{tr}^{rel} \delta_{ij} + \nu (\Pi_{ij} - RT_{tr}\delta_{ij}) }{\tau} ,
\label{eq:D_tr}
\end{equation}
\begin{equation}
    A_{rot} = -\frac{2}{\tau} \Big( \varepsilon_{rot} - RT_{rot}^{rel} \Big) ,
\end{equation}
\begin{equation}
    D_{rot} = \frac{2R T_{rot}^{rel} \varepsilon_{rot} }{\tau} ,
\end{equation}
\begin{equation}
    \omega_{I,J} = 
        \begin{cases}
            \dfrac{2 I}
                 {\tau (1 - \exp{(-\Theta_{vib}/T_{vib}^{rel})})} 
            & \text{if } J = I - 1, \\[1em]
            \dfrac{2 (I+1) \exp{(-\Theta_{vib}/T_{vib}^{rel})}}
                 {\tau (1 - \exp{(-\Theta_{vib}/T_{vib}^{rel})})} 
            & \text{if } J = I + 1, \\[1em]
            0 & \text{otherwise},
        \end{cases}
\end{equation}
where $\nu$ is a parameter for matching $\mathrm{Pr}$, $T^{rel}$ is the relaxation temperature, $\tau=2\mu(1-\nu)/p$ is the relaxation time of the FPM model, $\mu$ is the viscosity, $p=n k_B T_{tr}$ is the pressure, $n$ is the number density, and $k_B$ is the Boltzmann constant. The relaxation temperatures are defined to be consistent with the Landau-Teller equation:
\begin{equation}
    e_{tr}(T_{tr}^{rel}) = e_{tr}(T_{tr}) 
        + \frac{\tau}{2\tau_{rot}} \Big( e_{rot}(T_{rot}) - e_{rot}(T_{tr}) \Big) 
        + \frac{\tau}{2\tau_{vib}} \Big( e_{vib}(T_{vib}) - e_{vib}(T_{tr}) \Big) ,
\label{eq:T_tr_FPM}
\end{equation}
\begin{equation}
    e_{rot}(T_{rot}^{rel}) = e_{rot}(T_{rot}) 
        + \frac{\tau}{2\tau_{rot}} \Big( e_{rot}(T_{tr}) - e_{rot}(T_{rot}) \Big) ,
\label{eq:T_rot_FPM}
\end{equation}
\begin{equation}
    e_{vib}(T_{vib}^{rel}) = e_{vib}(T_{vib}) 
        + \frac{\tau}{2\tau_{vib}} \Big( e_{vib}(T_{tr}) - e_{vib}(T_{vib}) \Big) .
\label{eq:T_vib_FPM}
\end{equation}
The FPM model yields the following $\mathrm{Pr}$:
\begin{equation}
   \mathrm{Pr} = \frac{3}{2 (1-\nu)} .
\end{equation}
By setting $\nu = 1-\frac{3}{2\mathrm{Pr}}$, the correct $\mathrm{Pr}$ is recovered.

\subsection{\label{subsec:FPM_method} Particle evolution scheme in the FPM model}


For the numerical implementation of the FPM model, the stochastic trajectories of each particle’s position, velocity, rotational energy, and vibrational energy need to be tracked. The time evolution of position, velocity, and rotational energy, governed by the FP equation, is modeled using an equivalent Langevin equation \cite{Kim24-FPM}:
\begin{equation}
    dx_i=c_idt ,
\end{equation}
\begin{equation}
    dc_i=A_{tr,\, i} \, dt+\sqrt{2}d_{tr,\, ij} \, dW_{j} ,
\end{equation}
\begin{equation}
    d\varepsilon_{rot}=A_{rot} \, dt+\sqrt{2 D_{rot}} \, dW ,
\end{equation}
where $d_{tr,\, ij}$ is an element of a matrix satisfying $d_{tr,\, ik}\,d_{tr,\, jk}=D_{tr,\, ij}$, and $dW$ denotes the standard Wiener process. Based on Gorji's time integration scheme, the time evolution of the particle position and velocity is given as follows \cite{Gorji14_efficient-algorithm-FP}:
\begin{equation}
    x_i^{n+1}=x_i^n + c_i^n \Delta t ,
\end{equation}
\begin{equation}
    c_i^{n+1} 
    = U_i^n 
    + C_i^n \exp{(-\frac{\Delta t}{\tau})} 
    + \sqrt{\tau \Big( 1-\exp{(-\frac{2\Delta t}{\tau})} \Big)} d_{tr,\, ij} G_j,
\label{eq:conventionial-velo}
\end{equation}
where $G_j$ is a standard Gaussian random number, $\Delta t$ denotes the time step, and superscripts $n$ and $n+1$ represent values at times $t$ and $t + \Delta t$, respectively. To ensure the positivity of the rotational energy, the modified Milstein scheme is adopted \cite{Kim24-FPM}:
\begin{equation}
    \varepsilon_{rot}^{n+1}
    =
    \frac{RT_{rot}^{rel,\,n}}{2} \Big( 1 - \exp{(-\frac{2\Delta t}{\tau})} \Big)
    \,+\, 
    \Bigg(
        \sqrt{\varepsilon_{rot}^{n}} \, \exp{(-\frac{\Delta t}{\tau})}
    + \sqrt{\frac{RT_{rot}^{rel,\,n}}{2} \Big(1 - \exp{(-\frac{2\Delta t}{\tau})} \Big)} \, G
    \Bigg)^2 
    .
\label{eq:conventionial-erot}
\end{equation}
Gillespie's direct method describes the stochastic trajectory of discrete vibrational energy levels based on the master equation \cite{Gillespie76_direct}. At each time step, the cumulative time spent on vibrational updates, denoted by $t_{vib}$, is initialized to zero. The time to the next vibrational transition is sampled as $\Delta t_{vib} = -\ln(\mathbb{U}_1)/(\omega_{I,I+1} + \omega_{I,I-1})$, where $\mathbb{U}_1$ is a random number sampled from the uniform distribution $\mathrm{Unif}(0,1)$. If $t_{vib} + \Delta t_{vib} > \Delta t$, the process terminates. Otherwise, a second uniform random number $\mathbb{U}_2$ is sampled to determine the transition direction: the vibrational energy level increases to $I+1$ if $\mathbb{U}_2 < \omega_{I,I+1}/(\omega_{I,I+1} + \omega_{I,I-1})$, and decreases to $I-1$ otherwise. The cumulative time is then updated as $t_{vib} \leftarrow t_{vib} + \Delta t_{vib}$, and the procedure is repeated. In the following, these evolution schemes for particle velocity, rotational energy, and vibrational energy are referred to as the FPM method.

\subsection{\label{subsec:limitation} Limitations of the particle evolution scheme in the FPM model}


The FPM method accurately describes diatomic gas flows for sufficiently small time steps \cite{Kim24-FPM}. However, predictions for translational, rotational, and vibrational energies, as well as viscous stress, by the FPM method exhibit first-order temporal accuracy, leading to greater numerical errors for larger time steps. The changes in macroscopic moments after a time step $\Delta t$ are referred to as moment relaxations. For the FPM method, the moment relaxations for energy and heat flux in each mode, and for viscous stress, are derived in Appendix \ref{appendix:FPM} and summarized below:
\begin{multline}
    e_{tr}(T_{tr}^{n+1})
    = e_{tr} (T_{tr}^n) +
    \Bigg[
        \frac{ e_{rot}(T_{rot}^n) - e_{rot}(T_{tr}^n) }{\tau_{rot}}
        +
        \frac{ e_{vib}(T_{vib}^n) - e_{vib}(T_{tr}^n) }{\tau_{vib}}
    \Bigg]
    \Delta t
    \\[0.5em] 
    - \frac{1}{\tau}
    \Bigg[ 
        \frac{ e_{rot}(T_{rot}^n) - e_{rot}(T_{tr}^n) }{\tau_{rot}}
        +
        \frac{ e_{vib}(T_{vib}^n) - e_{vib}(T_{tr}^n) }{\tau_{vib}}
    \Bigg]
    \Delta t^2
    + \mathcal{O}(\Delta t^3)
    ,
\label{eq:FPM_etr} 
\end{multline}
\begin{equation}
    e_{rot}(T_{rot}^{n+1})
    =
    e_{rot}(T_{rot}^n)
    + 
    \Bigg[
        \frac{ e_{rot}(T_{tr}^n) - e_{rot}(T_{rot}^n) }{\tau_{rot}}
    \Bigg]
    \Delta t
    + 
    \Bigg[
        \frac{ e_{rot}(T_{tr}^n) - e_{rot}(T_{rot}^n) }{\tau \, \tau_{rot}}
    \Bigg]
    \Delta t^2
    +
    \mathcal{O} (\Delta t^3)
    ,
\end{equation}
\begin{equation}
    e_{vib}(T_{vib}^{n+1})
    =
    e_{vib}(T_{vib}^n)
    + 
    \Bigg[
        \frac{ e_{vib}(T_{tr}^n) - e_{vib}(T_{vib}^n) }{\tau_{vib}}
    \Bigg]
    \Delta t
    + 
    \Bigg[
        \frac{ e_{vib}(T_{tr}^n) - e_{vib}(T_{vib}^n) }{\tau \, \tau_{vib}}
    \Bigg]
    \Delta t^2
    +
    \mathcal{O} (\Delta t^3)
    ,
\end{equation}
\begin{equation}
    \sigma_{ij}^{n+1}
    = \sigma_{ij}^n
    \bigg( 
        1 
        - \frac{2(1-\nu)}{\tau} \Delta t
        + \frac{2(1-\nu)}{\tau^2} \Delta t^2
    \bigg)
    + \mathcal{O} (\Delta t^3)
    ,
\end{equation}
\begin{equation}
    q_{tr,\, i}^{n+1}
    = q_{tr,\, i}^n
    \Bigg[
        1
        - \frac{3}{\tau}\Delta t
        + \frac{9}{2\tau^2} \Delta t^2
    \Bigg]
    + \mathcal{O}(\Delta t^3)
        ,
\label{eq:FPM_qtr} 
\end{equation}
\begin{equation}
    q_{rot,\, i}^{n+1}
    = q_{rot,\, i}^n
    \Bigg[
        1
        - \frac{3}{\tau}\Delta t
        + \frac{9}{2\tau^2} \Delta t^2
    \Bigg]
    + \mathcal{O}(\Delta t^3)
        .
\label{eq:FPM_qrot} 
\end{equation}
\begin{equation}
    q_{vib,\, i}^{n+1}
    = q_{vib,\, i}^n
    \Bigg[
        1
        - \frac{3}{\tau}\Delta t
        + \frac{9}{2\tau^2} \Delta t^2
    \Bigg]
    + \mathcal{O}(\Delta t^3)
        .
\label{eq:FPM_qvib} 
\end{equation}
To assess the temporal accuracy of the FPM method, these moment relaxations are compared against the analytical time evolution of moments in the FPM model. The moment production terms are used to represent the time evolution of moments, since they satisfy the Landau–Teller equations, obey conservation laws, and preserve the transport properties in the Navier–Stokes limit \cite{Kim24-FPM}. Under homogeneous flow conditions, the time evolution of moments can be analytically obtained by performing a Taylor expansion based on the moment production terms, as detailed in Appendix \ref{appendix:Taylor}. The second-order Taylor expansions are presented for energy and heat flux in each mode, and for viscous stress:
\begin{multline}
    e_{tr}(T_{tr}^{n+1})
    =
    e_{tr}(T_{tr}^n)
    + 
    \Bigg[
        \frac{ e_{rot}(T_{rot}^n) - e_{rot}(T_{tr}^n) }{\tau_{rot}}
        +
        \frac{ e_{vib}(T_{vib}^n) - e_{vib}(T_{tr}^n) }{\tau_{vib}}
    \Bigg]
    \Delta t
    \\[0.5em] -
    \frac{1}{2}
    \Bigg[
        \bigg( 
            \frac{1}{\tau_{rot}^2} 
            +
            \frac{1}{\tau_{rot} ^2} 
            \frac{c_{v, \, rot}}{c_{v,\, tr}}
            +
            \frac{1}{\tau_{rot} \tau_{vib}} 
            \frac{c_{v, \, vib}(T_{tr}^n)}{c_{v,\, tr}}
        \bigg)
        \bigg( e_{rot}(T_{rot}^n) - e_{rot}(T_{tr}^n) \bigg)
        \\[0.5em]  +
        \bigg(
            \frac{1}{\tau_{vib} ^2} 
            +
            \frac{1}{\tau_{vib}^2} 
            \frac{c_{v, \, vib}(T_{tr}^n)}{c_{v,\, tr}}
            +
            \frac{1}{\tau_{rot} \tau_{vib}} 
            \frac{c_{v, \, rot}}{c_{v,\, tr}}
        \bigg)
        \bigg(      
             e_{vib}(T_{vib}^n) - e_{vib}(T_{tr}^n)
        \bigg)
    \Bigg]
    \Delta t^2
    +
    \mathcal{O}(\Delta t^3)
    ,
\label{eq:analytic_etr}
\end{multline}
\begin{equation}
\begin{aligned}
    e_{rot}(T_{rot}^{n+1})
    =
    e_{rot}(T_{rot}^n)
     & + 
    \Bigg[
        \frac{ e_{rot}(T_{tr}^n) - e_{rot}(T_{rot}^n) }{\tau_{rot}}
    \Bigg]
    \Delta t
    \\[0.5em] & -
    \frac{1}{2}
    \Bigg[
        \bigg( 
            \frac{1}{\tau_{rot}^2} 
            +
            \frac{1}{\tau_{rot}^2} 
            \frac{c_{v, \, rot}}{c_{v,\, tr}}
        \bigg)
        \bigg( e_{rot}(T_{tr}^n) - e_{rot}(T_{rot}^n) \bigg)
        \\[0.5em] & +
        \bigg(
            \frac{1}{\tau_{rot} \tau_{vib}} 
            \frac{c_{v, \, rot}}{c_{v,\, tr}}
        \bigg)
        \bigg(      
             e_{vib}(T_{tr}^n) - e_{vib}(T_{vib}^n)
        \bigg)
    \Bigg]
    \Delta t^2
    +
    \mathcal{O}(\Delta t^3)
    ,
\end{aligned}
\end{equation}
\begin{equation}
\begin{aligned}
    e_{vib}(T_{vib}^{n+1})
    =
    e_{vib}(T_{vib}^n)
    & + 
    \Bigg[
        \frac{ e_{vib}(T_{tr}^n) - e_{vib}(T_{vib}^n) }{\tau_{vib}}
    \Bigg]
    \Delta t
    \\[0.5em] &  -
    \frac{1}{2}
    \Bigg[
        \bigg( 
            \frac{1}{\tau_{vib}^2} 
            +
            \frac{1}{\tau_{vib}^2} 
            \frac{ c_{v, \, vib} (T_{tr}^n) }{c_{v,\, tr}}
        \bigg)
        \bigg( e_{vib}(T_{tr}^n) - e_{vib}(T_{vib}^n) \bigg)
        \\[0.5em] &  +
        \bigg(
            \frac{1}{\tau_{rot} \tau_{vib}} 
            \frac{c_{v, \, vib} (T_{tr}^n)}{c_{v,\, tr}}
        \bigg)
        \bigg(      
             e_{rot}(T_{tr}^n) - e_{rot}(T_{rot}^n)
        \bigg)
    \Bigg]
    \Delta t^2
    +
    \mathcal{O}(\Delta t^3)
    ,
\end{aligned}
\end{equation}
\begin{equation}
    \sigma_{ij}^{n+1}
    =
    \sigma_{ij}^n
    \Bigg[
    1 -\frac{2(1-\nu)}{\tau} \Delta t + \frac{2(1-\nu)^2}{\tau^2} \Delta t^2
    \Bigg]
    + \mathcal{O}(\Delta t^3)
    ,
\label{eq:analytic_sigma}
\end{equation}
\begin{equation}
    q_{tr,\, i}^{n+1}
    =
    q_{tr,\, i}^n
    \Bigg[
        1 - \frac{3}{\tau}\Delta t + \frac{9}{2\tau^2} \Delta t^2
    \Bigg]
    + \mathcal{O}(\Delta t^3)
    ,
\label{eq:analytic_qtr}
\end{equation}
\begin{equation}
    q_{rot,\, i}^{n+1}
    =
    q_{rot,\, i}^n
    \Bigg[
        1 - \frac{3}{\tau}\Delta t + \frac{9}{2\tau^2} \Delta t^2
    \Bigg]
    + \mathcal{O}(\Delta t^3)
    ,
\label{eq:analytic_qrot}
\end{equation}
\begin{equation}
    q_{vib,\, i}^{n+1}
    =
    q_{vib,\, i}^n
    \Bigg[
        1 - \frac{3}{\tau}\Delta t + \frac{9}{2\tau^2} \Delta t^2
    \Bigg]
    + \mathcal{O}(\Delta t^3)
    .
\label{eq:analytic_qvib}
\end{equation}
A comparison between the moment relaxation terms from the FPM method, given in Equations (\ref{eq:FPM_etr})-(\ref{eq:FPM_qvib}), and the Taylor expansions derived from the moment production terms, presented in Equations (\ref{eq:analytic_etr})-(\ref{eq:analytic_qvib}), shows that the relaxations of each energy mode and viscous stress remain first-order accurate. Only the heat flux relaxations match the second-order Taylor expansions. This result highlights that the FPM method is limited to first-order temporal accuracy, and using a large time step may result in significant numerical inaccuracies in the evolution of energy and stress. It is also important to recognize that the first-order temporal accuracy originates from the assumption that moments remain constant during each time step and from the use of operator splitting in the FPM method \cite{Kim24_USP-FP, Cui25_MSP}. Temporal accuracy can be improved by employing exact time integration schemes in homogeneous flows \cite{Jenny10_linearFP, Cui25_MSP}, and by correcting numerical errors arising from the operator splitting \cite{Fei20_USP, Kim24_USP-FP, Pfeiffer25_CN-BGK}.


In the FPM method, cell-averaged macroscopic properties are directly used to update particles without accounting for spatial gradients, which leads to inaccurate estimation of the macroscopic properties at the actual positions of particles. As a result, the FPM method exhibits first-order spatial accuracy, and using a large cell can lead to significant numerical errors, particularly in a region with a strong gradient such as shock waves or near boundaries \cite{Kim24_FP_poly}. Spatial accuracy can be improved by accounting for the spatial gradients of macroscopic properties \cite{Kim24_USP-FP, Kim24_FP_poly}.

\section{\label{sec:usp-fpm}Unified stochastic particle Fokker–Planck–Master method for diatomic gas flows}

\subsection{\label{subsec:temporal}Temporal evolution}

\subsubsection{\label{subsubsec:moment}Second-order moment relaxations}


To achieve second-order temporal accuracy in diatomic gas flows, this study corrects the moment relaxations of the FPM method. Since moments up to the heat flux are sufficient to recover the Navier–Stokes equations in the continuum limit \cite{Gombosi94_Gaskinetic-book, Kim24_USP-FP}, the correction focuses on moments up to the heat flux. The moment relaxation terms are analytically derived using a kinetic model that ensures second-order temporal accuracy for diatomic gases. Among existing models, the unified stochastic particle BGK (USP-BGK) method is selected, as it is the only known particle-based kinetic model that achieves second-order accuracy for diatomic gases \cite{Fei22_USP-poly, Tian23_USP-poly-SPARTACUS}. However, the original USP-BGK method is built on Dauvois’s polyatomic ES-BGK model \cite{Dauvois21-ES-BGK-poly}, which does not satisfy the Landau–Teller equations. To address this limitation, this study reformulates the USP-BGK method using Mathiaud’s polyatomic ES-BGK model \cite{Mathiaud22-ES-BGK-poly}, which satisfies the Landau–Teller equations. This enables the USP-BGK method to achieve second-order temporal accuracy while correctly capturing internal energy relaxation. The resulting moment relaxation terms for energy, viscous stress, and heat flux are derived from the USP-BGK method and are detailed in Appendix \ref{appendix:USPBGK}. The final results are summarized as follows:
\begin{equation}
    e_{tr} (T_{tr}^{n+1}) 
    = 
    e_{tr}(T_{tr}^n)
    - \gamma_{tr-rot}  \Big(T_{tr}^n - T_{rot}^n \Big) 
    - \gamma_{tr-vib}  \Big(T_{tr}^n - T_{vib}^n \Big) 
    ,
\label{eq:e_tr_relax}
\end{equation}
\begin{equation}
    e_{rot} (T_{rot}^{n+1}) 
    = 
    e_{rot}(T_{rot}^n)
    + \gamma_{tr-rot} \Big(T_{tr}^n - T_{rot}^n \Big) 
    - \gamma_{rot-vib} \Big(T_{rot}^n - T_{vib}^n \Big) 
    ,
\label{eq:e_rot_relax}
\end{equation}
\begin{equation}
    e_{vib}(T_{vib}^{n+1}) 
    = 
    e_{vib}(T_{vib}^n)
    + \gamma_{tr-vib} \Big(T_{tr}^n - T_{vib}^n \Big) 
    + \gamma_{rot-vib} \Big(T_{rot}^n - T_{vib}^n \Big) 
    ,
\label{eq:e_vib_relax}
\end{equation}
\begin{equation}
    \sigma_{ij}^{n+1} 
    = 
    \sigma_{ij}^{n} 
    \,
    \Big( \frac{2\mu/p - \Delta t}{2\mu/p + \Delta t} \Big) 
    ,
\label{eq:sigma_relax}
\end{equation}
\begin{equation}
    q_{tr,\,i}^{n+1} 
    = 
    q_{tr,\,i}^{n} 
    \,
    \Big( \frac{2\mu/p - \mathrm{Pr} \Delta t}{2\mu/p + \mathrm{Pr}  \Delta t} \Big)
    ,
\label{eq:qtr_relax}
\end{equation}
\begin{equation}
    q_{rot,\,i}^{n+1} 
    = 
    q_{rot,\,i}^{n} 
    \,
    \Big( \frac{2\mu/p - \mathrm{Pr} \Delta t}{2\mu/p + \mathrm{Pr}  \Delta t} \Big)
    ,
\label{eq:qrot_relax}
\end{equation}
\begin{equation}
    q_{vib,\,i}^{n+1} 
    = q_{vib,\,i}^{n} \,\Big( \frac{2\mu/p - \mathrm{Pr} \Delta t}{2\mu/p + \mathrm{Pr}  \Delta t} \Big)
    ,
\label{eq:qvib_relax}
\end{equation}
where coefficients $\gamma_{tr-rot}$, $\gamma_{tr-vib}$, $\gamma_{rot-vib}$ are given below:
\begin{equation}
    \gamma_{tr-rot} = 
    \dfrac
    {
        \dfrac{2\Delta t}{\Delta t + 2\tau_{rot}}
        c_{v,\,tr} \, c_{v,\,rot}
    }
    {
        c_{v,\,tr} 
        + \dfrac{\Delta t}{\Delta t + 2\tau_{rot}} c_{v,\,rot} 
        + \dfrac{\Delta t}{\Delta t + 2\tau_{vib}} c_{v,\,vib}(T_1)
    }
    ,
\end{equation}
\begin{equation}
    \gamma_{tr-vib} = 
    \dfrac
        {
            \dfrac{2\Delta t}{\Delta t + 2\tau_{vib}} 
            c_{v,\,tr} \, c_{v,\,vib}(T_1)
        }
        {
            c_{v,\,tr} 
            + \dfrac{\Delta t}{\Delta t + 2\tau_{rot}} c_{v,\,rot} 
            + \dfrac{\Delta t}{\Delta t + 2\tau_{vib}} c_{v,\,vib}(T_1)
        }
    ,
\end{equation}
\begin{equation}
    \gamma_{rot-vib} = 
    \dfrac
    {
        \dfrac{2\Delta t}{\Delta t + 2\tau_{rot}}
        \dfrac{\Delta t}{\Delta t + 2\tau_{vib}}
        \, c_{v,\,rot} 
        \, c_{v,\,vib}(T_1)
    }
    {
        c_{v,\,tr} 
            + \dfrac{\Delta t}{\Delta t + 2\tau_{rot}} c_{v,\,rot} 
            + \dfrac{\Delta t}{\Delta t + 2\tau_{vib}} c_{v,\,vib}(T_1)
    }
    .
\end{equation}
The temperature $T_1$ satisfies the following equation
\begin{equation}
    c_{v,\,vib}(T_1)
    =
    \frac{e_{vib}(\check{T}_{tr})-e_{vib}(T_{vib})}{\check{T}_{tr}-T_{vib}}
    ,
\end{equation}
and the rescaled translational temperature $\check{T}_{tr}$ is defined by
\begin{equation}
    e_{tr}(\check{T}_{tr})
    = 
    e_{tr}({T}_{tr})
    +
    \frac{\gamma_{tr-rot}}{2}    
    \Big( {T}_{rot} - T_{tr} \Big)
    +
    \frac{\gamma_{tr-vib}}{2}
    \Big( {T}_{vib} - T_{tr} \Big)
    .
\end{equation}

\subsubsection{\label{subsubsec:evolution}Particle evolution scheme of the USP-FPM method}


To achieve second-order temporal accuracy for diatomic gases, the USP-FPM method aims to reproduce the second-order energy, viscous stress, and heat flux relaxation terms, as given in Equations (\ref{eq:e_tr_relax})-(\ref{eq:qvib_relax}). The time integration scheme for the particle velocity is given by
\begin{equation}
    C_i^{n+1} 
    = C_i^n \alpha_{tr}
    + \sqrt{\tau \Big( 1-\alpha_{tr}^2 ) \Big)} d_{tr,\, ij} G_j
    ,
\label{eq:update_ci}
\end{equation}
which reduces to the FPM method when $\alpha_{tr} = \exp{(-\Delta t/\tau)}$. By following the same calculations in Appendix \ref{appendix:FPM}, the translational energy, viscous stress, and translational heat flux relaxations are re-expressed as follows:
\begin{equation}
    e_{tr}(T_{tr}^{n+1})
    =
    e_{tr}(T_{tr}^{n})
    \alpha_{tr}^2
    +
    e_{tr}(T_{tr}^{rel,\,n})
    (1-\alpha_{tr}^2)
    ,
\end{equation}
\begin{equation}
    \sigma_{ij}^{n+1}
    =
    \sigma_{ij}^{n}
    \Big( 
        \alpha_{tr}^2
        +
        \nu 
        (1 - \alpha_{tr}^2)
    \Big)
    ,
\end{equation}
\begin{equation}
    q_{tr,\,i}^{n+1}
    =
    q_{tr,\,i}^{n}
    \, \alpha_{tr}^3
    .
\end{equation}
To reproduce the second-order relaxations of translational energy, viscous stress, and translational heat flux, the following set of equations needs to be satisfied:
\begin{equation}
    e_{tr}(T_{tr}^{n})
    \alpha_{tr}^2
    +
    e_{tr}(T_{tr}^{rel,\,n})
    (1-\alpha_{tr}^2)
    =
    e_{tr}(T_{tr}^n)
    - \gamma_{tr-rot}  \Big(T_{tr}^n - T_{rot}^n \Big) 
    - \gamma_{tr-vib}  \Big(T_{tr}^n - T_{vib}^n \Big) 
    ,
\end{equation}
\begin{equation}
    \sigma_{ij}^{n}
    \Big( 
        \alpha_{tr}^2
        +
        \nu 
        (1 - \alpha_{tr}^2)
    \Big)
    =
    \sigma_{ij}^{n} 
    \,
    \bigg( \frac{2\mu/p - \Delta t}{2\mu/p + \Delta t} \bigg) 
    ,
\end{equation}
\begin{equation}
    q_{tr,\,i}^{n}
    \, \alpha_{tr}^3
    =
    q_{tr,\,i}^{n} 
    \,
    \bigg( \frac{2\mu/p - \mathrm{Pr} \Delta t}{2\mu/p + \mathrm{Pr}  \Delta t} \bigg)
    .
\end{equation}
Consequently, $\alpha_{tr}$, $\nu$, and $T_{tr}^{rel}$ in the USP-FPM method are uniquely determined as follows:
\begin{equation}
    \alpha_{tr}
    =
    \bigg( \frac{2\mu/p - \mathrm{Pr} \Delta t}{2\mu/p + \mathrm{Pr}  \Delta t} \bigg) 
    ^{1/3}
    ,
\label{eq:USP-FPM_alpha_tr}
\end{equation}
\begin{equation}
    \nu
    =
    \frac{1}{1-\alpha_{tr}^2}
    \bigg( \frac{2\mu/p - \Delta t}{2\mu/p + \Delta t} - \alpha_{tr}^2 \bigg)
    ,
\label{eq:USP-FPM_nu}
\end{equation}
\begin{equation}
    e_{tr}(T_{tr}^{rel})
    =
    e_{tr}(T_{tr})
    - \frac{\gamma_{tr-rot}}{1-\alpha_{tr}^2}  \Big(T_{tr} - T_{rot} \Big) 
    - \frac{\gamma_{tr-vib}}{1-\alpha_{tr}^2}  \Big(T_{tr} - T_{vib} \Big) 
    .
\label{eq:T_tr_USP}
\end{equation}


Similarly, the time integration scheme for the rotational energy is expressed as
\begin{equation}
    \varepsilon_{rot}^{n+1}
    =
    \frac{RT_{rot}^{rel,\,n}}{2} \Big( 1 - \alpha_{rot}^2 \Big)
    \,+\, 
    \bigg(
        \sqrt{\varepsilon_{rot}^{n}} \, \alpha_{rot}
    + \sqrt{\frac{RT_{rot}^{rel,\,n}}{2} \Big(1 - \alpha_{rot}^2 \Big)} \, G
    \bigg)^2 
    .
\label{eq:update_erot}
\end{equation}
The choice $\alpha_{rot}=\exp{(-\Delta t/\tau)}$ recovers the FPM method. The rotational energy and rotational heat flux can be written as:
\begin{equation}
    e_{rot}(T_{rot}^{n+1})
    =
    e_{rot}(T_{rot}^{rel,\,n}) \Big( 1 - \alpha_{rot}^2 \Big)
    +
    e_{rot}(T_{rot}^{n}) \, \alpha_{rot}^2
    ,
\end{equation}
\begin{equation}
    q_{rot,\, i}^{n+1}
    = q_{rot,\, i}^n \, \alpha_{tr}\,\alpha_{rot}^2
    .
\end{equation}
To match the second-order relaxations of rotational energy and heat flux, the following set of equations needs to be satisfied:
\begin{equation}
    e_{rot}(T_{rot}^{rel,\,n}) \Big( 1 - \alpha_{rot}^2 \Big)
    +
    e_{rot}(T_{rot}^{n}) \, \alpha_{rot}^2
    = 
    e_{rot}(T_{rot}^n)
    + \gamma_{tr-rot} \Big(T_{tr}^n - T_{rot}^n \Big) 
    - \gamma_{rot-vib} \Big(T_{rot}^n - T_{vib}^n \Big) 
    ,
\end{equation}
\begin{equation}
    q_{rot,\, i}^n \, \alpha_{tr}\,\alpha_{rot}^2
    = 
    q_{rot,\,i}^{n} 
    \,
    \Big( \frac{2\mu/p - \mathrm{Pr} \Delta t}{2\mu/p + \mathrm{Pr}  \Delta t} \Big)
    .
\end{equation}
Correspondingly, $\alpha_{rot}$ and $T_{rot}^{rel}$ are determined as follows:
\begin{equation}
    \alpha_{rot}
    =
    \bigg( \frac{2\mu/p - \mathrm{Pr} \Delta t}{2\mu/p + \mathrm{Pr}  \Delta t} \bigg) 
    ^{1/3}
    ,
\label{eq:USP-FPM_alpha_rot}
\end{equation}
\begin{equation}
    e_{rot}(T_{rot}^{rel})
    =
    e_{rot}(T_{rot})
    + \frac{\gamma_{tr-rot}}{1-\alpha_{rot}^2} \Big(T_{tr} - T_{rot} \Big) 
    - \frac{\gamma_{rot-vib}}{1-\alpha_{rot}^2} \Big(T_{rot} - T_{vib} \Big) 
    .
\label{eq:T_rot_USP}
\end{equation}


Within each time step $\Delta t$, transitions between discrete vibrational energy levels may occur. In the FPM method, Gillespie's direct method provides an exact time integration scheme for the vibrational energy and offers an analytic realization of the FPM model \cite{Gillespie07_SSA_exact}. However, because the direct method exactly simulates individual transitions based on the master equation, the resulting macroscopic moments cannot be explicitly adjusted without modifying the equation or algorithm. This makes it difficult to reproduce the second-order moment relaxations. Unlike the direct method, the tau-leaping method approximates the number of transitions over a given time interval, while preserving key statistical properties \cite{Gillespie01_tau-leaping}. In this study, a modified tau-leaping method is employed to approximate the stochastic evolution of the vibrational energy and to reproduce the second-order moment relaxations \cite{Chatterjee05_modified_tau-leaping}. The downward transitions are modeled using a binomial distribution to prevent transition to negative energy levels, consistent with the modified tau-leaping method of Chatterjee et al. \cite{Chatterjee05_modified_tau-leaping}. In contrast, the upward transitions in vibrational energy levels are modeled using a negative binomial distribution to capture the overdispersed vibrational energy distribution, deviating from Chatterjee's method. The update scheme for the vibrational energy level is given by:
\begin{equation}
    I^{n+1}=I^{n} + \textrm{NB}(r_{NB}, \,p_{NB}) - \textrm{B}(n_{B}, \, p_{B})
    ,
\label{eq:update_evib}
\end{equation}
where $\textrm{NB}(r_{NB},\, p_{NB})$ denotes samples from a negative binomial distribution with parameters $r_{NB}$ and $p_{NB}$, and $\textrm{B}(n_{B},\, p_{B})$ denotes samples from a binomial distribution with parameters $n_{B}$ and $p_{B}$. These parameters are defined as follows:
\begin{equation}
    r_{NB}
    =
    \Big( I^n + 1 \Big)
    ,
\end{equation}
\begin{equation}
    p_{NB}
    =
    \Bigg( 
        1 
        +
        \frac{  1 }
             { \exp{(\Theta_{vib}/T_{vib}^{rel,\,n})} - 1 }
        \, \Big(1 - \alpha_{vib}^2 \Big)
    \Bigg) ^{-1}
    ,
\end{equation}
\begin{equation}
    n_{B}
    =
    I^n 
    ,
\end{equation}
\begin{equation}
    p_{N}
    =
    \frac{ \exp{(\Theta_{vib}/T_{vib}^{rel,\,n})}     }
         { \exp{(\Theta_{vib}/T_{vib}^{rel,\,n})} - 1 }
    \, \Big(1 - \alpha_{vib}^2 \Big)
    .
\end{equation}
The vibrational energy and vibrational heat flux relaxations are obtained as follows:
\begin{equation}
\begin{aligned}
    e_{vib}(T_{vib}^{n+1})
    &=
    \langle \varepsilon_{vib}^{n+1} \rangle
    =
    \langle I^{n+1} R\Theta_{vib} \rangle
    \\[0.5em] & =
    \langle 
        I^{n} R\Theta_{vib}  
        +
        \mathbb{E} [\textrm{NB}(r_{NB}, \,p_{NB})] \, R\Theta_{vib}  
        -
        \mathbb{E} [\textrm{B}(n_{B}, \, p_{B})] \, R\Theta_{vib} 
    \rangle
    \\[0.5em] & =
    \langle 
        I^{n} R\Theta_{vib}  
        +
        r_{NB} \frac{1 - p_{NB}}{p_{NB}} R\Theta_{vib}  
        -
        n_{B} p_{B} R\Theta_{vib} 
    \rangle  
    \\[0.5em] & =
    \Big< 
        I^{n} R\Theta_{vib} \alpha_{vib}^2
        +
        \frac{ R\Theta_{vib} }
             { \exp{(\Theta_{vib}/T_{vib}^{rel,\,n})} - 1 }
        \, \Big(1 - \alpha_{vib}^2 \Big)
    \Big>  
    \\[0.5em] & =
    e_{vib}(T_{vib}^{n}) 
    \, \alpha_{vib}^2
    +
    e_{vib}(T_{vib}^{rel,\,n}) 
    \, \Big(1 - \alpha_{vib}^2 \Big)
    ,
\end{aligned}
\end{equation}
\begin{equation}
\begin{aligned}
    q_{vib,\,i}^{n+1}
    &=
    \langle C_i^{n+1} \varepsilon_{vib}^{n+1} \rangle
    \\[0.5em] &=
    \Bigg<
    \bigg( 
        C_i^n \alpha_{tr}
        + \sqrt{\tau \Big( 1-\alpha_{tr}^2 ) \Big)} d_{tr,\, ij} G_j
    \bigg)
    \bigg( 
        \varepsilon_{vib}^{n} 
        \, \alpha_{vib}^2
        +
        e_{vib}(T_{vib}^{rel,\,n}) 
        \, \Big(1 - \alpha_{vib}^2 \Big)
    \bigg)
    \Bigg>
    \\[0.5em] &=
    q_{vib,\,i}^{n}
    \, \alpha_{tr} 
    \, \alpha_{vib}^2
    ,
\end{aligned}
\end{equation}
where $\mathbb{E}[\cdot]$ denotes a expectation value. The following set of equations should be satisfied to reproduce the second-order relaxations of vibrational energy and heat flux:
\begin{equation}
    e_{vib}(T_{vib}^{n}) 
    \, \alpha_{vib}^2
    +
    e_{vib}(T_{vib}^{rel,\,n}) 
    \, \Big(1 - \alpha_{vib}^2 \Big)
    = 
    e_{vib}(T_{vib}^n)
    + \gamma_{tr-vib} \Big(T_{tr}^n - T_{vib}^n \Big) 
    + \gamma_{rot-vib} \Big(T_{rot}^n - T_{vib}^n \Big) 
    ,
\end{equation}
\begin{equation}
    q_{vib,\,i}^{n}
    \, \alpha_{tr} 
    \, \alpha_{vib}^2
    = 
    q_{vib,\,i}^{n} 
    \,
    \Big( \frac{2\mu/p - \mathrm{Pr} \Delta t}{2\mu/p + \mathrm{Pr}  \Delta t} \Big)
    .
\end{equation}
Accordingly, $\alpha_{vib}$ and $T_{vib}^{rel}$ are given by:
\begin{equation}
    \alpha_{vib}
    =
    \bigg( \frac{2\mu/p - \mathrm{Pr} \Delta t}{2\mu/p + \mathrm{Pr}  \Delta t} \bigg) 
    ^{1/3}
    ,
\label{eq:USP-FPM_alpha_vib}
\end{equation}
\begin{equation}
    e_{vib}(T_{vib}^{rel}) 
    =
    e_{vib}(T_{vib}^n)
    + \frac{\gamma_{tr-vib}}{1-\alpha_{rot}^2} \Big(T_{tr}^n - T_{vib}^n \Big) 
    + \frac{\gamma_{rot-vib}}{1-\alpha_{rot}^2} \Big(T_{rot}^n - T_{vib}^n \Big) 
    .
\label{eq:T_vib_USP}
\end{equation}
The two parameters $p_{NB}$ and $p_{B}$ must lie within the interval $[0, 1]$, as required by the definition of their respective distributions. The parameter $p_{NB}$ satisfies this condition regardless of the time step $\Delta t$. In contrast, for sufficiently large $\Delta t$, the value $p_{B}$ may exceed unity. Although this issue is avoided in this study by choosing $\Delta t$ to satisfy $p_{B} \leq 1$, further investigation is required to address this limitation, for example, by developing an alternative method that remains valid in this time step regime.

\subsubsection{\label{subsubsec:positivity}Discussion on positivity of diffusion tensor}


In the FPM model, the diffusion tensor must remain positive definite to prevent unphysical behavior \cite{Kim24-FPM}. The rotational diffusion tensor is always positive definite. In contrast, the translational diffusion tensor may not remain positive definite under highly anisotropic flow conditions. The condition for positive definiteness of the translational diffusion tensor is given by \cite{Kim24-FPM}:
\begin{equation}
    -\frac{RT_{tr}^{\mathrm{rel}}}{\lambda_{\mathrm{max}}-RT_{tr}} 
    <\nu<
    \frac{RT_{tr}^{\mathrm{rel}}}{RT_{tr} - \lambda_{\mathrm{min}}} 
    ,
\end{equation}
where $\lambda_{max}$ and $\lambda_{min}$ represent maximum and minimum eigenvalues of $\boldsymbol{\Pi}$. When the flow is near equilibrium, the value of $\nu$ defined by Equation (\ref{eq:USP-FPM_nu}) lies within this admissible range, yielding the correct $\mathrm{Pr}$ and maintaining positive definiteness. However, in strongly anisotropic flows, the value of $\nu$ may fall outside this range, resulting in a loss of positive definiteness. In such cases, $\nu$ needs to be adjusted to prioritize positive definiteness over the exact preservation of $\mathrm{Pr}$:
\begin{equation}
    \nu = 
    \begin{cases}
        \max{\bigl( 
            \nu_{\mathrm{ref}}, 
            -\frac{RT_{tr}^{\mathrm{rel}}}{\lambda_{\mathrm{max}}-RT_{tr}} 
        \bigr)} & \text{if } \nu_{\mathrm{ref}} \leq 0 , \\[1em]
        \min{\bigl( 
            \nu_{\mathrm{ref}}, 
            \frac{RT_{tr}^{\mathrm{rel}}}{RT_{tr} - \lambda_{\mathrm{min}}}  
        \bigr)} & \text{otherwise}.
    \end{cases}
    ,
\label{eq:nu_USP_positivity}
\end{equation}
where $\nu_{ref}$ is computed from Equation (\ref{eq:USP-FPM_nu}). Although the FPM model cannot guarantee an exact $\mathrm{Pr}$ for every flow, second-order temporal accuracy can be maintained by adjusting $\alpha_{tr}$ as follows:
\begin{equation}
    \alpha_{tr}^2 
    = 
    \frac{ 1 }{\nu - 1} 
    \Bigl(
        \nu 
        - \frac{2\mu/p - \Delta t}{2\mu/p + \Delta t} 
    \Bigr)
    .
\label{eq:alpha_USP_positivity_1}
\end{equation}
This yields two possible values for $\alpha_{tr}$,
\begin{equation}
    \alpha_{tr,1} = \sqrt{\alpha_{tr}^2} 
    \text{ or } 
    \alpha_{tr,2} = -\sqrt{\alpha_{tr}^2} 
    .
\label{eq:alpha_USP_positivity_2}
\end{equation}
Each $\alpha$ leads to a distinct value of $\mathrm{Pr}$: 
\begin{equation}
    \mathrm{Pr}_1 = \frac{2\mu/p \, (1 - \alpha_{tr,1}^3)}{\Delta t \, (1 + \alpha_{tr,1}^3)} 
    \text{ or } 
    \mathrm{Pr}_2 = \frac{2\mu/p \, (1 - \alpha_{tr,2}^3)}{\Delta t \, (1 + \alpha_{tr,2}^3)}  
    ,
\label{eq:alpha_USP_positivity_3}
\end{equation}
Between two candidates, the value of $\alpha_{tr}$ that minimizes the deviation from the target $\mathrm{Pr}$ is selected. The values of $\alpha_{rot}$ and $\alpha_{vib}$ are also adjusted to be consistent with the selected $\alpha_{tr}$:
\begin{equation}
    \alpha_{rot}=\alpha_{vib}=\alpha_{tr}
    .
\label{eq:alpha_USP_positivity_4}
\end{equation}

\subsection{\label{subsec:spatial}Spatial reconstruction}


\begin{figure*}[b]
	\centering
	\begin{subfigure}{0.495\textwidth}
		\centering
		\includegraphics[width=1.\linewidth]{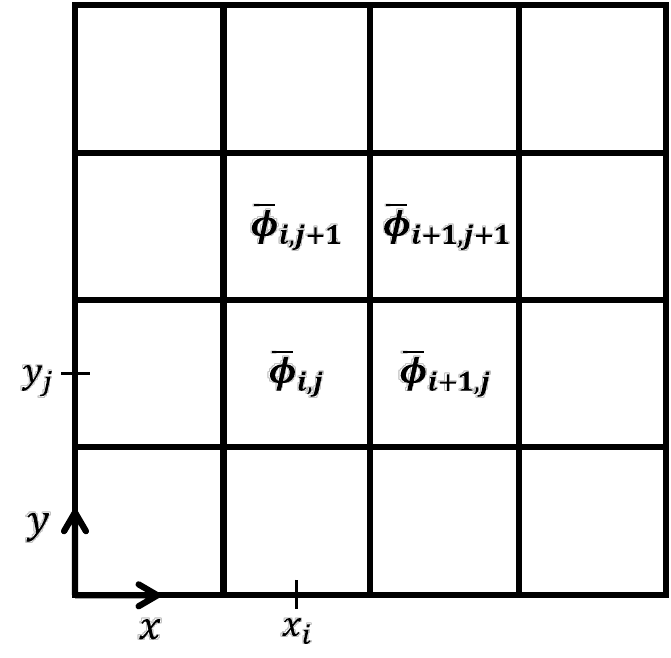}
            \caption{\label{subfig:poly_uniform} Uniform grid.}
	\end{subfigure}
	\begin{subfigure}{0.495\textwidth}
		\centering
		\includegraphics[width=1.\linewidth]{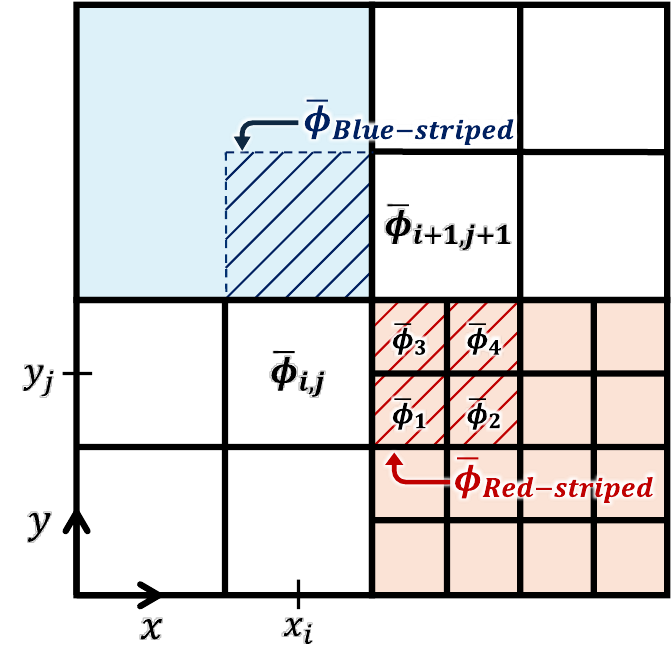}
            \caption{\label{subfig:poly_nonuniform} Non-uniform grid.}
	\end{subfigure}
	\caption{\label{fig:grid} Reconstruction stencils for uniform and non-uniform grids.}
\end{figure*}

To achieve second-order spatial accuracy in the USP-FPM method, the first-order polynomial reconstruction method is employed \cite{Kim24_FP_poly}. Figure \ref{subfig:poly_uniform} illustrates a uniform grid, where cell $\{i,j\}$ refers to the cell centered at $(x_i,y_j)$, and $\overline{\phi}_{i,j}$ represents the cell-averaged value of a macroscopic property $\phi$. In the uniform grid, the polynomial reconstruction method reconstructs a first-order polynomial $\phi(x,y)$ within each cell to approximate the macroscopic properties, centered at the target cell's centroid $(x_i,y_i)$:
\begin{equation}
    \phi(x,y)=C_{00} + C_{01} (x-x_i) + C_{10} (y - y_j) + C_{11} (x-x_i)(y-y_j) 
    .
\end{equation}
To determine the polynomial coefficients, four cell-averaged values, $\overline{\phi}_{i,j}$, $\overline{\phi}_{i+1,j}$, $\overline{\phi}_{i,j+1}$, and $\overline{\phi}_{i+1,j+1}$, are used, taken from the target cell and its three neighboring cells. This group of four cells is referred to as the stencil. To ensure that these four cell-averaged values are conserved, the following equation is derived \cite{Kim24_FP_poly}:
\begin{equation}
    \begin{aligned}
    \begin{bmatrix}
        C_{00} \\
        C_{10} \\
        C_{01} \\
        C_{11} \\
    \end{bmatrix}
    & =
    \begin{bmatrix}
          1 
        & 0 
        & 0 
        & 0 \\
          \frac{-1}{\Delta x} 
        & \frac{ 1}{\Delta x}
        & 0 
        & 0 \\
          \frac{-1}{\Delta y}  
        & 0 
        & \frac{ 1}{\Delta y} 
        & 0 \\
          \frac{ 1}{\Delta x \Delta y}  
        & \frac{-1}{\Delta x \Delta y}  
        & \frac{-1}{\Delta x \Delta y} 
        & \frac{ 1}{\Delta x \Delta y}  \\
    \end{bmatrix}
    \begin{bmatrix}
        \overline{\phi}_{i  ,j  } \\
        \overline{\phi}_{i+1,j  } \\
        \overline{\phi}_{i  ,j+1} \\
        \overline{\phi}_{i+1,j+1} \\
    \end{bmatrix} 
    ,
    \end{aligned}
\end{equation}
where $\Delta x$ and $\Delta y$ denote the target cell sizes in x- and y-directions, respectively. To reduce directional bias, four independent first-order reconstructions are performed in each quadrant around the cell center. When a boundary or surface is present, the stencil is selected from the opposite direction. Once the polynomial is reconstructed, macroscopic properties are interpolated to each particle's location. The interpolated macroscopic quantities include $U_i$, $T_{tr}$, $T_{rot}$, $T_{vib}$, and $\Pi_{ij}$, following the approach of Kim et al \cite{Kim24_FP_poly}.


When mesh refinement occurs, the sizes of neighboring cells may differ from the size of the target cell. Figure \ref{subfig:poly_nonuniform} illustrates a non-uniform grid obtained by applying mesh refinement to the uniform grid shown in Figure \ref{subfig:poly_uniform}. The region shaded in light-blue represents a coarsened cell, created by merging four original cells into one. The region shaded in light-orange represents a refined area, where each original cell is subdivided into four refined cells. The blue-striped region corresponds to the lower-right quadrant of the coarsened cell, which spatially coincides with the region covered by cell $\{i,j+1\}$ in the uniform grid. Similarly, the red-striped region corresponds to the area originally associated with cell $\{i+1,j\}$. For simplicity, the same reconstruction formula used in the uniform grid is applied to the non-uniform grid. It requires average values over the same spatial regions as defined in the uniform grid. If a neighboring cell is coarsened, the spatial region defined in the uniform grid, which corresponds to the blue-striped region in Figure \ref{subfig:poly_nonuniform}, occupies a portion of the coarsened cell. The average value $\overline{\phi}_{{blue-striped}}$ is approximated by integrating the reconstructed polynomial of the coarsened cell $\phi_{{coarsened}}(x,y)$ over the blue-striped region and dividing by its area:
\begin{equation}
    \overline{\phi}_{{blue-striped}}
    =
    \frac{1}{\Delta x \Delta y}
       \int_{y_i+\Delta x/2}^{y_i+3\Delta x/2}
       \int_{x_i-\Delta x/2}^{x_i+\Delta x/2}
       \phi_{{coarsened}}(x,y) \; dxdy
    .
\end{equation} Conversely, if a neighboring cell is refined, the spatial region defined in the uniform grid, which corresponds to the red-striped region in Figure \ref{subfig:poly_nonuniform}, is covered by multiple refined cells. The average value over the red-striped region $\overline{\phi}_{{red-striped}}$ is obtained by taking a weighted average of the four refined cells,
\begin{equation}
    \overline{\phi}_{{red-striped}}
    =
    (
          N_{C,1} \, \overline{\phi}_1 
        + N_{C,2} \, \overline{\phi}_2 
        + N_{C,3} \, \overline{\phi}_3 
        + N_{C,4} \, \overline{\phi}_4
    ) / (
          N_{C,1} 
        + N_{C,2} 
        + N_{C,3} 
        + N_{C,4}
    )
    ,
\end{equation}
where the weights $N_{C,i}$ is the number of particles in each fine cell $i$. After rescaling, the reconstruction is carried out in the same manner as for uniform grids.

\section{\label{sec:implementation}Numerical implementation}

\subsection{\label{subsec:vss}Viscosity and mean collision time from the Chapman–Enskog theory}


For the USP-FPM method, the viscosity must be specified for particle evolution. According to Chapman–Enskog theory, the viscosity of the variable soft sphere (VSS) model is defined as \cite{Bird_94_DSMC-book}:
\begin{equation}
    \mu = \mu_{ref} \bigg( \frac{T}{T_{ref}} \bigg)^{c_\omega} ,
\end{equation}
where $\mu_{ref}$ is a reference viscosity, $T_{ref}$ is a reference temperature, and $c_\omega$ is the viscosity index. The reference viscosity of the VSS model is defined as follows:
\begin{equation}
    \mu_{\mathrm{ref}} 
    = 
    \frac
    {
        5 
        (c_\alpha + 1) 
        (c_\alpha + 2) 
        \sqrt{mk_BT_{\mathrm{ref}}/\pi}
    }
    {
        4 
        c_\alpha 
        (5-2c_\omega)
        (7-2c_\omega)
        d_{\mathrm{ref}}^2
    } 
    ,
\end{equation}
where $m$ is a molecular mass, $d_{ref}$ is a reference diameter, and $c_\alpha$ is the angular scattering parameter. The mean collision time determines the rotational and vibrational relaxation times. The mean collision time of the VSS model is given by \cite{Bird_94_DSMC-book}:
\begin{equation}
    \tau_c 
    = 
    \frac
    {
        c_\alpha
        (5-2c_\omega)
        (7-2c_\omega)
    }
    {
        5
        (c_\alpha+1)
        (c_\alpha+2)
    } \frac{\mu}{p} 
    .
\label{eq:tau_c}
\end{equation}

\subsection{\label{subsec:eucken}Prandtl Number Using the Eucken Formula}


For diatomic gases, the Eucken formula provides a theoretical expression for $\mathrm{Pr}$ \cite{Boyd17_Neq-book}:
\begin{equation}
    \mathrm{Pr} = \frac{14+2\xi_{vib}}{19+2\xi_{vib}},
\end{equation}
where the number of vibrational degrees of freedom $\xi_{vib}$ is defined as:
\begin{equation}
    \xi_{vib} = \frac{\Theta_{vib}/T}{\exp{(\Theta_{vib}/T)}-1} .
\end{equation}

\subsection{\label{subsec:conserve}Momentum and energy conservation scheme}


During each collision step, momentum and total energy should be conserved, and each energy mode is expected to relax toward the target energies given by Equations (\ref{eq:e_tr_relax})-(\ref{eq:e_vib_relax}). However, the use of a finite number of particles in the USP-FPM method introduces statistical noise, thereby preventing exact conservation of momentum and total energy, and disrupting the expected relaxation of internal energy modes. To reduce this statistical noise, the correction scheme proposed by Kim et al. is employed \cite{Kim24-FPM}. For the vibrational energy, a summation-based correction scheme is applied. The difference between the target and actual vibrational energy levels is calculated as: 
\begin{equation}
    I_{\Delta} = \mathrm{int} 
    \Bigg( 
        N_p \cdot \frac{e_{vib}(T_{vib}^{n+1}) - e_{vib}(T_{vib}^*)}
             {R\Theta_{vib}} 
        + \mathbb{U} 
    \Bigg) 
      ,
\end{equation}
where $\mathrm{int}()$ denotes the integer part, $N_p$ is the total number of particles in the cell, $\mathbb{U}$ is a uniform random number sampled from $\mathrm{Unif}(0,1)$, and the superscript $*$ represents the post-collision values before applying a correction. After computing $I_\Delta$, a particle within the cell is randomly selected to apply the correction. If $I_\Delta > 0$, the vibrational level of the selected particle is increased by 1 and $I_\Delta$ is decreased by 1. Conversely, if $I_\Delta < 0$, the vibrational level is decreased by 1 and $I_\Delta$ is increased by 1. This particle selection and adjustment procedure is repeated until $I_\Delta = 0$. For the rotational energy, a scaling-based correction scheme is employed. In this scheme, the rotational energy of each particle is rescaled as:
\begin{equation}
    \varepsilon_{rot}^{n+1}
    = \bigg( \frac{e_{rot}(T_{rot}^{n+1})}{e_{rot}(T_{rot}^*)} \bigg)
    \cdot\varepsilon_{rot}^* 
    \; .
\end{equation}
To conserve momentum and total energy, particle velocities are adjusted using the following scaling-based correction scheme:
\begin{equation}
    c_i^{n+1}
    = U_i^n
    + \sqrt{\dfrac{
        e_{tr}(T_{tr}^{n}) 
        + e_{rot}(T_{rot}^{n}) 
        + e_{vib}(T_{vib}^{n}) 
        - e_{rot}(T_{rot}^{n+1}) 
        - e_{vib}(T_{vib}^{n+1}) 
        }{e_{tr}(T_{tr}^*)}}
     \cdot  (c_i^* - U_i^*)
    \; .
\end{equation}

\subsection{\label{subsec:algorithm}Algorithm Description for the USP-FPM method}


\begin{figure*}[t!htp]
	\centering
	\begin{subfigure}{0.55\textwidth}
		\centering
		\includegraphics[width=1.\linewidth]{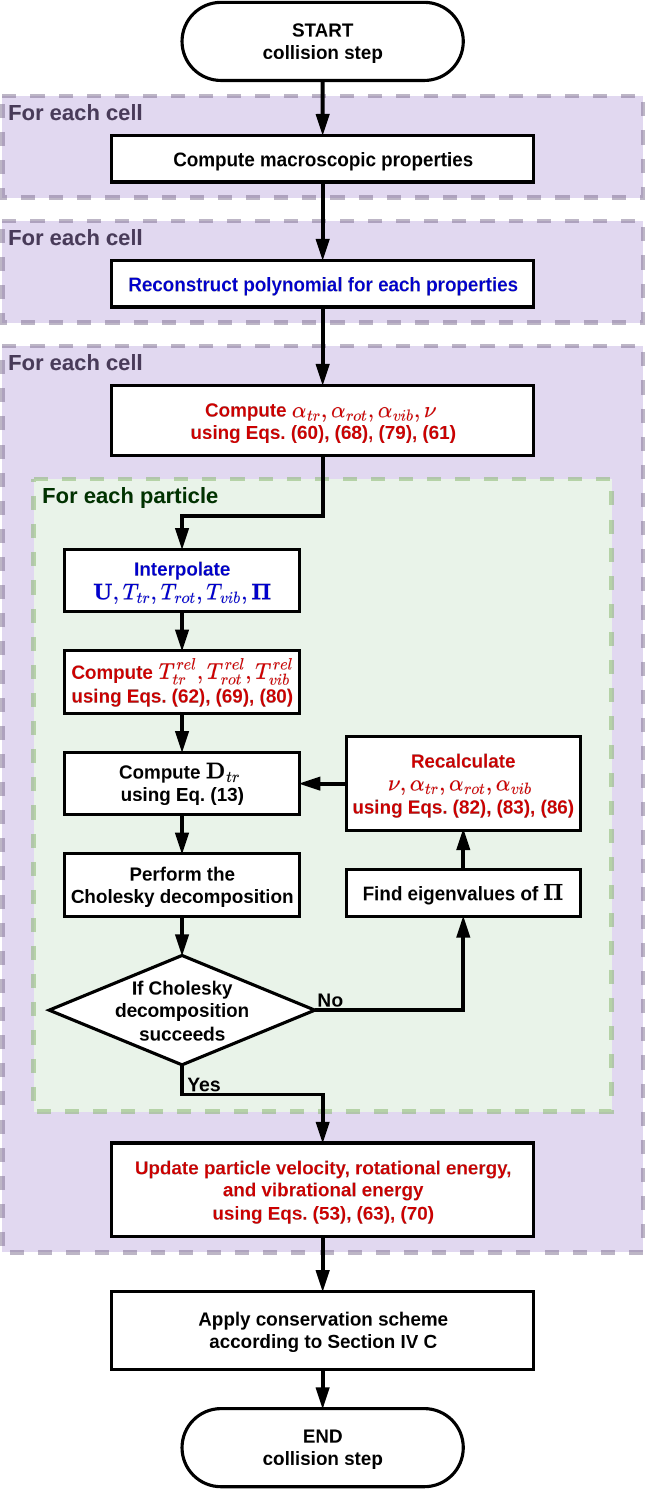}
	\end{subfigure}
	\caption{\label{fig:flowchart}A flow chart for the collision step of the USP-FPM method.}
\end{figure*}

Figure \ref{fig:flowchart} illustrates the flow chart for the collision step of the USP-FPM method. The purple box indicates the loop over each cell, and the green box indicates the loop over each particle within a cell. It is worth noting that the USP-FPM algorithm extends the structure of the FPM method \cite{Kim24-FPM}. Accordingly, the blue text highlights additional procedures introduced to improve spatial accuracy, while the red text indicates modified parameters for enhancing temporal accuracy. The collision step of the USP-FPM method begins by computing the cell-averaged macroscopic properties. Then, each cell reconstructs a polynomial for each macroscopic property using the polynomial reconstruction method. Next, for each cell, the coefficients $\alpha_{tr}$, $\alpha_{rot}$, $\alpha_{vib}$, and $\nu$ are estimated using Equations (\ref{eq:USP-FPM_alpha_tr}), (\ref{eq:USP-FPM_alpha_rot}), (\ref{eq:USP-FPM_alpha_vib}), and (\ref{eq:USP-FPM_nu}), respectively. For each particle within a cell, the macroscopic quantities, namely $\boldsymbol{U}$, $T_{tr}$, $T_{rot}$, $T_{vib}$, and $\boldsymbol{\Pi}$, are interpolated to each particle's position. After interpolation, the relaxation temperatures $T_{tr}^{rel}$, $T_{rot}^{rel}$, and $T_{vib}^{rel}$ are computed using Equations (\ref{eq:T_tr_USP}), (\ref{eq:T_rot_USP}), and (\ref{eq:T_vib_USP}), respectively. Subsequently, the translational diffusion tensor $\boldsymbol{D}_{tr}$ is calculated according to Equation (\ref{eq:D_tr}) and its positivity is checked via Cholesky decomposition. If Cholesky decomposition fails, $\nu$, $\alpha_{tr}$, $\alpha_{rot}$, and $\alpha_{vib}$ are recalculated using Equations (\ref{eq:nu_USP_positivity}), (\ref{eq:alpha_USP_positivity_1}), and (\ref{eq:alpha_USP_positivity_4}), to preserve the positive definiteness of $\boldsymbol{D}_{tr}$. The $\boldsymbol{D}_{tr}$ is then recalculated and tested again for positivity. If Cholesky decomposition succeeds, the particle velocity, rotational energy, and vibrational energy are updated using Equations (\ref{eq:update_ci}), (\ref{eq:update_erot}), and (\ref{eq:update_evib}), respectively. After updating all particles, the conservation scheme described in Section \ref{subsec:conserve} is applied. The USP-FPM method is implemented within the SPARTA code, an open-source DSMC solver developed by Sandia National Laboratories \cite{Plimpton19_SPARTA}.

\section{\label{sec:results}Results and discussion}


To evaluate the accuracy and efficiency of the USP-FPM method, three test cases are considered. First, a homogeneous relaxation problem is performed to assess the temporal accuracy of the USP-FPM method with varying time step sizes. This homogeneous relaxation problem includes two sub-cases: relaxation of viscous stress and heat flux, and relaxation from a thermal non-equilibrium state where the translational, rotational, and vibrational temperatures differ. Second, the Poiseuille flow is performed to evaluate the temporal and spatial accuracy of the USP-FPM method by varying the time step and cell size. Finally, the hypersonic flow around a cylinder is considered to evaluate the accuracy and computational efficiency of the USP-FPM method in a two-dimensional flow involving strong shock waves and significant thermodynamic non-equilibrium. Table \ref{tab:case} summarizes the test cases and their objectives. All cases use nitrogen gas with a molecular mass of $6.63\times10^{-26} \,\rm{kg}$ and a characteristic vibrational temperature of $3371 \,\rm{K}$. Molecular parameters based on the VSS model are applied in all cases except for the hypersonic flow around a cylinder. For the hypersonic flow around a cylinder, molecular parameters from the VHS model are adopted to match those used in the DSMC study by Lofthouse \cite{Lofthouse08_cylinder}. The molecular parameters are summarized in Table \ref{tab:parameter}. Since the original DSMC code does not satisfy the Landau-Teller equation, the prohibiting double relaxation method is applied \cite{Zhang13_prohibitting}.

\begin{table}[t]
\caption{\label{tab:case}Summary of test cases for evaluating the USP-FPM method.}
\begin{ruledtabular}
\begin{tabular}{l l l}
    Case &
    Sub-case &
    Objective \\
    \hline 
    \hline 
    \multirow{2}{*}{\makecell[l]{\\[-1em] \textbf{A.} Homogeneous \\ relaxation}} &
    \makecell[l]{\textbf{A.1.} Relaxation of viscous \; \\ stress and heat flux} &
    \multirow{2}{*}{\makecell[l]{\\[-1em] To evaluate temporal accuracy of the \\ USP-FPM method}} \\
    \cline{2-2} 
    &
    \makecell[l]{\textbf{A.2.} Relaxation toward \\ thermal equilibrium} &
    \\
    \hline
    \makecell[l]{\textbf{B.} Poiseuille flow} &
     &
    \makecell[l]{To evaluate spatial and temporal accuracy \\
                 of the USP-FPM method} \\[0.5em]
    \hline
    \makecell[l]{\textbf{C.} Hypersonic flow \\ around a cylinder} &
     &
    \makecell[l]{To evaluate spatio-temporal accuracy and \\ 
                 efficiency of the USP-FPM method} \\[0.5em]
\end{tabular}
\end{ruledtabular}
\end{table}

\begin{table}[t]
\caption{\label{tab:parameter}Molecular parameters of nitrogen gas.}
\begin{ruledtabular}
\begin{tabular}{l llll l}
    Model &
    $T_{ref} \,[\rm{K}]$ &
    $d_{ref} \,[\rm{m}]$ &
    $c_\omega $ &
    $c_\alpha $ &
    Applied section \\
    \hline
    VSS model & $273.15$ & $4.11\times 10^{-10}$ & $0.74$ & $1.36$ &
    \ref{subsec:relax}, \ref{subsec:poiseuille} \\
    VHS model & $290.00$ & $4.11\times 10^{-10}$ & $0.70$ & $1.00$ &
    \ref{subsec:cylinder} \\
\end{tabular}
\end{ruledtabular}
\end{table}

\subsection{\label{subsec:relax} Homogeneous relaxations}


\subsubsection{\label{subsubsec:relax_moment}Relaxation of viscous stress and heat flux in a homogeneous flow}


A homogeneous relaxation of viscous stress and heat flux is investigated to assess the temporal accuracy of the USP-FPM method in comparison to the FPM method. To impose initial viscous stress and heat fluxes, initial particle velocity, rotational energy, and vibrational energy are sampled from Grad's 17-moment distribution function at an equilibrium temperature $T_{eq}$, which is given by \cite{McCormack68_G17}:
\begin{multline}
    \mathcal{F}_{G17}
    =\mathcal{F}_{M} 
    \Bigg( 
        1
        + 
        \frac{\sigma_{ij}(0)}{2pRT_{eq}} 
        \Big(C_iC_j - \frac{1}{3}C^2\delta_{ij} \Big)
        + 
        \frac{q_{tr,i}(0) C_i}{5p (RT_{eq})^2 } 
        \Bigl( C^2 - 5 RT_{eq} \Bigr)
        \\[0.5em] + 
        \frac{q_{rot,i}(0) C_i}{p(RT_{eq})^2 } 
        \Bigl( \varepsilon_{rot} - RT_{eq} \Bigr)
        + 
        \frac{q_{vib,i}(0) C_i}{p(RT_{eq})^2 } 
        \Bigl( \frac{\sinh{( \Theta_{vib}/2T_{eq} )}}{\Theta_{vib}/2T_{eq} } \Bigr)^2
        \Bigl( \varepsilon_{vib} - \frac{R \Theta_{vib}}{\exp{\big( {\Theta_{vib}/T_{eq}} \big)} -1}  \Bigr)
    \Bigg)
    .
\end{multline}
The initial viscous stress is given by $\sigma_{ij}(0)=0.1\rho RT_{eq}$, and the initial translational, rotational, and vibrational heat fluxes are given by $q_{tr,i}(0) = q_{rot,i}(0) = q_{vib,i}(0) = 0.1\rho (RT_{eq})^{1.5}$. The translational, rotational, and vibrational temperatures are all set to $T_{tr} = T_{rot} = T_{vib} = T_{eq} = 4000\,\rm{K}$. The initial number density of nitrogen is initialized as $n=10^{24} \,\rm{m^{-3}}$. Fixed rotational and vibrational collision numbers of $Z_{rot}=10$ and $Z_{vib}=50$ are employed. The computational domain consists of a single cell of size $\Delta x = 0.001 \,\rm{m}$, containing $10^7$ computational particles. Four different time steps are employed: $\Delta t/\tau_c=\{ 0.1, 0.5, 1.0, 2.0 \}$. The mean collision time is estimated as $\tau_c=1.4573\times10^{-9} \,\rm{s}$ based on the initial number density and temperature. The analytical solutions provided in Eqs. (\ref{eq:analytic_sigma_full}) and (\ref{eq:analytic_qtr_full})-(\ref{eq:analytic_qvib_full}) serve as reference solutions.


\begin{figure*}[b]
	\centering
	\begin{subfigure}{0.495\textwidth}
		\centering
		\includegraphics[width=1.\linewidth]{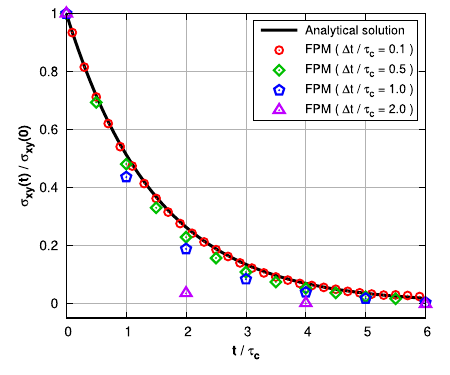}
            \caption{\label{subfig:Relax_stress_FPM}FPM method.}
	\end{subfigure}
	\begin{subfigure}{0.495\textwidth}
		\centering
		\includegraphics[width=1.\linewidth]{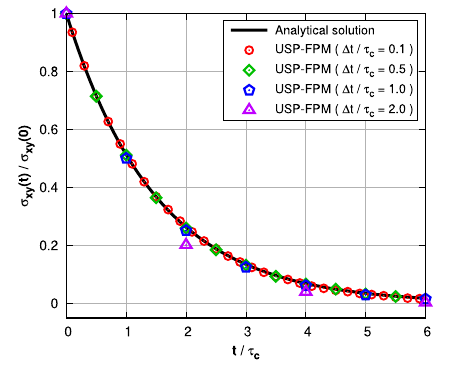}
            \caption{\label{subfig:Relax_stress_USP}USP-FPM method.}
	\end{subfigure}
	\caption{\label{fig:Relax_stress}Normalized viscous stress relaxations using different time steps.}
\end{figure*}

Figure \ref{fig:Relax_stress} illustrates the relaxation of viscous stress in a homogeneous flow. Figures \ref{subfig:Relax_stress_FPM} and \ref{subfig:Relax_stress_USP} show the results from the FPM and USP-FPM methods, respectively. The viscous stress decays as the nitrogen gas relaxes toward equilibrium, which corresponds to the Maxwellian distribution with zero viscous stress \cite{Bird_94_DSMC-book}. To quantitatively compare the results with the analytical solution, a normalized L2-norm error of viscous stress is employed. This error is defined as:
\begin{equation}
    E_{L2}(\sigma_{ij})
    =
    \frac{\sqrt{ \langle (\sigma_{ij} - \sigma_{ref,\, ij} )^2 \rangle}}
         {\sqrt{ \langle (\sigma_{ref,\, ij} )^2 \rangle}}
    ,
\end{equation}
where $\sigma_{ref,\,ij}$ denotes the viscous stress from the analytical solution. The resulting normalized L2-norm errors are summarized in Table \ref{tab:L2norm_sigma}. In Figure \ref{subfig:Relax_stress_FPM}, the FPM method shows good agreement with the analytical solution at $\Delta t/\tau_c=0.1$, yielding an error of $1.3\,\%$. As the time step increases from $\Delta t/\tau_c=0.5$ to $2.0$, the viscous stress decays more rapidly than predicted by the analytical solution, resulting in errors rising from $8.2\,\%$ to $87.4\,\%$. This behavior is attributed to the first-order temporal accuracy of the viscous stress relaxation in the FPM model. Conversely, as shown in Figure \ref{subfig:Relax_stress_USP}, the USP-FPM method maintains good agreement with the analytical solution up to $\Delta t/\tau_c=1.0$, with errors below $3.7\,\%$. When the time step increases to $\Delta t/\tau_c=2.0$, the viscous stress exhibits a noticeably faster decay, leading to an error of $25.6\,\%$. This is because, despite its second-order temporal accuracy, the USP-FPM method becomes less accurate as the time step becomes excessively large.

\begin{table}[t]
\caption{\label{tab:L2norm_sigma}Normalized L2-norm errors of the viscous stress.}
\begin{ruledtabular}
    \begin{tabular}{l cccc}
            \multirow{2}{*}{Method} &
            \multicolumn{4}{c}{Normalized L2-norm errors of viscous stress} \\
            \cline{2-5} 
             &
            $\Delta t / \tau_c = 0.1$ &
            $\Delta t / \tau_c = 0.5$ &
            $\Delta t / \tau_c = 1.0$ &
            $\Delta t / \tau_c = 2.0$ \\
        \hline
            FPM     & 0.013 & 0.082 & 0.209 & 0.874 \\
            USP-FPM & 0.007 & 0.008 & 0.037 & 0.256 \\
    \end{tabular}
\end{ruledtabular}
\end{table}


\begin{figure*}[t]
	\centering
	\begin{subfigure}{0.495\textwidth}
		\centering
		\includegraphics[width=1.\linewidth]{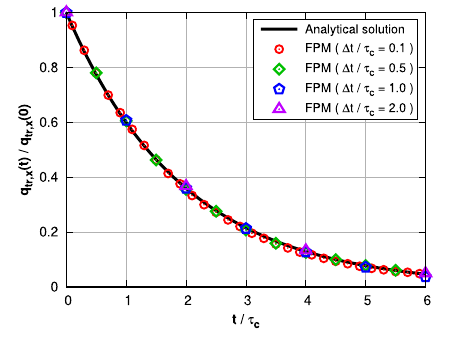}
            \caption{\label{subfig:Relax_qtr_FPM}Translational mode from FPM method.}
	\end{subfigure}
	\begin{subfigure}{0.495\textwidth}
		\centering
		\includegraphics[width=1.\linewidth]{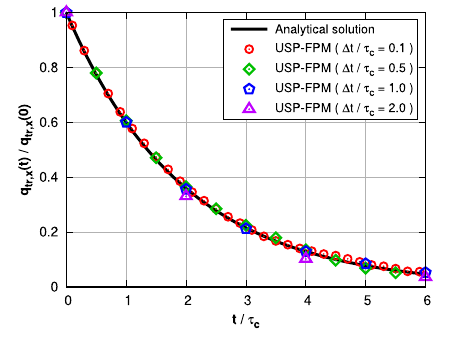}
            \caption{\label{subfig:Relax_qtr_USP}Translational mode from USP-FPM method.}
	\end{subfigure}
	\centering
	\begin{subfigure}{0.495\textwidth}
		\centering
		\includegraphics[width=1.\linewidth]{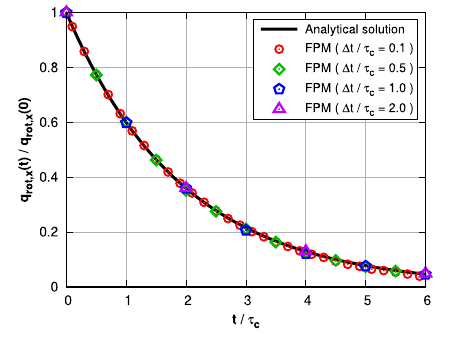}
            \caption{\label{subfig:Relax_qrot_FPM}Rotational mode from FPM method.}
	\end{subfigure}
	\begin{subfigure}{0.495\textwidth}
		\centering
		\includegraphics[width=1.\linewidth]{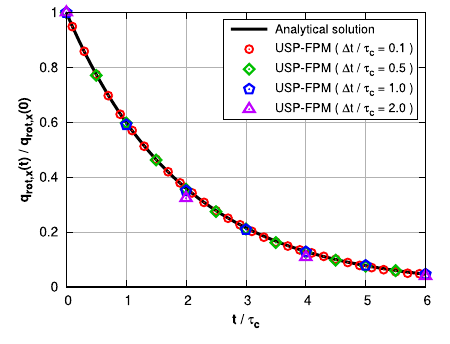}
            \caption{\label{subfig:Relax_qrot_USP}Rotational mode from USP-FPM method.}
	\end{subfigure}
        \centering
	\begin{subfigure}{0.495\textwidth}
		\centering
		\includegraphics[width=1.\linewidth]{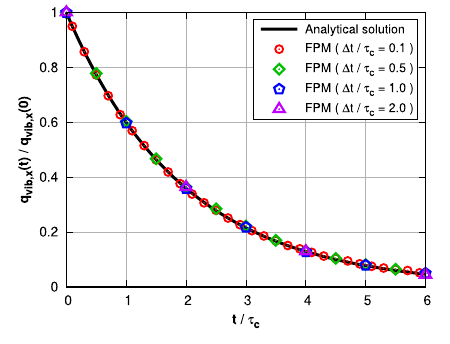}
            \caption{\label{subfig:Relax_qvib_FPM}Vibrational mode from FPM method.}
	\end{subfigure}
	\begin{subfigure}{0.495\textwidth}
		\centering
		\includegraphics[width=1.\linewidth]{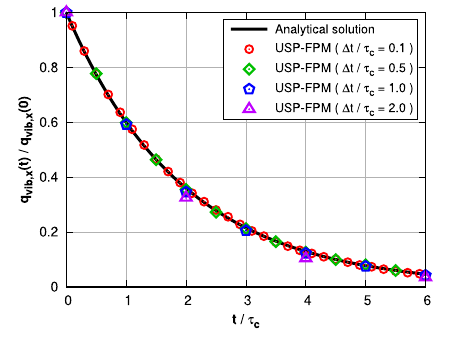}
            \caption{\label{subfig:Relax_qvib_USP}Vibrational mode from USP-FPM method.}
	\end{subfigure}
	\caption{\label{fig:Relax_q} Normalized heat flux relaxations using different time steps.}
\end{figure*}

Figure \ref{fig:Relax_q} presents the relaxation of heat fluxes in a homogeneous flow. Figures \ref{subfig:Relax_qtr_FPM}-\ref{subfig:Relax_qtr_USP}, \ref{subfig:Relax_qrot_FPM}-\ref{subfig:Relax_qrot_USP}, and \ref{subfig:Relax_qvib_FPM}-\ref{subfig:Relax_qvib_USP} represent the relaxation of translational, rotational, and vibrational heat fluxes, respectively. Figures \ref{subfig:Relax_qtr_FPM}, \ref{subfig:Relax_qrot_FPM}, and \ref{subfig:Relax_qvib_FPM} show the results obtained from the FPM method, while Figures \ref{subfig:Relax_qtr_USP}, \ref{subfig:Relax_qrot_USP}, and \ref{subfig:Relax_qvib_USP} present the results from the USP-FPM method. In all energy modes, the heat flux exhibits exponential decay toward zero as the nitrogen gas relaxes to a Maxwellian distribution, which has zero heat flux \cite{Bird_94_DSMC-book}. The normalized L2-norm errors of heat fluxes with respect to the analytical solution are summarized in Table \ref{tab:L2norm_q}. The FPM method is consistent with the analytical solution across all time steps, maintaining errors below $2.1\,\%$. This is because the FPM method exactly reproduces the analytical relaxation behavior of the translational and rotational heat fluxes, as given by Equations (\ref{eq:FPM-qtr-relax-exp}) and (\ref{eq:FPM-qrot-relax-exp}). The FPM method does not provide a closed-form expression for the vibrational heat flux relaxation, as the master equation involves an infinite number of vibrational energy levels. Nevertheless, an approximated expression for the vibrational heat flux relaxation, given in Equation (\ref{eq:FPM-qvib-relax-exp}), indicates that the FPM method reproduces the analytical behavior with at least third-order temporal accuracy. The USP-FPM method maintains good agreement with the analytical solution in predicting heat flux relaxations across all modes up to $\Delta t/\tau_c=1.0$, with a maximum error of $2.7\,\%$. At $\Delta t/\tau_c=2.0$, a faster decay of heat fluxes in all energy modes is observed compared to the analytical solution, resulting in errors exceeding $10.8\,\%$. Notably, at this time step, the error of the USP-FPM method is larger than the error of the FPM method. This behavior occurs because the USP-FPM method is designed to achieve second-order accuracy in all three relaxations involving energy, viscous stress, and heat flux. While the USP-FPM method improves temporal accuracy in energy and viscous stress relaxations, it sacrifices temporal accuracy in the heat flux relaxation, which was better in the FPM method. Thus, when the time step becomes sufficiently large, the heat flux predicted by the USP-FPM method deviates from the analytical solution.

\begin{table}[t!]
\caption{\label{tab:L2norm_q}Normalized L2-norm errors of heat fluxes.}
\begin{ruledtabular}
    \begin{tabular}{l cccc}
            \multirow{2}{*}{Method} &
            \multicolumn{4}{c}{Normalized L2-norm errors of heat fluxes} \\
            \cline{2-5} 
             &
            $\Delta t / \tau_c = 0.1$ &
            $\Delta t / \tau_c = 0.5$ &
            $\Delta t / \tau_c = 1.0$ &
            $\Delta t / \tau_c = 2.0$ \\
        \hline
            \multicolumn{5}{c}{Translational heat flux} \\
            FPM     & 0.015 & 0.012 & 0.019 & 0.021 \\
            USP-FPM & 0.018 & 0.021 & 0.014 & 0.108 \\
        \hline
            \multicolumn{5}{c}{Rotational heat flux} \\
            FPM     & 0.011 & 0.010 & 0.016 & 0.005 \\
            USP-FPM & 0.005 & 0.008 & 0.018 & 0.111 \\
        \hline
            \multicolumn{5}{c}{Vibrational heat flux} \\
            FPM     & 0.006 & 0.013 & 0.006 & 0.012 \\
            USP-FPM & 0.006 & 0.008 & 0.027 & 0.117 \\
    \end{tabular}
\end{ruledtabular}
\end{table}

\subsubsection{\label{subsubsec:relax_temp}Relaxation toward thermal equilibrium in a homogeneous flow}


A homogeneous relaxation toward thermal equilibrium is examined to evaluate the temporal accuracy of the USP-FPM method compared to the FPM method. The viscous stress and heat fluxes are initialized to zero by sampling computational particles from the Maxwellian distribution. Thermal non-equilibrium is introduced by setting the initial temperatures to $T_{tr}=6000\,\rm{K}$, $T_{rot}=4000\,\rm{K}$, and $T_{vib}=2000\,\rm{K}$. A number density of $n=10^{24} \,\rm{m^{-3}}$ is assigned, with fixed collision numbers $Z_{rot}=10$ and $Z_{vib}=50$. The computational domain consists of a single cell of size $\Delta x = 0.001 \,\rm{m}$, containing $10^7$ computational particles. Four different time steps are employed: $\Delta t/\tau_c=\{ 0.1, 0.5, 1.0, 2.0 \}$. The mean collision time is calculated as $\tau_c=1.3132\times10^{-9} \,\rm{s}$. Since there is no analytical solution to the Landau-Teller equation, the DSMC method using a small time step of $\Delta t/\tau_c=0.01$ serves as a reference solution.


\begin{figure*}[b]
	\centering
	\begin{subfigure}{0.495\textwidth}
		\centering
		\includegraphics[width=1.\linewidth]{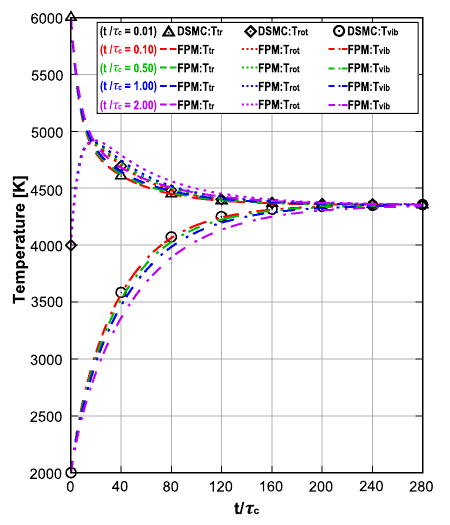}
            \caption{\label{subfig:Relax_temp_FPM}FPM method.}
	\end{subfigure}
	\begin{subfigure}{0.495\textwidth}
		\centering
		\includegraphics[width=1.\linewidth]{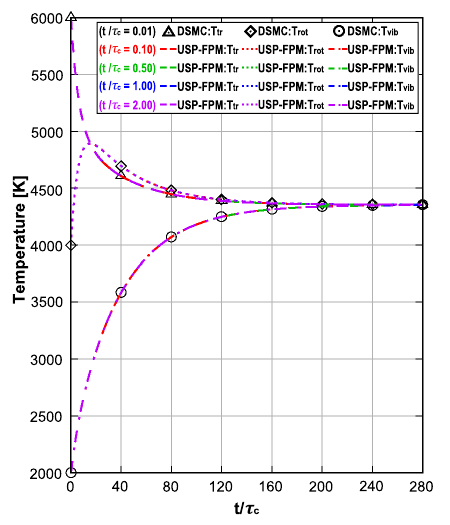}
            \caption{\label{subfig:Relax_temp_USP}USP-FPM method.}
	\end{subfigure}
	\caption{\label{fig:Relax_temp}Translational, rotational, and vibrational temperature relaxations using different time steps.}
\end{figure*}

\begin{table}[t]
\caption{\label{tab:L2norm_temp}Normalized L2-norm errors of the translational, rotational, and vibrational temperatures.}
\begin{ruledtabular}
    \begin{tabular}{l cccc}
            \multirow{2}{*}{Method} &
            \multicolumn{4}{c}{Normalized L2-norm errors of temperatures} \\
            \cline{2-5} 
             &
            $\Delta t / \tau_c = 0.1$ &
            $\Delta t / \tau_c = 0.5$ &
            $\Delta t / \tau_c = 1.0$ &
            $\Delta t / \tau_c = 2.0$ \\
        \hline
            \multicolumn{5}{c}{Translational temperature} \\
            FPM     & 0.001 & 0.003 & 0.005 & 0.010 \\
            USP-FPM & 0.000 & 0.000 & 0.000 & 0.000 \\
        \hline
            \multicolumn{5}{c}{Rotational temperature} \\
            FPM     & 0.001 & 0.003 & 0.006 & 0.012 \\
            USP-FPM & 0.000 & 0.000 & 0.000 & 0.001 \\
        \hline
            \multicolumn{5}{c}{Vibrational temperature} \\
            FPM     & 0.002 & 0.008 & 0.015 & 0.030 \\
            USP-FPM & 0.001 & 0.001 & 0.001 & 0.000 \\
    \end{tabular}
\end{ruledtabular}
\end{table}

Figure \ref{fig:Relax_temp} illustrates the temperature relaxations toward thermal equilibrium in a homogeneous flow. Figures \ref{subfig:Relax_temp_FPM} and \ref{subfig:Relax_temp_USP} present the results from the FPM and USP-FPM methods, respectively. The translational and rotational temperatures equilibrate first, followed by a gradual equilibration with the vibrational temperature. This behavior arises because the rate of internal energy relaxation is inversely proportional to the collision number \cite{Boyd17_Neq-book}. Since the rotational collision number is lower than the vibration collision number in this case, the rotational energy relaxes more rapidly. Table \ref{tab:L2norm_temp} summarizes the L2-norm errors of the translational, rotational, and vibrational temperatures. As shown in Figure \ref{subfig:Relax_temp_FPM}, the FPM method shows good agreement with the reference DSMC solution for all temperatures when $\Delta t/\tau_c=0.1$, with errors below $0.2\,\%$. As the time step increases from $\Delta t/\tau_c = 0.5$ to $2.0$, the translational temperature relaxes more slowly than in the reference DSMC solution, with errors increasing from $0.3\,\%$ to $1.0\,\%$. Because of the first-order temporal accuracy of the FPM method, the translational relaxation temperature is overestimated when the time step exceeds $\Delta t/\tau_c = 0.5$. This overestimation leads to insufficient energy exchange and consequently delays the relaxation of the translational temperature. The rotational temperature exhibits slower relaxation toward the equilibrium temperature when the time step exceeds $\Delta t/\tau_c=0.5$. This slower relaxation becomes evident once the rotational temperature surpasses the translational temperature, with the corresponding error increasing from $0.3\,\%$ to $1.2\,\%$. Similar to the translational temperature, this slower relaxation results from the overestimation of the rotational relaxation temperature, which arises due to the first-order temporal accuracy of the FPM method. The vibrational temperature also shows slower relaxation compared to the reference DSMC solution as the time step increases beyond $\Delta t/\tau_c=0.5$, with errors ranging from $0.8\,\%$ to $3.0\,\%$. The vibrational temperature shows the largest deviation from the reference DSMC solution among the three energy modes. This largest deviation results from accumulated errors, as the underestimated vibrational relaxation temperature persists over a longer relaxation period due to the higher collision number. In contrast, as illustrated in Figure \ref{subfig:Relax_temp_USP}, the USP-FPM method remains in close agreement with the reference DSMC solution for all temperatures even at $\Delta t/\tau_c=2.0$, maintaining errors under $0.1\,\%$. These results highlight the second-order temporal accuracy of the USP-FPM method.

\subsection{\label{subsec:poiseuille}Poiseuille flow}


Poiseuille flow is selected as a fundamental one-dimensional case to evaluate the accuracy of the USP-FPM method in time and space, and to compare its performance with the DSMC and FPM methods. The nitrogen gas is initialized with a number density of $n=1.33245\times 10^{26}\,\rm{m^{-3}}$, and uniform temperatures of $T_{tr}= T_{rot} = T_{vib}=273.15\,\rm{K}$. Because the initial temperature differs from that of the homogeneous relaxation cases, the collision numbers are chosen as $Z_{rot}=5$ and $Z_{vib}=10^{14}$ to appropriately reflect the internal energy relaxation at this temperature. Both walls remain stationary and have fully diffusive boundary conditions at a constant temperature of $T_{wall}=273\,\rm{K}$. The distance between two walls is defined as $L=10^{-5}\,\rm{m}$, resulting in a Knudsen number of $\mathrm{Kn}=0.001$. A pressure gradient of $dp/dx=5\times10^7\,\rm{Pa/m}$ is imposed, resulting in a maximum velocity of $U_{max}=225.52\,\rm{m/s}$. The reference solution is obtained using the DSMC method with a cell size of $\Delta x_{ref}=3.125\times10^{-9}\,\rm{m}$ and a time step of $\Delta t_{ref}=2.0\times10^{-12}\,\rm{s}$. This condition corresponds to $0.3125\lambda$ and $0.1\tau_c$, where $\lambda$ is a mean free path. Two groups of simulations are conducted using the DSMC, FPM, and USP-FPM methods. Group A evaluates temporal accuracy by varying the time step as $\Delta t /\Delta t_{ref} \in \{10, 20, 40, 80, 160, 320\}$ under a fixed cell size of $\Delta x_{ref}$, corresponding to cases A-1 through A-6. Group B investigates spatial accuracy by varying the cell size as $\Delta x/\Delta x_{ref} \in \{3.2, 6.4, 16, 32, 64, 160, 320, 640\}$ under a fixed time step of $\Delta t_{ref}$, corresponding to cases B-1 through B-8. The cases A-4 to A-6 and B-5 to B-8 are used to evaluate the temporal and spatial accuracy of the USP-FPM method. For comparison, the temporal and spatial accuracy of the DSMC and FPM methods is assessed using cases A-1 to A-4 and B-1 to B-5. In all cases, the number of particles per cell is set to 5000. The numerical parameters for each case are summarized in Table \ref{tab:case_poiseuille}.

\begin{table}[t]
\caption{\label{tab:case_poiseuille}Numerical parameters for the Poiseuille flow.}
\begin{ruledtabular}
\begin{tabular}{cl ccl}
    Group &
    Case &
    $\Delta x$ &
    $\Delta t$ &
    Applied Method \\
    \hline
    \multicolumn{2}{c}{Reference} &
    $3.125 \times 10^{-9}\,\rm{m} \,(\Delta x_{ref})$ &
    $2.0 \times 10^{-12}\,\rm{s} \,(\Delta t_{ref})$ &
    DSMC \\[0.5em]
    \multirow{6}{*}{A}  & A-1 &     $\Delta x_{ref}$ &  $10 \Delta t_{ref}$ & DSMC, FPM \\
                        & A-2 &     $\Delta x_{ref}$ &  $20 \Delta t_{ref}$ & DSMC, FPM \\
                        & A-3 &     $\Delta x_{ref}$ &  $40 \Delta t_{ref}$ & DSMC, FPM \\
                        & A-4 &     $\Delta x_{ref}$ &  $80 \Delta t_{ref}$ & DSMC, FPM, USP-FPM \\
                        & A-5 &     $\Delta x_{ref}$ & $160 \Delta t_{ref}$ & USP-FPM \\
                        & A-6 &     $\Delta x_{ref}$ & $320 \Delta t_{ref}$ & USP-FPM \\[0.5em]
    \multirow{9}{*}{B}  & B-1 & $3.2 \Delta x_{ref}$ &     $\Delta t_{ref}$ & DSMC, FPM \\
                        & B-2 & $6.4 \Delta x_{ref}$ &     $\Delta t_{ref}$ & DSMC, FPM \\
                        & B-3 &  $16 \Delta x_{ref}$ &     $\Delta t_{ref}$ & DSMC, FPM \\
                        & B-4 &  $32 \Delta x_{ref}$ &     $\Delta t_{ref}$ & DSMC, FPM \\
                        & B-5 &  $64 \Delta x_{ref}$ &     $\Delta t_{ref}$ & DSMC, FPM, USP-FPM \\
                        & B-6 & $160 \Delta x_{ref}$ &     $\Delta t_{ref}$ & USP-FPM \\
                        & B-7 & $320 \Delta x_{ref}$ &     $\Delta t_{ref}$ & USP-FPM \\
                        & B-8 & $640 \Delta x_{ref}$ &     $\Delta t_{ref}$ & USP-FPM \\
\end{tabular}
\end{ruledtabular}
\end{table}


\begin{figure*}[t]
    \centering
	\begin{subfigure}{0.495\textwidth}
		\centering
		\includegraphics[width=1.\linewidth]{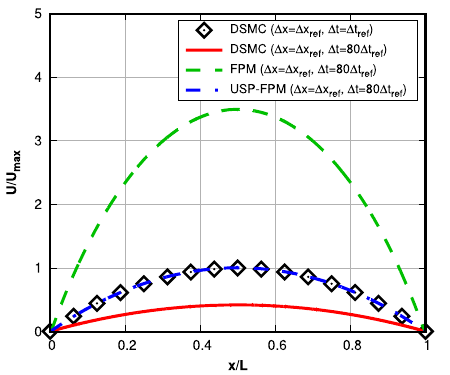}
            \caption{\label{subfig:Poi_Velo_dt}Case A-4.}
	\end{subfigure}
	\begin{subfigure}{0.495\textwidth}
		\centering
		\includegraphics[width=1.\linewidth]{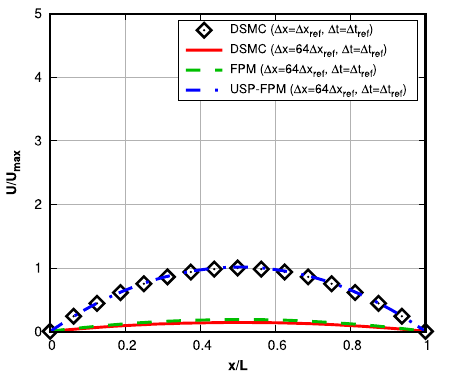}
            \caption{\label{subfig:Poi_Velo_dx}Case B-5.}
	\end{subfigure}
	\caption{\label{fig:Poi_Velo}Bulk velocity profiles in the Poiseuille flow.}
\end{figure*}

Figure \ref{fig:Poi_Velo} illustrates the bulk velocity profiles in Poiseuille flow. Figure \ref{subfig:Poi_Velo_dt} presents the profiles obtained from case A-4, while Figure \ref{subfig:Poi_Velo_dx} shows the profiles obtained from case B-5. The nitrogen gas develops a parabolic bulk velocity profile due to the balance between the pressure gradient and viscous force from the walls. As shown in Figure \ref{subfig:Poi_Velo_dt}, the DSMC method in case A-4 underestimates the bulk velocity relative to the reference DSMC solution. As the time step exceeds the mean collision time, particles in the DSMC method may travel distances greater than the mean free path without undergoing collisions. As a result, particles located far from the surface may reach and interact with the surface in a single time step, which leads to excessive momentum transfer to the wall and underestimation of the bulk velocity. The FPM method in case A-4 overestimates the bulk velocity relative to the reference DSMC solution due to the excessive decay of viscous stress at larger time steps. The USP-FPM method in case A-4 shows good agreement with the reference DSMC solution. For case B-5, the DSMC method underestimates the bulk velocity relative to the reference DSMC solution, as illustrated in Figure \ref{subfig:Poi_Velo_dx}. As the cell size exceeds the mean free path, collision partners may be selected over distances greater than the mean free path. In such cases, a particle close to the surface may gain excessive momentum from a distant collision partner and transfer it to the wall, leading to unphysical momentum exchange and an underestimation of the bulk velocity. The FPM method also underestimates the bulk velocity relative to the reference DSMC solution. When the cell size increases, the cell-averaged velocity near the surface becomes higher because particles located farther from the surface are included. Since particles near the surface adopt this overestimated cell-averaged velocity, excessive momentum is transferred to the surface, resulting in an underestimation of the overall bulk velocity. The USP-FPM method matches well with the reference DSMC solution.


\begin{figure*}[t]
    \centering
	\begin{subfigure}{0.495\textwidth}
		\centering
		\includegraphics[width=1.\linewidth]{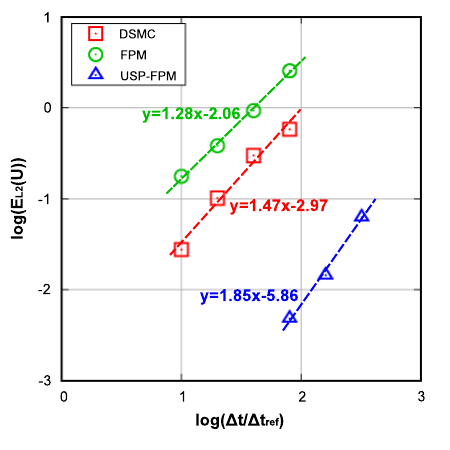}
            \caption{\label{subfig:Poi_l2_dt}Group A.}
	\end{subfigure}
	\begin{subfigure}{0.495\textwidth}
		\centering
		\includegraphics[width=1.\linewidth]{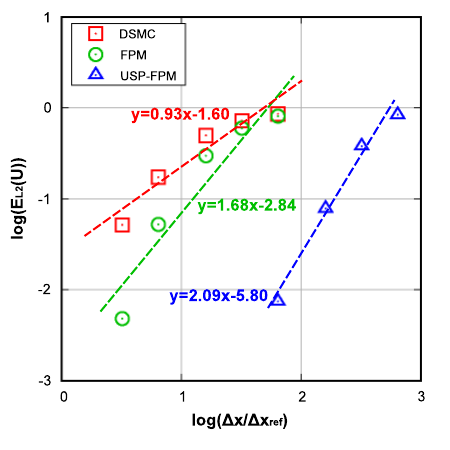}
            \caption{\label{subfig:Poi_l2_dx}Group B.}
	\end{subfigure}
	\caption{\label{fig:Poi_l2}The L2-norm errors of bulk velocity in the Poiseuille flow.}
\end{figure*}

To determine the temporal and spatial accuracy of the USP-FPM method, the L2-norm error of the bulk velocity is evaluated. Figures \ref{subfig:Poi_l2_dt} and \ref{subfig:Poi_l2_dx} illustrate the L2-norm error with respect to the time step and cell size, corresponding to the temporal and spatial cases of Group A and Group B, respectively. For group A, the DSMC and FPM methods exhibit convergence orders of 1.47 and 1.28, respectively, consistent with first-order accuracy under the tested conditions. The USP-FPM method exhibits lower L2-norm errors and achieves a convergence order of 1.85, indicating second-order temporal accuracy for the tested range. For group B, the DSMC and FPM methods display nonlinear error curves, with overall convergence orders of 0.93 and 1.68, respectively. As discussed before, the momentum of particles located farther from the wall is transferred directly to the surface due to larger cell sizes, which flattens the velocity profile. This flattening causes the slope of the error curve to decrease as the cell size increases, resulting in nonlinear convergence behavior. Although the FPM method shows lower L2-norm errors than the DSMC method as the cell size decreases, both methods exhibit similar error magnitudes over the tested resolution range. In contrast, the USP-FPM method shows noticeably lower L2-norm errors and a convergence order of 2.09, consistent with second-order spatial accuracy at the tested resolutions.

\subsection{\label{subsec:cylinder}Hypersonic flow around a cylinder}


\begin{figure*}[b]
    \centering
	\begin{subfigure}{0.8\textwidth}
		\centering
        \includegraphics[width=1.\linewidth]{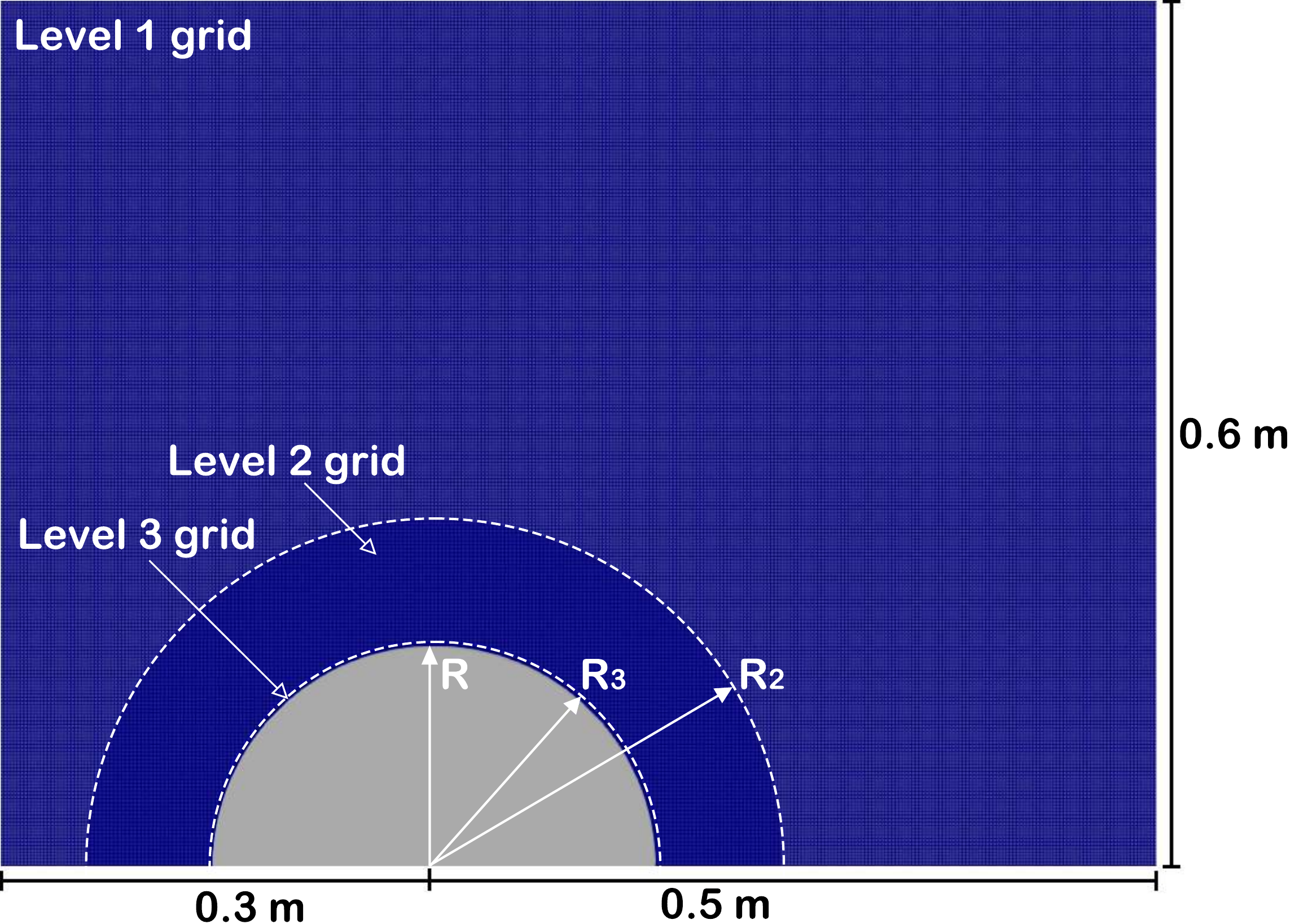}
	\end{subfigure}
	\caption{\label{fig:cylinder_domain}Computational domain of the hypersonic flow around a cylinder.}
\end{figure*}

The hypersonic flow around a cylinder exhibits a strong shock structure and significant non-equilibrium effects, as previously studied by Lofthouse \cite{Lofthouse08_cylinder}. This case is employed to evaluate the spatio-temporal accuracy and efficiency of the USP-FPM method in comparison with the DSMC and FPM methods. The freestream consists of nitrogen gas with a number density of $n_{\infty}=2.124\times10^{21}\,\rm{m^{-3}}$, temperatures of $T_{tr,\infty}=T_{rot,\infty}=T_{vib,\infty}=200\,\rm{K}$, and a velocity of $U_{\infty}=2883\,\rm{m/s}$, yielding a Mach number of $\mathrm{Ma}=10$. The cylinder has a radius of $R=0.1524\,\rm{m}$, corresponding to the Knudsen number of $\mathrm{Kn}=0.002$. To be consistent with the study by Lofthouse, molecular parameters based on the VHS model and the temperature-dependent collision numbers are employed. The Parker model is used for the rotational collision number, and the high-temperature corrected Millikan-White model is adopted for the vibrational collision number \cite{Bird_94_DSMC-book, Boyd17_Neq-book}:
\begin{equation}
    Z_{rot}=\frac{Z_{rot}^{\infty}}
                 {1 + (\pi^{3/2}/2)(T^*/T_{tr})^{1/2} + (\pi+\pi^2/4)(T^*/T_{tr})}
    ,
\end{equation}  
\begin{equation}
    Z_{vib}
    = 
    Z_{vib}^{MH} + Z_{vib}^{HT}
    = 
    \Big(C_1/T_{tr}^{c_{\omega}} \Big) \exp{\Big( C_2 T_{tr}^{-1/3} \Big)}
    +
    \Big( \pi d_{ref}^2 / \sigma_{vib}^{Haas,Boyd} \Big) \Big( T_{ref} / T_{tr} \Big)^{c_{\omega}-0.5}
    ,
\end{equation}
with parameters $C_1=9.1$, $C_2=220.0$, $Z_{rot}^\infty=23.0$, $T^*=91.5\,\rm{K}$, and $\sigma_{vib}^{Haas,Boyd}=5.81\times10^{-21}\,\rm{m^2}$.The cylinder surface is maintained at a temperature of $T_{wall}=500\,\rm{K}$, with a fully diffusive boundary condition. A half-body simulation is performed with a reflective boundary condition along the lower x-axis. The computational domain measures $0.8\,\rm{m} \times 0.6\,\rm{m}$. A three-level mesh refinement is employed to efficiently resolve flow features. The Level 1 grid covers the entire computational domain and captures the freestream region. Around the cylinder, a Level 2 grid is used to resolve the shock wave, applied within a circular region of radius $R_2=0.24\,\rm{m}$. Closer to the surface, a finer Level 3 grid is applied to resolve the high-density region within a smaller circular area of radius $R_3=0.155\,\rm{m}$. Each level doubles the grid resolution. The computational domain is illustrated in Figure \ref{fig:cylinder_domain}. The DSMC method with a fine spatio-temporal resolution acts as a reference, with a Level 1 grid cell size of $\Delta x_{ref} = 0.25 \times 10^{-4} \,\rm{m}$ and time step of $\Delta t_{ref} = 1.6 \times 10^{-8} \,\rm{s}$. A coarse resolution is used for the DSMC, FPM, and USP-FPM methods, using a Level 1 grid cell size of $\Delta x=5\Delta x_{ref}$ and time step of $\Delta t=10\Delta t_{ref}$. In all cases, 100 computational particles per Level 1 grid cell are initialized. The simulation time to reach a steady state is set to $400000\Delta t_{ref}$, which corresponds to $N_{steady}=400000$ iterations for the fine case and $N_{steady}=40000$ for the coarse case. The number of iterations for sampling is fixed at $N_{sampling}=100000$ for all cases. The numerical parameters for each case are summarized in Table \ref{tab:case_cylinder}.

\begin{table}[t]
\caption{\label{tab:case_cylinder}Numerical parameters for the hypersonic flow around a cylinder.}
\begin{ruledtabular}
    \begin{tabular}{l ccccl}
        Case &
        $\Delta x$ for Level 1 grid &
        $\Delta t$ &
        $N_{steady}$ &
        $N_{sampling}$ &
        Applied Method \\
        \hline
        Fine &
        \makecell[c]{$2.5 \times 10^{-4}\,\rm{m}$ $(\Delta x_{ref})$} &
        \makecell[c]{$1.6 \times 10^{-8}\,\rm{m}$ $(\Delta t_{ref})$} &
        \makecell[r]{$400,000$} &
        \makecell[r]{$100,000$} &
        DSMC \\[0.5em]
        Coarse &
        $5 \Delta x_{ref}$ &
        $10 \Delta t_{ref}$ &
        \makecell[r]{$40,000$} &
        \makecell[r]{$100,000$} &
        \makecell[l]{DSMC, FPM, \\ USP-FPM} \\
    \end{tabular}
\end{ruledtabular}
\end{table}


\begin{figure*}[b]
    \centering
	\begin{subfigure}{0.32\textwidth}
		\centering
        \includegraphics[width=1.\linewidth]{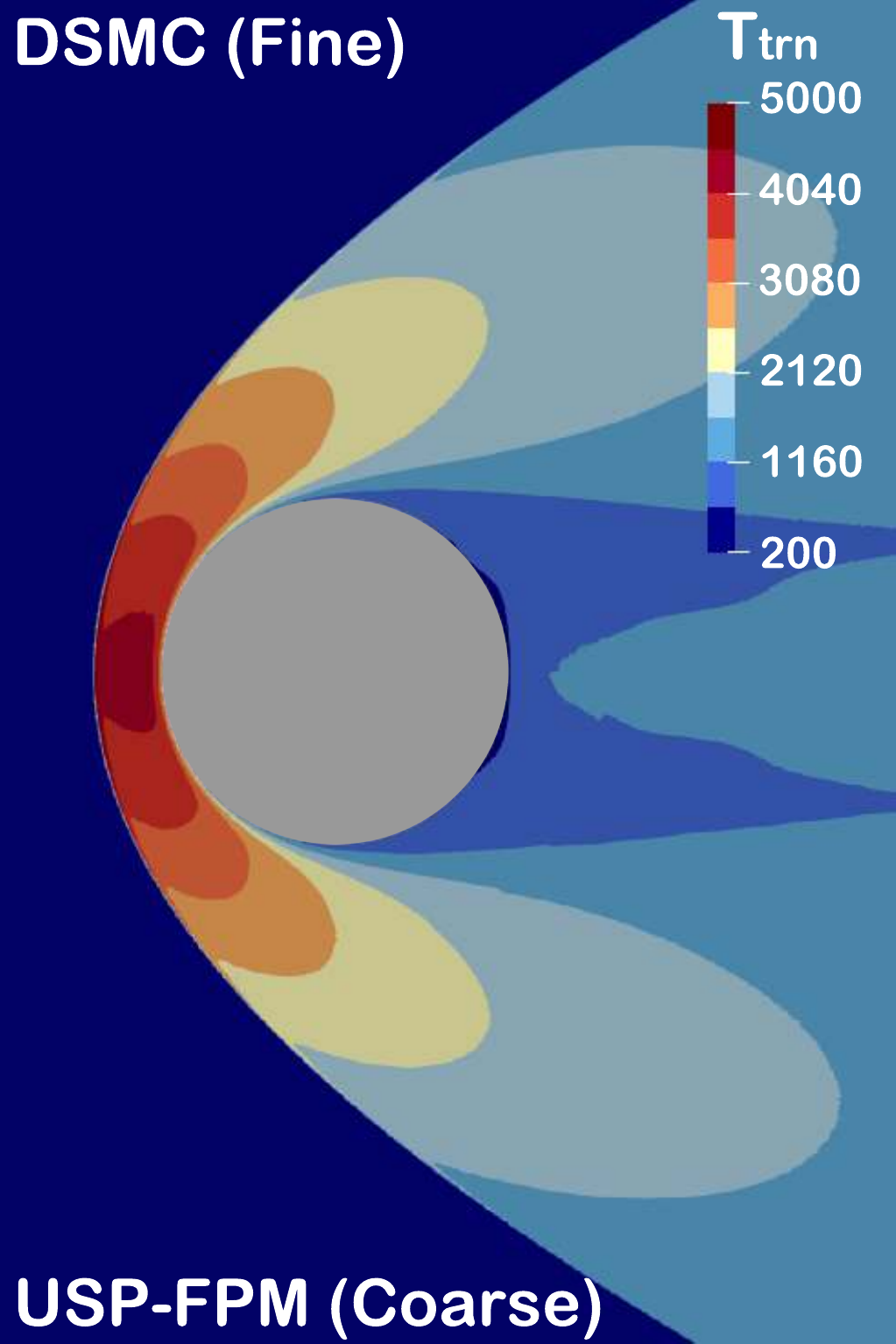}
            \caption{\label{subfig:cylinder_contour_Ttr}Translational tempearture.}
	\end{subfigure}
	\begin{subfigure}{0.32\textwidth}
		\centering
        \includegraphics[width=1.\linewidth]{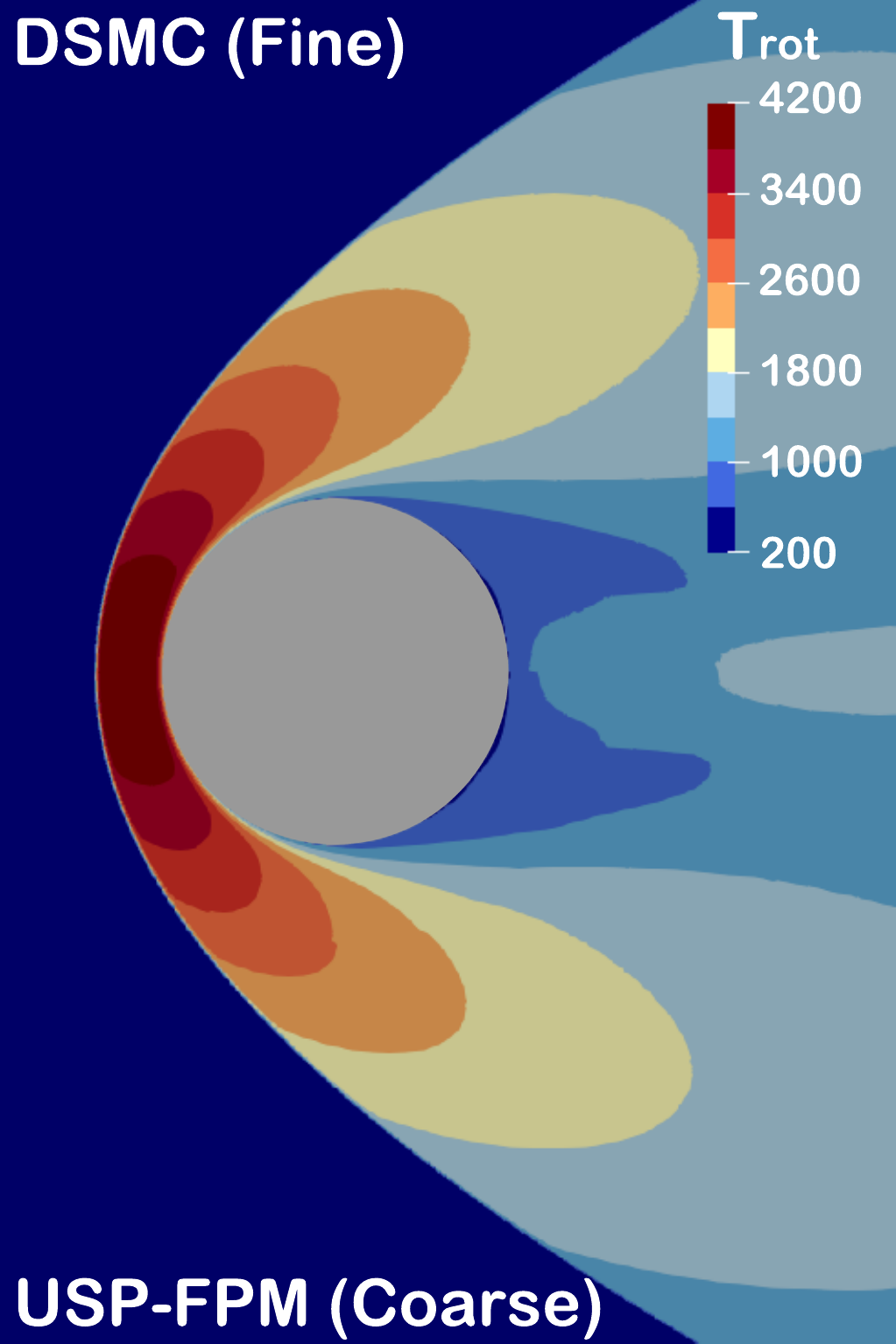}
            \caption{\label{subfig:cylinder_contour_Trot}Rotational tempearture.}
	\end{subfigure}
	\begin{subfigure}{0.32\textwidth}
		\centering
        \includegraphics[width=1.\linewidth]{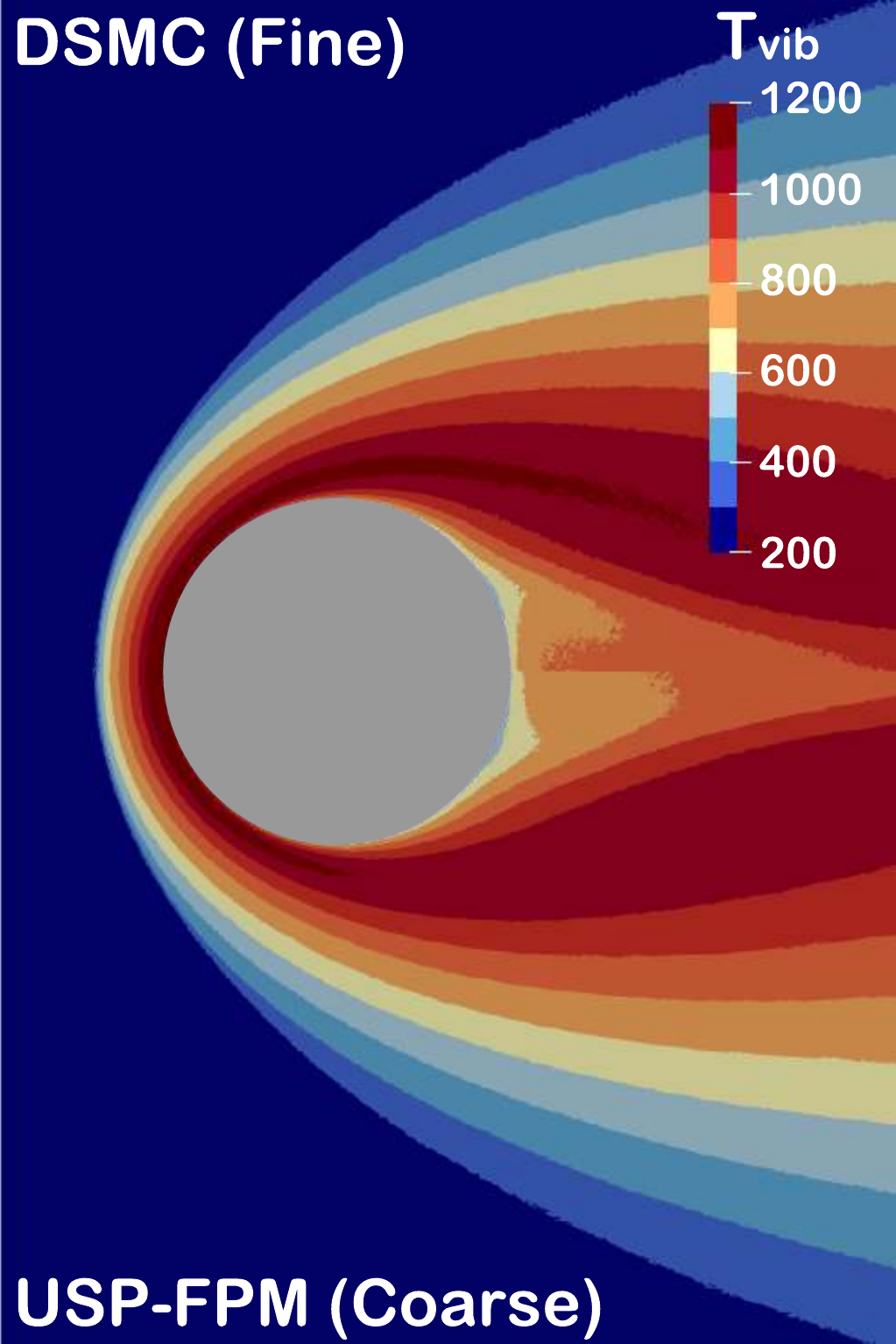}
            \caption{\label{subfig:cylinder_contour_Tvib}Vibrational tempearture.}
	\end{subfigure}
	\caption{\label{fig:cylinder_contour}Temperature contours around a cylinder.}
\end{figure*}

Figures \ref{subfig:cylinder_contour_Ttr}, \ref{subfig:cylinder_contour_Trot}, and \ref{subfig:cylinder_contour_Tvib} show the translational, rotational, and vibrational temperature contours around a cylinder, respectively. As shown in Figure \ref{fig:cylinder_contour}, a strong bow shock is formed in front of the cylinder. After the shock, thermal nonequilibrium is observed. The USP-FPM method with the coarse resolution generally shows good agreement compared to the DSMC method with the fine resolution.


\begin{figure*}[b]
    \centering
	\begin{subfigure}{0.495\textwidth}
		\centering
		\includegraphics[width=1.\linewidth]{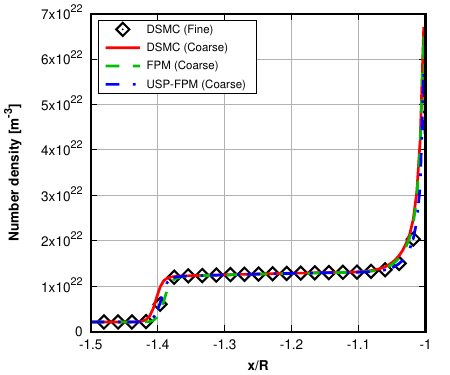}
            \caption{\label{subfig:cylinder_nrho}Number density.}
	\end{subfigure}
	\begin{subfigure}{0.495\textwidth}
		\centering
		\includegraphics[width=1.\linewidth]{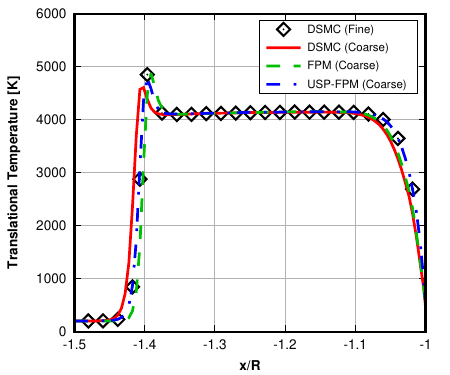}
            \caption{\label{subfig:cylinder_Ttr}Translational temperature.}
	\end{subfigure}
    \centering
	\begin{subfigure}{0.495\textwidth}
		\centering
		\includegraphics[width=1.\linewidth]{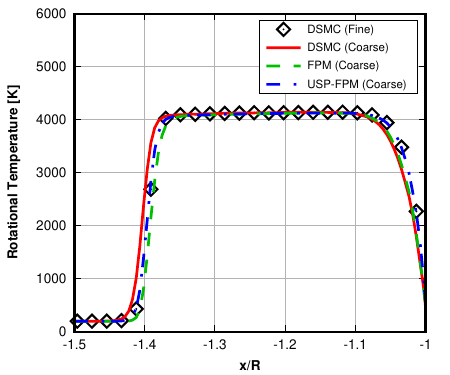}
            \caption{\label{subfig:cylinder_Trot}Rotational temperature.}
	\end{subfigure}
	\begin{subfigure}{0.495\textwidth}
		\centering
		\includegraphics[width=1.\linewidth]{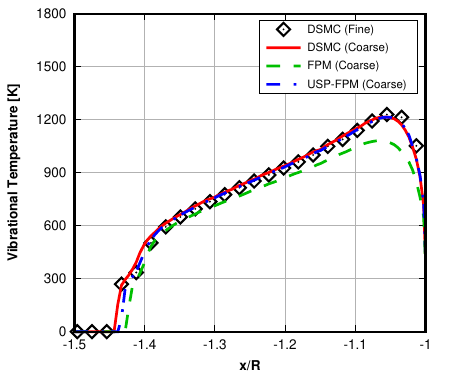}
            \caption{\label{subfig:cylinder_Tvib}Vibrational temperature.}
	\end{subfigure}
	\caption{\label{fig:cylinder_plot}The number density and temperature profiles along the stagnation line.}
\end{figure*}

For quantitative comparison, number density and temperature profiles along the stagnation line are presented in Figure \ref{fig:cylinder_plot}. Figure \ref{subfig:cylinder_nrho} shows the number density profile, and Figures \ref{subfig:cylinder_Ttr}, \ref{subfig:cylinder_Trot}, and \ref{subfig:cylinder_Tvib} present the translational, rotational, and vibrational temperature profiles, respectively. The flow passes through the shock near $x/R=-1.4$, resulting in a sharp rise in the number density, translational temperature, and rotational temperature. This region of rapid transition is referred to as the shock front. As the flow proceeds, the number density remains constant, and the translational and rotational temperatures equilibrate and remain near $4000\,\rm{K}$. The vibrational temperature begins to rise near $x/R=-1.4$, but increases more slowly than the rotational temperature and does not reach the translational temperature. This is due to the higher vibrational collision number, as the relaxation rate is inversely proportional to the collision number \cite{Boyd17_Neq-book}. Near the surface, the number density further increases, while all temperatures decrease toward the surface temperature. Compared to the DSMC method with the fine resolution, the DSMC method with the coarse resolution exhibits an upstream shift of the shock front, as seen in Figures \ref{subfig:cylinder_nrho}-\ref{subfig:cylinder_Tvib}. This shift occurs because the large cell size allows unphysical collisions between pre-shock and post-shock particles. The DSMC method with the coarse resolution shows a thicker boundary layer, since particles reflected from the surface can travel longer distances and collide with particles far from the surface. The FPM method with the coarse resolution predicts a downstream shift of the shock front, as shown in Figures \ref{subfig:cylinder_nrho}-\ref{subfig:cylinder_Tvib}. This downstream shift results from post-shock particles acquiring excessive velocity and undergoing longer displacement due to the coarser spatial and temporal resolution. As shown in Figure \ref{subfig:cylinder_Tvib}, the FPM method with the coarse resolution underestimates the vibrational temperature, due to the first-order accuracy in the vibrational temperature relaxation. The FPM method with the coarse resolution shows a thicker boundary layer near the surface, as the large cell size leads to overestimated cell-averaged temperatures for particles near the surface, increasing energy transfer to the surface. The USP-FPM method with the coarse resolution shows good agreement with the DSMC method with the fine resolution.


\begin{figure*}[t]
    \centering
	\begin{subfigure}{0.495\textwidth}
		\centering
		\includegraphics[width=1.\linewidth]{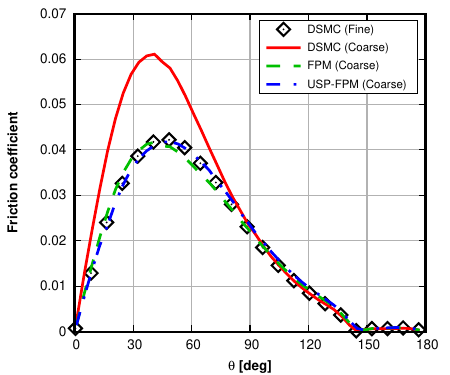}
            \caption{\label{subfig:cylinder_Cf}Friction coefficient.}
	\end{subfigure}
	\begin{subfigure}{0.495\textwidth}
		\centering
		\includegraphics[width=1.\linewidth]{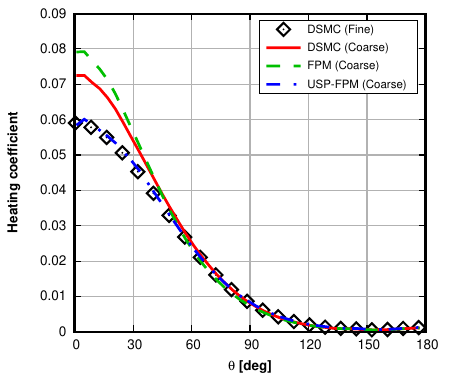}
            \caption{\label{subfig:cylinder_Ch}Heating coefficient.}
	\end{subfigure}
	\caption{\label{fig:cylinder_surf}The surface coefficient distribution on the surface.}
\end{figure*}

Accurate estimation of surface properties is an important part of practical engineering. Figure \ref{fig:cylinder_surf} illustrates the distribution of surface properties along the cylinder. In Figures \ref{subfig:cylinder_Cf} and \ref{subfig:cylinder_Ch}, the friction coefficient $C_f$ and heating coefficient $C_h$ are plotted as functions of the surface angle $\Theta$. The angle $\Theta$ is measured from the stagnation point and increases clockwise. The friction and heating coefficients are defined as follows:
\begin{equation}
    C_f = \frac{\sigma}{\frac{1}{2}\rho_{\infty} U_{\infty}^2}
    ,
\end{equation}
\begin{equation}
    C_h = \frac{q}{\frac{1}{2}\rho_{\infty} U_{\infty}^3}
    ,
\end{equation}
where $\rho_{\infty}$ is the freestream density. Figure \ref{subfig:cylinder_Cf} indicates that the friction coefficient increases and reaches its peak near $\Theta = 45 \,\rm{deg}$, since the flow accelerates along the surface, increasing the velocity gradient near the surface. Beyond this peak, the growth of the boundary layer reduces the near-wall velocity gradient, resulting in a decrease in the friction coefficient. Compared to the DSMC method with the fine resolution, the DSMC method with the coarse resolution overestimates the friction coefficient due to unphysical collisions near the surface. The FPM method with the coarse resolution underestimates the friction coefficient beyond $\Theta = 45\,\rm{deg}$. This is because the FPM method with the coarse resolution predicts a thicker boundary layer, leading to lower near-wall velocity and reduced tangential momentum transfer to the surface. The USP-FPM method with the coarse resolution shows good agreement with the DSMC method with the fine resolution. As illustrated in Figure \ref{subfig:cylinder_Ch}, the heating coefficient peaks at the stagnation point, where the maximum pressure occurs. The DSMC and FPM methods with the coarse resolution overestimate the peak heat flux compared to the DSMC method with the fine resolution, because particles near the surface are assigned excessive energy and collide with the surface. In contrast, the USP-FPM method with the coarse resolution shows close agreement with the DSMC method with the fine resolution. To highlight differences in total aerodynamic and thermal loads, the total drag and peak heat flux are summarized in Table \ref{tab:cylinder_property}, along with the corresponding values obtained by Lofthouse  \cite{Lofthouse08_cylinder}. The total drag and peak heat flux obtained from the DSMC method with the fine resolution closely match those obtained by Lofthouse. Small discrepancies are attributed to the fact that their spatio-temporal resolution was not sufficiently fine \cite{Kim24_USP-FP}. The total drag values obtained from the DSMC, FPM, and USP-FPM methods with the coarse resolution show good agreement with the DSMC method with the fine resolution, with errors below $1.07\,\%$. Meanwhile, the DSMC and FPM methods with coarse resolution overestimate the peak heat flux by $22.9\,\%$ and $34.4\,\%$, respectively, compared to the DSMC method with the fine resolution. In contrast, the USP-FPM method predicts the peak heat flux with only a $0.6\,\%$ error, demonstrating excellent consistency with the DSMC result with the fine resolution.

\begin{table}[!t]
\caption{\label{tab:cylinder_property}Surface properties in the hypersonic flow around a cylinder.}
\begin{ruledtabular}
    \begin{tabular}{ll ll}
        \multirow{2}{*}{Method} & 
        \multirow{2}{*}{Case}   & 
        Total drag $[\rm{N/m}]$ & 
        Peak heat flux $[\rm{kW/m^2}]$ \\
                                 &
                                 &
         (\% difference)         &
         (\% difference)         \\
        \hline
        DSMC (Lofthouse) & -      & 162.4           & 69.88           \\
        DSMC                    & Fine   & 161.4 ( - )     & 69.89 ( - )     \\
        DSMC                    & Coarse & 163.2 (1.07 \%) & 85.89 (22.9 \%) \\
        FPM                     & Coarse & 161.3 (0.07 \%) & 93.90 (34.4 \%) \\
        USP-FPM                 & Coarse & 161.6 (0.10 \%) & 69.46 (0.6 \%)  \\
    \end{tabular}
\end{ruledtabular}
\end{table}


\begin{table}[t]
\caption{\label{tab:cylinder_CPU}Computational parameters at the end of the simulation and total CPU time for the hypersonic flow around a cylinder.}

\begin{ruledtabular}
    \begin{tabular}{ll rrr}
        Method  &   Case &  Number of cells & \makecell[r]{Number of \\ computational particles} & \makecell[r]{Total CPU time $[\rm{h}]$ \\ (speed-up ratio)} \\
        \hline
        DSMC    & Fine   & $19,281,597$ & $1,050,882,185$ & $64,971 \;(1)\;\;$ \\
        DSMC    & Coarse &    $773,325$ &    $42,367,770$ &    $853 \;(76)$ \\
        FPM     & Coarse &    $773,325$ &    $42,167,327$ &  $1,082 \;(60)$ \\
        USP-FPM & Coarse &    $773,325$ &    $42,044,423$ &  $2,347 \;(28)$ \\
    \end{tabular}
\end{ruledtabular}
\end{table}

Table \ref{tab:cylinder_CPU} presents the computational parameters at the end of the simulation along with the total CPU time. All simulations were conducted using 192 cores on an AMD EPYC 9654 processor with a clock speed of $2.4 \, {\rm{GHz}}$. Compared to the DSMC method with the fine resolution, the USP-FPM method with the coarse resolution achieves a speed-up by a factor of $28$, using only $4\,\%$ of the grid cells and computational particles. While the USP-FPM method is slower than the DSMC and FPM methods under the same spatio-temporal resolution due to the interpolation of macroscopic properties and the evaluation of drift and diffusion coefficients for each particle, it maintains high accuracy even with coarser cell sizes and larger time steps, thereby offering a significant computational advantage.

\section{\label{sec:conclusions}Conclusion}


This study presents the USP-FPM method for diatomic gas flows, which achieves second-order accuracy in both time and space. The USP-FPM method attains second-order temporal accuracy by reproducing the second-order relaxation behavior of energy, viscous stress, and heat flux. The second-order spatial accuracy is achieved through a first-order polynomial reconstruction method. The USP-FPM method is evaluated across three test cases: homogeneous relaxation, Poiseuille flow, and hypersonic flow around a cylinder. The accuracy and efficiency of the USP-FPM method are assessed through comparisons with the DSMC and FPM methods across various cell sizes and time steps. The results demonstrate that the USP-FPM method yields accurate solutions compared to the DSMC and FPM methods with the same coarse cell sizes and large time steps. In the hypersonic flow around a cylinder, the USP-FPM method with a coarse resolution achieves a computational speed-up of a factor of 28 compared to the DSMC method with a fine resolution, while maintaining high accuracy. Future work includes extending the USP-FPM method to polyatomic gases and mixtures by accounting for interactions between vibrational modes and momentum exchange between species.

\begin{acknowledgments}

This work was supported by the National Research Foundation of Korea(NRF) grant funded by the Korea government(MSIT) (N01250638). This work was also supported by the National Supercomputing Center with supercomputing resources including technical support(KSC-2025-CRE-0427).

\end{acknowledgments}


\section*{Data Availability}
The data that support the findings of this study are available from the corresponding author upon reasonable request.

\appendix

\section{\label{appendix:FPM}Moment relaxations in the FPM method for diatomic gases}



The moment relaxations in the FPM method are derived from the update schemes in Equations (\ref{eq:conventionial-velo}) and (\ref{eq:conventionial-erot}). The analysis begins with the translational energy relaxation. Multiplying $C_j^{n+1}$ with $C_j^{n+1}/2$ yields the following expression:
\begin{equation}
\begin{aligned}
    \frac{1}{2} C_j^{n+1} C_j^{n+1} 
    = \; & 
    \frac{1}{2} 
    \bigg( C_j^n \exp{(-\frac{\Delta t}{\tau})} 
    + \sqrt{\tau \Big( 1-\exp{(-\frac{2\Delta t}{\tau})} \Big)} d_{tr,\, jk} \, G_k
    \bigg)
    \\[0.2em] & \cdot
    \bigg( C_j^n \exp{(-\frac{\Delta t}{\tau})} 
     + \sqrt{\tau \Big( 1-\exp{(-\frac{2\Delta t}{\tau})} \Big)} d_{tr,\, jl} \, G_l
    \bigg)
    \\[0.5em] = \; & 
    \frac{1}{2} 
    C_j^n C_j^n \exp{(-2\frac{\Delta t}{\tau})} 
    + 
    \frac{\tau}{2} 
     \Big( 1-\exp{(-\frac{2\Delta t}{\tau})} \Big) \Big( d_{tr,\, jk} \, d_{tr,\, jl} \, G_k \, G_l \Big)
    \\[0.2em] & + 
    \frac{1}{2} 
      \sqrt{\tau \Big( \exp{(-\frac{2\Delta t}{\tau})} - \exp{(-\frac{4\Delta t}{\tau}}) \Big)} 
      \Big( d_{tr,\, jk} \, C_j^n \,  G_k + d_{tr,\, jl} \, C_j^n \, G_l \Big)
    .
\end{aligned}
\end{equation}
Taking the ensemble average on both sides yields the following equation:
\begin{equation}
    \langle \frac{1}{2} C_j^{n+1} C_j^{n+1} \rangle 
    = 
    \langle \frac{1}{2} 
    C_j^n C_j^n \rangle \exp{(-\frac{2\Delta t}{\tau})} 
    + 
    \frac{\tau}{2} 
     \Big( 1-\exp{(-\frac{2\Delta t}{\tau})} \Big) D_{tr,\, jj}
    ,
\end{equation}
where $\langle C_j^n \,  G_k \rangle = \langle C_j^n \, G_l \rangle = 0$ and $d_{tr,\, jk} \, d_{tr,\, jl} \, \langle G_k \, G_l \rangle = d_{tr,\, jk} \, d_{tr,\, jl} \, \delta_{kl} = D_{tr, \, jj}$. The translational energy relaxation can be expressed as:
\begin{equation}
\begin{aligned}
    e_{tr}(T_{tr}^{n+1})
    =  
    e_{tr}(T_{tr}^{n}) & \exp{(-\frac{2\Delta t}{\tau})} 
    + 
        e_{tr}(T_{tr}^{rel, \,n}) \Big( 1-\exp{(-\frac{2\Delta t}{\tau})} \Big) 
    \\[0.5em] =  
    e_{tr}(T_{tr}^{n}) 
    & + 
    \Big( 1-\exp{(-\frac{2\Delta t}{\tau})} \Big) \,
    \frac{\tau}{2\tau_{rot}} \Big( e_{rot}(T_{rot}^n) - e_{rot}(T_{tr}^n) \Big) 
    \\[0.2em] & +
    \Big( 1-\exp{(-\frac{2\Delta t}{\tau})} \Big) \,
    \frac{\tau}{2\tau_{vib}} \Big( e_{vib}(T_{vib}^n) - e_{vib}(T_{tr}^n) \Big)
    ,
\end{aligned}
\end{equation}
A Taylor expansion of the translational energy relaxation yields:
\begin{equation}
\begin{aligned}
    e_{tr}(T_{tr}^{n+1})
    = e_{tr} (T_{tr}^n)
    & \Bigg[
        \frac{ e_{rot}(T_{rot}^n) - e_{rot}(T_{tr}^n) }{\tau_{rot}}
        +
        \frac{ e_{vib}(T_{vib}^n) - e_{vib}(T_{tr}^n) }{\tau_{vib}}
    \Bigg]
    \Delta t
    \\[0.5em] 
    - \frac{1}{\tau}
    & \Bigg[ 
        \frac{ e_{rot}(T_{rot}^n) - e_{rot}(T_{tr}^n) }{\tau_{rot}}
        +
        \frac{ e_{vib}(T_{vib}^n) - e_{vib}(T_{tr}^n) }{\tau_{vib}}
    \Bigg]
    \Delta t^2
    + \mathcal{O}(\Delta t^3)
    .
\end{aligned}
\end{equation}
The viscous stress relaxation is evaluated by taking the ensemble average of $(C_iC_j-\frac {1}{3} C^2\delta_{ij})$:
\begin{equation}
\begin{aligned}
    \sigma_{ij}^{n+1} = \; &
    \langle C_i^{n+1} C_j^{n+1} - \frac{1}{3}C_k^{n+1} C_k^{n+1} \delta_{ij} \rangle
    \\ = \; & \Bigg<
    \bigg( C_i^n \exp{(-\frac{\Delta t}{\tau})} 
    + \sqrt{\tau \Big( 1-\exp{(-\frac{2\Delta t}{\tau})} \Big)} d_{tr,\, il} \, G_l
    \bigg)
    \\ & \cdot
    \bigg( C_j^n \exp{(-\frac{\Delta t}{\tau})} 
    + \sqrt{\tau \Big( 1-\exp{(-\frac{2\Delta t}{\tau})} \Big)} d_{tr,\, jm} \, G_m
    \bigg) \Bigg>
    \\ &-\Bigg< \frac{1}{3} \delta_{ij}
    \bigg( C_k^n \exp{(-\frac{\Delta t}{\tau})} 
    + \sqrt{\tau \Big( 1-\exp{(-\frac{2\Delta t}{\tau})} \Big)} d_{tr,\, kp} \, G_p
    \bigg)
    \\ & \cdot
    \bigg( C_k^n \exp{(-\frac{\Delta t}{\tau})} 
     + \sqrt{\tau \Big( 1-\exp{(-\frac{2\Delta t}{\tau})} \Big)} d_{tr,\, kq} \, G_q
    \bigg) \Bigg>
    \\[0.5em] =\; &
    \langle C_i^n C_j^n - \frac{1}{3} C_k^n C_k^n \rangle \exp{(-\frac{2\Delta t}{\tau})} 
    + 
    \tau 
     \Big( 1-\exp{(-\frac{2\Delta t}{\tau})} \Big) \Big( D_{tr,\, ij} - \frac{1}{3} D_{tr,\, kk} \; \delta_{ij} \Big)
    \\[0.5em] =\; &
    \sigma_{ij}^{n} \bigg( \exp{(-\frac{2\Delta t}{\tau})} 
    + 
    \nu \Big( 1-\exp{(-\frac{2\Delta t}{\tau})} \Big) \bigg)
    .
\end{aligned}
\end{equation}
The Taylor expansion of the viscous stress relaxation is expressed as:
\begin{equation}
    \sigma_{ij}^{n+1}
    = \sigma_{ij}^n
    \bigg( 
        1 
        - \frac{2(1-\nu)}{\tau} \Delta t
        + \frac{2(1-\nu)}{\tau^2} \Delta t^2
    \bigg)
    + \mathcal{O} (\Delta t^3)
    .
\end{equation}
The translational heat flux relaxation results from the ensemble average of $C_iC_jC_j/2$:
\begin{equation}
\begin{aligned}
    q_{tr,\, i}^{n+1}
    = \; & \langle \frac{1}{2} C_i^{n+1} C_j^{n+1} C_j^{n+1} \rangle
    \\[0.5em] = \; & \Bigg< \frac{1}{2}
    \bigg( C_i^n \exp{(-\frac{\Delta t}{\tau})} 
    + \sqrt{\tau \Big( 1-\exp{(-\frac{2\Delta t}{\tau})} \Big)} d_{tr,\, il} \, G_l
    \bigg)
    \\ & \cdot
    \bigg( C_j^n \exp{(-\frac{\Delta t}{\tau})} 
    + \sqrt{\tau \Big( 1-\exp{(-\frac{2\Delta t}{\tau})} \Big)} d_{tr,\, jm} \, G_m
    \bigg) 
    \\ & \cdot
    \bigg( C_j^n \exp{(-\frac{\Delta t}{\tau})} 
    + \sqrt{\tau \Big( 1-\exp{(-\frac{2\Delta t}{\tau})} \Big)} d_{tr,\, jn} \, G_n
    \bigg) \Bigg>
    \\[0.5em] = \; & 
    \langle \frac{1}{2}  C_i^n  C_j^n  C_j^n \rangle \exp{(-\frac{3 \Delta t}{\tau})} 
    \\[0.5em] = \; & q_{tr,\, i}^n
        \exp{(-\frac{3\Delta t}{\tau})} 
        .
\label{eq:FPM-qtr-relax-exp}
\end{aligned}
\end{equation}
The corresponding Taylor expansion is given by:
\begin{equation}
    q_{tr,\, i}^{n+1}
    = q_{tr,\, i}^n
    \Bigg[
        1
        - \frac{3}{\tau}\Delta t
        + \frac{9}{2\tau^2} \Delta t^2
    \Bigg]
    + \mathcal{O}(\Delta t^3)
        .
\end{equation}
The rotational energy relaxation is derived by taking the ensemble average of Equation (\ref{eq:conventionial-erot}):
\begin{equation}
\begin{aligned}
    \langle \varepsilon_{rot}^{n+1} \rangle
    = \; &
    \Bigg<
    \frac{RT_{rot}^{rel,\,n}}{2} \Big( 1 - \exp{(-\frac{2\Delta t}{\tau})} \Big)
    \\ & + 
    \Bigg(
        \sqrt{\varepsilon_{rot}^{n}} \, \exp{(-\frac{\Delta t}{\tau})}
    + \sqrt{\frac{RT_{rot}^{rel,\,n}}{2} \Big(1 - \exp{(-\frac{2\Delta t}{\tau})} \Big)} \, G
    \Bigg)^2 
    \Bigg>
    \\[0.5em] = \; &
    \langle \varepsilon_{rot}^{n} \rangle \, \exp{(-\frac{2 \Delta t}{\tau})}
    +
    \frac{RT_{rot}^{rel,\,n}}{2} \Big( 1 - \exp{(-\frac{2\Delta t}{\tau})} \Big) \Big( 1 + \langle G^2 \rangle \Big)
    \\[0.2em] & +
    2 \langle \sqrt{\varepsilon_{rot}^{n}} G \rangle \sqrt{\frac{RT_{rot}^{rel,\,n}}{2} \Big(\exp{(-\frac{2\Delta t}{\tau})} - \exp{(-\frac{4\Delta t}{\tau})} \Big)}
    ,
\end{aligned}
\end{equation}
where $\langle \sqrt{\varepsilon_{rot}^{n}} G \rangle = 0$ and $\langle G^2\rangle = 1$. The rotational energy relaxation is written as:
\begin{equation}
\begin{aligned}
    e_{rot} (T_{rot}^{n+1})
    & =
    e_{rot} (T_{rot}^{n}) \, \exp{(-\frac{2 \Delta t}{\tau})}
    +
    e_{rot} (T_{rot}^{rel,\,n}) \Big( 1 - \exp{(-\frac{2\Delta t}{\tau})} \Big)
    \\ & = e_{rot} (T_{rot}^n)
    +   \Big( e_{rot} (T_{tr}^n) - e_{rot} (T_{rot}^n) \Big) 
        \frac{\tau}{2\tau_{rot}}
        \Big(1 - \exp{(-\frac{2\Delta t}{\tau})} \Big) 
    .
\end{aligned}
\end{equation}
The Taylor expansion of the rotational energy relaxation yields:
\begin{equation}
    e_{rot}(T_{rot}^{n+1})
    =
    e_{rot}(T_{rot}^n)
    + 
    \Bigg[
        \frac{ e_{rot}(T_{tr}^n) - e_{rot}(T_{rot}^n) }{\tau_{rot}}
    \Bigg]
    \Delta t
    + 
    \Bigg[
        \frac{ e_{rot}(T_{tr}^n) - e_{rot}(T_{rot}^n) }{\tau \, \tau_{rot}}
    \Bigg]
    \Delta t^2
    +
    \mathcal{O} (\Delta t^3)
    .
\end{equation}
The rotational heat flux is obtained from the ensemble average of $C_i \varepsilon_{rot}$:
\begin{equation}
\begin{aligned}
    q_{rot,\, i}^{n+1}
    = \; & \langle C_i^{n+1} \varepsilon_{rot}^{n+1} \rangle
    \\[0.5em] = \; &
    \Bigg<
    \bigg[ C_i^n \exp{(-\frac{\Delta t}{\tau})} 
    + \sqrt{\tau \Big( 1-\exp{(-\frac{2\Delta t}{\tau})} \Big)} d_{tr,\, ik} \, G_k
    \bigg]
    \\ & \cdot 
    \Bigg[
    \frac{RT_{rot}^{rel,\,n}}{2} \Big( 1 - \exp{(-\frac{2\Delta t}{\tau})} \Big)
    \\ & + 
    \Bigg(
        \sqrt{\varepsilon_{rot}^{n}} \, \exp{(-\frac{\Delta t}{\tau})}
    + \sqrt{\frac{RT_{rot}^{rel,\,n}}{2} \Big(1 - \exp{(-\frac{2\Delta t}{\tau})} \Big)} \, G
    \Bigg)^2 \Bigg] 
    \Bigg>
    \\[0.5em] = \; & 
    \langle C_i^n \varepsilon_{rot}^{n} \rangle \exp{(-\frac{3\Delta t}{\tau})} 
    \\[0.5em] = \; & 
    q_{rot,\, i}^n  \exp{(-\frac{3\Delta t}{\tau})} 
        .
\end{aligned}
\label{eq:FPM-qrot-relax-exp}
\end{equation}
The Taylor expansion of the rotational heat flux relaxation is:
\begin{equation}
    q_{rot,\, i}^{n+1}
    = q_{rot,\, i}^n
    \Bigg[
        1
        - \frac{3}{\tau}\Delta t
        + \frac{9}{2\tau^2} \Delta t^2
    \Bigg]
    + \mathcal{O}(\Delta t^3)
        .
\end{equation}
Unlike the particle velocity and rotational energy, there is no analytic expression for the evolution of the vibrational energy, because there is an infinite number of vibrational energy levels and they are coupled \cite{Gillespie76_direct}. The vibrational energy is assumed to exactly follow Equation (\ref{eq:LT-evib}) under the assumption of constant moments \cite{Gillespie07_SSA_exact}. Integration of Equation (\ref{eq:LT-evib}) yields the following vibrational energy relaxation:
\begin{equation}
    e_{vib}(T_{vib}^{n+1})
    = e_{vib} (T_{vib}^{n})
    +   \Big( e_{vib} (T_{tr}^n) - e_{vib} (T_{vib}^n) \Big) 
        \frac{\tau}{2\tau_{vib}}
        \Big(1 - \exp{(-\frac{2\Delta t}{\tau})} \Big) 
        .
\end{equation}
The Taylor expansion of the vibrational energy relaxation yields:
\begin{equation}
    e_{vib}(T_{vib}^{n+1})
    =
    e_{vib}(T_{vib}^n)
    + 
    \Bigg[
        \frac{ e_{vib}(T_{tr}^n) - e_{vib}(T_{vib}^n) }{\tau_{vib}}
    \Bigg]
    \Delta t
    + 
    \Bigg[
        \frac{ e_{vib}(T_{tr}^n) - e_{vib}(T_{vib}^n) }{\tau \, \tau_{vib}}
    \Bigg]
    \Delta t^2
    +
    \mathcal{O} (\Delta t^3)
    ,
\end{equation}
The vibrational heat flux is defined as the ensemble average of $C_i \varepsilon_{vib}$:
\begin{equation}
\begin{aligned}
    q_{vib,\,i}^{n+1}
    &=
    \langle C_i^{n+1} \varepsilon_{vib}^{n+1} \rangle
    \\& =
    \langle 
        \bigg( C_i^n \exp{(-\frac{\Delta t}{\tau})} 
        + \sqrt{\tau \Big( 1-\exp{(-\frac{2\Delta t}{\tau})} \Big)} d_{tr,\, ij} \, G_j
        \bigg)
        \bigg(  \varepsilon_{vib}^{n} +\Delta {\varepsilon}_{vib}^n \bigg)
    \rangle
    \\& =
    \langle 
        \Big(  C_i^n \exp{(-\frac{\Delta t}{\tau})} \Big)
        \Big(  \varepsilon_{vib}^{n} +\Delta {\varepsilon}_{vib}^n \Big)
    \rangle
    ,
\label{eq:qvib_appendix}
\end{aligned}
\end{equation}
where $\Delta {\varepsilon}_{vib}^n$ is the change in the vibrational energy after a time step $\Delta t$. The change in the vibrational energy can be decomposed into a deterministic component and a stochastic fluctuation. The stochastic fluctuation is assumed to be statistically independent of the thermal velocity, since it originates from a uniform random number in the Gillespie direct method. Thus, its contribution to the vibrational heat flux can be neglected. Assuming the particle is in a specific vibrational energy level $I$, the time derivative of the vibrational energy is obtained from the master equation as:
\begin{equation}
\begin{aligned}
    \frac{d {\varepsilon}_{vib}^n}{dt}
    & = R\Theta_{vib} \Big(\omega_{I,I+1}- \omega_{I,I-1} \Big)
    \\[0.5em] & = R\Theta_{vib} 
    \Big(
        (I+1) \frac{2}{\tau} \frac{\exp{(-\Theta_{vib}/T_{vib}^{rel})}}{1 - \exp{(-\Theta_{vib}/T_{vib}^{rel})}}  
        - I \frac{2}{\tau} \frac{1}{1 - \exp{(-\Theta_{vib}/T_{vib}^{rel})}}  
    \Big)
    \\[0.5em] & = R\Theta_{vib} 
    \Big(
        - I \frac{2}{\tau}  
        + \frac{2}{\tau} \frac{\exp{(-\Theta_{vib}/T_{vib}^{rel})}}{1 - \exp{(-\Theta_{vib}/T_{vib}^{rel})}}  
    \Big)
    \\[0.5em] & = - \varepsilon_{vib}^n \frac{2}{\tau} + e_{vib} (T_{vib}^{rel}) \frac{2}{\tau} 
    .
\end{aligned}
\end{equation}
By assuming that the vibrational relaxation temperature is constant over $\Delta t$, higher-order time derivatives can be recursively computed. The change in the vibrational energy can be approximated using a Taylor expansion:
\begin{equation}
\begin{aligned}
    \Delta {\varepsilon}_{vib}^n
    = &
    \frac{d {\varepsilon}_{vib}^n}{dt} \Delta t
    +
    \frac{1}{2} \frac{d^2 {\varepsilon}_{vib}^n}{dt^2} \Delta t^2
    +
    \frac{1}{6} \frac{d^3 {\varepsilon}_{vib}^n}{dt^3} \Delta t^3
    +
    \mathcal{O}(\Delta t^4)
    \\[0.5em] = &
    \bigg( - \varepsilon_{vib}^n \frac{2}{\tau} + e_{vib} (T_{vib}^{rel}) \frac{2}{\tau}  \bigg) \Delta t
    \\[0.5em] & +
    \frac{1}{2} \bigg( \varepsilon_{vib}^n \frac{4}{\tau^2} - e_{vib} (T_{vib}^{rel}) \frac{4}{\tau^2}  \bigg) \Delta t^2
    \\[0.5em] & +
    \frac{1}{6} \bigg( - \varepsilon_{vib}^n \frac{8}{\tau^3} + e_{vib} (T_{vib}^{rel}) \frac{8}{\tau^3}  \bigg) \Delta t^3
    +
    \mathcal{O}(\Delta t^4)
    .
\end{aligned}
\end{equation}
By substituting this equation into Equation (\ref{eq:qvib_appendix}), the vibrational heat flux can be approximated as:
\begin{equation}
\begin{aligned}
    q_{vib,\,i}^{n+1}
    = & \;
    \Bigg< 
        \bigg( C_i^n \Big( 1 - \frac{\Delta t}{\tau} +  \frac{\Delta t^2}{2\tau^2}  - \frac{\Delta t^3}{6\tau^3} + \mathcal{O}(\Delta t^4)  \Big)
        \bigg)
        \bigg( \varepsilon_{vib}^{n} 
             + 
            \bigg( - \varepsilon_{vib}^n \frac{2}{\tau} + e_{vib} (T_{vib}^{rel}) \frac{2}{\tau}  \bigg) \Delta t
            \\[0.5em] & +
            \frac{1}{2} \bigg( \varepsilon_{vib}^n \frac{4}{\tau^2} - e_{vib} (T_{vib}^{rel}) \frac{4}{\tau^2}  \bigg) \Delta t^2
             +
            \frac{1}{6} \bigg( - \varepsilon_{vib}^n \frac{8}{\tau^3} + e_{vib} (T_{vib}^{rel}) \frac{8}{\tau^3}  \bigg) \Delta t^3
             +
            \mathcal{O}(\Delta t^4)
    \bigg) 
    \Bigg>
    \\[0.5em] =  &
    \langle C_i^n \varepsilon_{vib}^n \rangle
        \bigg( 1 - \frac{\Delta t}{\tau} +  \frac{\Delta t^2}{2\tau^2}  - \frac{\Delta t^3}{6\tau^3} + \mathcal{O}(\Delta t^4)  \bigg)
        \bigg( 1 -  \frac{2\Delta t}{\tau} + \frac{1}{2}\frac{4\Delta t^2}{\tau^2} - \frac{1}{6} \frac{8\Delta t^3}{\tau^3} + \mathcal{O}(\Delta t^4) \bigg)
    \\[0.5em] = &
    q_{vib,\,i}^{n}
        \bigg( 1 - \frac{3\Delta t}{\tau} +  \frac{9\Delta t^2}{2\tau^2}  - \frac{27\Delta t^3}{6\tau^3} + \mathcal{O}(\Delta t^4)  \bigg) 
    . 
\end{aligned}
\label{eq:FPM-qvib-relax-exp}
\end{equation}


\section{\label{appendix:Taylor}Taylor expansion of the moment production terms for homogeneous flows in the FPM model}



In homogeneous flows, the moment equations can be derived by averaging quantities $\phi$ over the kinetic equation:
\begin{equation}
    \sum_{I=0}^{\infty} \int_{\mathbb{R}^+} \int_{\mathbb{R}^3}
    \phi \, \frac{\partial \mathcal{F}}{\partial t} \, dc_i d\varepsilon_{rot}
    =
    \sum_{I=0}^{\infty} \int_{\mathbb{R}^+} \int_{\mathbb{R}^3}
    \phi \, \mathcal{S}(\mathcal{F}) \, dc_i d\varepsilon_{rot}
    ,
\label{eq:moment-eq}
\end{equation}
where $\mathcal{S}(\mathcal{F})$ is the collision term. The collision term of the FPM model is given as follows:
\begin{multline}
    \mathcal{S}(\mathcal{F})
    =
    - \frac{\partial}{\partial c_i} (A_{tr,\, i} \, \mathcal{F})
    + \frac{\partial^2}{\partial c_i \partial c_j} (D_{tr,\, ij} \, \mathcal{F}) \\
    - \frac{\partial}{\partial \varepsilon_{rot}} (A_{rot} \, \mathcal{F})
    + \frac{\partial^2}{\partial \varepsilon_{rot}} (D_{rot} \, \mathcal{F})
    + \sum_{J=0}^{\infty} (\omega_{J,I} \, \mathcal{F}_J - \omega_{I,J} \, \mathcal{F}_I) 
    .
\end{multline}
The production term refers to the right-hand side of Equation \ref{eq:moment-eq}, and is denoted as $\mathcal{P}_{\mathcal{S}}(\phi)$. Equation \ref{eq:moment-eq} can be rewritten as:
\begin{equation}
    \frac{\partial \, \langle \phi \mathcal{F} \rangle}{\partial t}
    =
    \mathcal{P}_{\mathcal{S}}(\phi)
    = 
    \sum_{I=0}^{\infty} \int_{\mathbb{R}^+} \int_{\mathbb{R}^3}
    \phi \, \mathcal{S}(\mathcal{F}) \, dc_i d\varepsilon_{rot}
\end{equation}


The quantities $\phi \in \{ C^2/2, \varepsilon_{rot}, \varepsilon_{vib} \}$ are associated with translational, rotational, and vibrational energies, respectively:
\begin{equation}
\begin{aligned}
    \frac{\partial \, \langle \frac{1}{2} C^2\, \mathcal{F} \rangle}{\partial t}
    &=
    \frac{\partial \, \big( \rho \, e_{tr} (T_{tr}) \big) }{\partial t}
    \\[0.5em] &=
    \mathcal{P}_{\mathcal{S}}( \frac{1}{2} C^2 )
    \\[0.5em] &=
    \frac{ \rho \, e_{rot}(T_{rot}) -\rho \, e_{rot}(T_{tr}) }{\tau_{rot}}
    +
    \frac{ \rho \, e_{vib}(T_{vib}) -\rho \, e_{vib}(T_{tr}) }{\tau_{vib}}
    .
\end{aligned}
\end{equation}
The first-order time derivative of the translational specific energy is obtained by dividing both sides by $\rho$:
\begin{equation}
    \frac{\partial e_{tr} (T_{tr}) }{\partial t}
    =
    \frac{ e_{rot}(T_{rot}) - e_{rot}(T_{tr}) }{\tau_{rot}}
    +
    \frac{ e_{vib}(T_{vib}) - e_{vib}(T_{tr}) }{\tau_{vib}}
    .
\label{eq:d1-et-Tt}
\end{equation}
In the same way, the first-order time derivatives of the rotational and vibrational specific energies are given by:
\begin{equation}
    \frac{\partial e_{rot} (T_{rot}) }{\partial t}
    =
    \frac{ e_{rot}(T_{tr}) - e_{rot}(T_{rot}) }{\tau_{rot}}
    ,
\label{eq:d1-er-Tr}
\end{equation}
\begin{equation}
    \frac{\partial  e_{vib} (T_{vib}) }{\partial t}
    =
    \frac{ e_{vib}(T_{tr}) - e_{vib}(T_{vib}) }{\tau_{vib}}
    .
\label{eq:d1-ev-Tv}
\end{equation}
By differentiating the above expressions with respect to time, the second-order time derivatives of the specific energies can be obtained.
\begin{equation}
\begin{aligned}
    \frac{\partial^2 e_{tr} (T_{tr}) }{\partial t^2}
    & =
    \frac{1}{\tau_{rot}} 
    \bigg( \frac{ \partial e_{rot}(T_{rot})}{\partial t} - \frac{ \partial e_{rot}(T_{tr})}{\partial t} \bigg)
    +
    \frac{1}{\tau_{vib}} 
    \bigg( \frac{ \partial e_{vib}(T_{vib})}{\partial t} - \frac{ \partial e_{vib}(T_{tr})}{\partial t} \bigg)
    ,
    \\[1.0em] &=
    \frac{1}{\tau_{rot}} 
    \bigg( 
        \frac{ \partial e_{rot}(T_{rot})}{\partial t} 
        -
        \frac{c_{v, \, rot}}{c_{v,\, tr}} \cdot  \frac{ \partial e_{tr}(T_{tr})}{\partial t} 
    \bigg)
    \\[0.5em] & \;\;\;\;\; +
    \frac{1}{\tau_{vib}} 
    \bigg( 
        \frac{ \partial e_{vib}(T_{vib})}{\partial t} 
        -
        \frac{c_{v, \, vib}(T_{tr})}{c_{v,\, tr}} \cdot \frac{ \partial e_{tr}(T_{tr})}{\partial t} 
    \bigg)
    \\[1.0em] &= -
    \bigg( 
        \frac{1}{\tau_{rot}^2} 
        +
        \frac{1}{\tau_{rot} ^2} 
        \frac{c_{v, \, rot}}{c_{v,\, tr}}
        +
        \frac{1}{\tau_{rot} \tau_{vib}} 
        \frac{c_{v, \, vib}(T_{tr})}{c_{v,\, tr}}
    \bigg)
    \bigg( e_{rot}(T_{rot}) - e_{rot}(T_{tr}) \bigg)
    \\[0.5em] & \;\;\; -
    \bigg(
        \frac{1}{\tau_{vib} ^2} 
        +
        \frac{1}{\tau_{vib}^2} 
        \frac{c_{v, \, vib}(T_{tr})}{c_{v,\, tr}}
        +
        \frac{1}{\tau_{rot} \tau_{vib}} 
        \frac{c_{v, \, rot}}{c_{v,\, tr}}
    \bigg)
    \bigg(      
         e_{vib}(T_{vib}) - e_{vib}(T_{tr})
    \bigg)
    ,\\[0.1em]
\end{aligned}
\label{eq:d2-et-Tt-1}
\end{equation}
\begin{equation}
\begin{aligned}
    \frac{\partial^2 e_{rot} (T_{rot}) }{\partial t^2}
    & =
    \frac{1}{\tau_{rot}} 
    \bigg( \frac{ \partial e_{rot}(T_{tr})}{\partial t} - \frac{ \partial e_{rot}(T_{rot})}{\partial t} \bigg)
    \\[1.0em] & = -
    \bigg( 
        \frac{1}{\tau_{rot}^2} 
        +
        \frac{1}{\tau_{rot}^2} 
        \frac{c_{v, \, rot}}{c_{v,\, tr}}
    \bigg)
    \bigg( e_{rot}(T_{tr}) - e_{rot}(T_{rot}) \bigg)
    \\[0.5em] & \;\;\; -
    \bigg(
        \frac{1}{\tau_{rot} \tau_{vib}} 
        \frac{c_{v, \, rot}}{c_{v,\, tr}}
    \bigg)
    \bigg(      
         e_{vib}(T_{tr}) - e_{vib}(T_{vib})
    \bigg)
    ,\\[0.1em]
\end{aligned}
\label{eq:d2-er-Tr}
\end{equation}
\begin{equation}
\begin{aligned}
    \frac{\partial^2 e_{vib} (T_{vib}) }{\partial t^2}
    & =
    \frac{1}{\tau_{vib}} 
    \bigg( \frac{ \partial e_{vib}(T_{tr})}{\partial t} - \frac{ \partial e_{vib}(T_{vib})}{\partial t} \bigg)
    \\[1.0em] & = -
    \bigg( 
        \frac{1}{\tau_{vib}^2} 
        +
        \frac{1}{\tau_{vib}^2} 
        \frac{ c_{v, \, vib} (T_{tr}) }{c_{v,\, tr}}
    \bigg)
    \bigg( e_{vib}(T_{tr}) - e_{vib}(T_{vib}) \bigg)
    \\[0.5em] & \;\;\; -
    \bigg(
        \frac{1}{\tau_{rot} \tau_{vib}} 
        \frac{c_{v, \, vib} (T_{tr})}{c_{v,\, tr}}
    \bigg)
    \bigg(      
         e_{rot}(T_{tr}) - e_{rot}(T_{rot})
    \bigg)
    .
\end{aligned}
\label{eq:d2-ev-Tv}
\end{equation}
Using first and second derivatives, Taylor expansions of specific energies are given by:
\begin{multline}
    e_{tr}(T_{tr}^{n+1})
    = 
    e_{tr}(T_{tr}^n)
    + 
    \frac{\partial e_{tr}(T_{tr}^n)}{\partial t} \Delta t
    + 
    \frac{1}{2!}
    \frac{\partial^2 e_{tr}(T_{tr}^n)}{\partial t^2} \Delta t^2
    +
    \mathcal{O}(\Delta t^3)
    \\[1.0em] =
    e_{tr}(T_{tr}^n)
    + 
    \Bigg[
        \frac{ e_{rot}(T_{rot}^n) - e_{rot}(T_{tr}^n) }{\tau_{rot}}
        +
        \frac{ e_{vib}(T_{vib}^n) - e_{vib}(T_{tr}^n) }{\tau_{vib}}
    \Bigg]
    \Delta t
    \\[0.5em] -
    \frac{1}{2}
    \Bigg[
        \bigg( 
            \frac{1}{\tau_{rot}^2} 
            +
            \frac{1}{\tau_{rot} ^2} 
            \frac{c_{v, \, rot}}{c_{v,\, tr}}
            +
            \frac{1}{\tau_{rot} \tau_{vib}} 
            \frac{c_{v, \, vib}(T_{tr}^n)}{c_{v,\, tr}}
        \bigg)
        \bigg( e_{rot}(T_{rot}^n) - e_{rot}(T_{tr}^n) \bigg)
        \\[0.5em]  +
        \bigg(
            \frac{1}{\tau_{vib} ^2} 
            +
            \frac{1}{\tau_{vib}^2} 
            \frac{c_{v, \, vib}(T_{tr}^n)}{c_{v,\, tr}}
            +
            \frac{1}{\tau_{rot} \tau_{vib}} 
            \frac{c_{v, \, rot}}{c_{v,\, tr}}
        \bigg)
        \bigg(      
             e_{vib}(T_{vib}^n) - e_{vib}(T_{tr}^n)
        \bigg)
    \Bigg]
    \Delta t^2
    +
    \mathcal{O}(\Delta t^3)
    ,
\end{multline}
\begin{multline}
    e_{rot}(T_{rot}^{n+1})
    =
    e_{rot}(T_{rot}^n)
    + 
    \Bigg[
        \frac{ e_{rot}(T_{tr}^n) - e_{rot}(T_{rot}^n) }{\tau_{rot}}
    \Bigg]
    \Delta t
    \\[0.5em] -
    \frac{1}{2}
    \Bigg[
        \bigg( 
            \frac{1}{\tau_{rot}^2} 
            +
            \frac{1}{\tau_{rot}^2} 
            \frac{c_{v, \, rot}}{c_{v,\, tr}}
        \bigg)
        \bigg( e_{rot}(T_{tr}^n) - e_{rot}(T_{rot}^n) \bigg)
        \\[0.5em]  +
        \bigg(
            \frac{1}{\tau_{rot} \tau_{vib}} 
            \frac{c_{v, \, rot}}{c_{v,\, tr}}
        \bigg)
        \bigg(      
             e_{vib}(T_{tr}^n) - e_{vib}(T_{vib}^n)
        \bigg)
    \Bigg]
    \Delta t^2
    +
    \mathcal{O}(\Delta t^3)
    ,
\end{multline}
\begin{multline}
    e_{vib}(T_{vib}^{n+1})
    =
    e_{vib}(T_{vib}^n)
    + 
    \Bigg[
        \frac{ e_{vib}(T_{tr}^n) - e_{vib}(T_{vib}^n) }{\tau_{vib}}
    \Bigg]
    \Delta t
    \\[0.5em]  -
    \frac{1}{2}
    \Bigg[
        \bigg( 
            \frac{1}{\tau_{vib}^2} 
            +
            \frac{1}{\tau_{vib}^2} 
            \frac{ c_{v, \, vib} (T_{tr}^n) }{c_{v,\, tr}}
        \bigg)
        \bigg( e_{vib}(T_{tr}^n) - e_{vib}(T_{vib}^n) \bigg)
        \\[0.5em]  +
        \bigg(
            \frac{1}{\tau_{rot} \tau_{vib}} 
            \frac{c_{v, \, vib} (T_{tr}^n)}{c_{v,\, tr}}
        \bigg)
        \bigg(      
             e_{rot}(T_{tr}^n) - e_{rot}(T_{rot}^n)
        \bigg)
    \Bigg]
    \Delta t^2
    +
    \mathcal{O}(\Delta t^3)
    .
\end{multline}


The quantity $\phi = (C_iC_j - C^2/3 \delta_{ij})$ is associated with viscous stress:
\begin{equation}
\begin{aligned}
    \frac{\partial \, \langle (C_iC_j - C^2/3 \delta_{ij})\, \mathcal{F} \rangle}{\partial t}
    &=
    \frac{\partial \, \big( \sigma_{ij} \big) }{\partial t}
    \\[0.5em] &=
    \mathcal{P}_{\mathcal{S}}( C_iC_j - C^2/3 \delta_{ij} )
    \\[0.5em] &=
    -\frac{2(1-\nu)}{\tau} \sigma_{ij}
    .
\end{aligned}
\end{equation}
The analytic solution of the viscous stress can be obtained as:
\begin{equation}
\begin{aligned}
    \sigma_{ij}^{n+1}
    =
    \sigma_{ij}^n
    \; \exp{\Big( -\frac{2(1-\nu)}{\tau} \Delta t \Big)}
    .
\end{aligned}
\label{eq:analytic_sigma_full}
\end{equation}
The Taylor expansion of the viscous stress is given by:
\begin{equation}
    \sigma_{ij}^{n+1}
    =
    \sigma_{ij}^n
    \Bigg[
    1 -\frac{2(1-\nu)}{\tau} \Delta t + \frac{2(1-\nu)^2}{\tau^2} \Delta t^2
    \Bigg]
    + \mathcal{O}(\Delta t^3)
    .
\end{equation}


The quantities $\phi \in \{ C_iC^2/2, C_i\varepsilon_{rot}, C_i\varepsilon_{vib} \}$ are associated with translational, rotational, and vibrational heat fluxes, respectively:
\begin{equation}
\begin{aligned}
    \frac{\partial \, \langle \frac{1}{2}C_iC^2 \, \mathcal{F} \rangle}{\partial t}
    =
    \frac{\partial \, \big( q_{tr,\, i} \big) }{\partial t}
    =
    \mathcal{P}_{\mathcal{S}}( \frac{1}{2}C_iC^2  )
    =
    -\frac{3}{\tau} q_{tr,\, i}
    ,
\end{aligned}
\end{equation}
\begin{equation}
    \frac{\partial \, \langle C_i\varepsilon_{rot} \, \mathcal{F} \rangle}{\partial t}
    =
    \frac{\partial \, \big( q_{rot,\, i} \big) }{\partial t}
    =
    \mathcal{P}_{\mathcal{S}}( C_i\varepsilon_{rot}  )
    =
    -\frac{3}{\tau} q_{rot,\, i}
    ,
\end{equation}
\begin{equation}
    \frac{\partial \, \langle C_i\varepsilon_{vib} \, \mathcal{F} \rangle}{\partial t}
    =
    \frac{\partial \, \big( q_{vib,\, i} \big) }{\partial t}
    =
    \mathcal{P}_{\mathcal{S}}( C_i\varepsilon_{vib}  )
    =
    -\frac{3}{\tau} q_{vib,\, i}
    .
\end{equation}
The analytic solutions of heat fluxes are expressed as:
\begin{equation}
    q_{tr,\, i}^{n+1}
    =
    q_{tr,\, i}^n
     \exp{\Big( -\frac{3}{\tau} \Delta t \Big)}
    ,
\label{eq:analytic_qtr_full}
\end{equation}
\begin{equation}
    q_{rot,\, i}^{n+1}
    =
    q_{rot,\, i}^n
     \exp{\Big( -\frac{3}{\tau} \Delta t \Big)}
    ,
\label{eq:analytic_qrot_full}
\end{equation}
\begin{equation}
    q_{vib,\, i}^{n+1}
    =
    q_{vib,\, i}^n
     \exp{\Big( -\frac{3}{\tau} \Delta t \Big)}
    .
\label{eq:analytic_qvib_full}
\end{equation}
The Taylor expansions of the heat fluxes are given by:
\begin{equation}
    q_{tr,\, i}^{n+1}
    =
    q_{tr,\, i}^n
    \Bigg[
        1
        - \frac{3}{\tau}\Delta t
        + \frac{9}{2\tau^2} \Delta t^2
    \Bigg]
    + \mathcal{O}(\Delta t^3)
    ,
\end{equation}
\begin{equation}
    q_{rot,\, i}^{n+1}
    =
    q_{rot,\, i}^n
    \Bigg[
        1
        - \frac{3}{\tau}\Delta t
        + \frac{9}{2\tau^2} \Delta t^2
    \Bigg]
    + \mathcal{O}(\Delta t^3)
    ,
\end{equation}
\begin{equation}
    q_{vib,\, i}^{n+1}
    =
    q_{vib,\, i}^n
    \Bigg[
        1
        - \frac{3}{\tau}\Delta t
        + \frac{9}{2\tau^2} \Delta t^2
    \Bigg]
    + \mathcal{O}(\Delta t^3)
    .
\end{equation}


\section{\label{appendix:USPBGK}Moment relaxations in the USP-BGK method for diatomic gases}



In the USP-BGK method, the solution of the collision step is given as \cite{Fei22_USP-poly, Tian23_USP-poly-SPARTACUS}:
\begin{equation}
    \mathcal{F}^{n+1}
    =
    \mathcal{F}^{n}
    \exp{(-\frac{\Delta t}{ \tau' })}
    +     
    \Big( 1 - \exp{(-\frac{\Delta t}{ \tau' })} \Big)
    \check{{\mathcal{F}}}_{U}^{n}
    ,
\label{eq:USP-BGK-collision}
\end{equation}
where $\tau'=(\mu/p)/\mathrm{Pr}$ is the relaxation time of the USP-BGK method, $\mathcal{F}_U$ represents the target distribution function for the USP-BGK method, and a PDF with a check mark $\check{\mathcal{F}}$ denotes a PDF in which all moments are replaced by rescaled moment $\check{\phi}$. The target distribution function $\mathcal{F}_U$ is defined as:
\begin{equation}
    \check{{\mathcal{F}}}_U
    =
    \check{{\mathcal{F}}}_G
    +
    \Bigg[ 
    \frac{\Delta t}{2}
    \frac{1 + \exp{(- \Delta t / \tau')}}
         {1 - \exp{(- \Delta t / \tau')}}
    -
    \tau'
    \Bigg]
    \check{\mathcal{Q}}_C
    ,
\end{equation}
where $\mathcal{F}_G$ is the Gaussian distribution function, $\mathcal{Q}_C$ is a Grad distribution function. The Gaussian distribution is defined as follows:
\begin{equation}
    \mathcal{F}_G
    =
    \rho\,
    \mathcal{G}_{tr}(\boldsymbol{c})\,
    \mathcal{G}_{rot}(\varepsilon_{rot})\,
    \mathcal{G}_{vib}(\varepsilon_{vib})
    ,
\end{equation}
with
\begin{equation}
    \mathcal{G}_{tr}(\boldsymbol{c})
    =
    \frac{1}{\sqrt{ \det{(2\pi \boldsymbol{\Gamma} )} }}
    \exp{ \Big( -\frac{1}{2} \boldsymbol{c}^T \boldsymbol{\Gamma} \boldsymbol{c} \Big) }
    ,
\end{equation}
\begin{equation}
    \mathcal{G}_{rot}(\varepsilon_{rot})
    =
    \frac{1}{ RT_{rot}^{rel} }
    \exp{ \Big( -\frac{\varepsilon_{rot}}{RT_{rot}^{rel}} \Big) }
    ,
\end{equation}
\begin{equation}
    \mathcal{G}_{vib}(\varepsilon_{vib})
    =
    \bigg[1 - \exp{\Big( -\frac{\Theta_{vib}}{T_{vib}^{rel}} \Big)} \bigg]
    \exp{ \Big( -\frac{\varepsilon_{vib}}{R T_{vib}^{rel}}  \Big) }
    ,
\end{equation}
where $\boldsymbol{\Gamma_{ij}}=RT_{tr}^{rel}\delta_{ij}+(1-1/\mathrm{Pr})(\Pi_{ij}-RT_{tr}\delta_{ij})$ is the relaxation tensor. The expansion coefficients of the $\mathcal{Q}_C$ are determined to ensure that first nine moments, $\phi = $\{$1$, $C_i$, $C^2/2$, $\varepsilon_{rot}$, $\varepsilon_{vib}$, $C_iC_j - C^2/3\,\delta_{ij}$, $C_i C^2/2$, $C_i\varepsilon_{rot}$, $C_i\varepsilon_{vib}$\}, are equal to those of the BGK collision term:
\begin{equation}
    \langle \phi \, \mathcal{Q}_C \rangle 
    = 
    \langle \phi \, \mathcal{S}_{BGK}(\mathcal{F}) \rangle
    .
\end{equation}
In this paper, the right-hand side is evaluated using Mathiaud's ES-BGK model \cite{Mathiaud22-ES-BGK-poly}. The final result of $\mathcal{Q}_C$ is given as:
\begin{equation}
\begin{aligned}
    \mathcal{Q}_C
    = & 
    -\frac{1}{\tau'} \mathcal{F}_M
    \Bigg[ 
        \frac{1}{2\rho R T_{tr} \mathrm{Pr}} 
        \bigg( 
            \frac{C_iC_j}{RT_{tr}}
            -\frac{1}{3} \frac{C^2}{RT_{tr}} \delta_{ij}
        \bigg)
        \sigma_{ij}
        \\[0.5em] & + 
        \frac{2}{5\rho RT_{tr}} 
        \bigg(
            \frac{C^2}{2RT_{tr}}
            - \frac{5}{2}
        \bigg)
        \frac{C_i}{RT_{tr}}
        q_{tr,\, i}
        \\[0.5em] & + 
        \frac{1}{\rho RT_{rot}} 
        \bigg(
            \frac{\varepsilon_{rot}}{RT_{rot}}
            - 
            1
        \bigg)
        \frac{C_i}{RT_{tr}}
        q_{rot,\, i}
        \\[0.5em] & + 
        \frac{1}{\rho \, c_{v,\,vib}(T_{vib}) \,T_{vib}}
        \bigg(
            \varepsilon_{vib} \frac{\Theta_{vib}}{T_{vib}}
            - 
            \frac{\delta_{vib}}{2}
        \bigg)
        \frac{C_i}{RT_{tr}}
        q_{vib,\, i}
        \\[0.5em] & + 
        \frac{2}{3RT_{tr}}
        \bigg(
            \frac{1}{\tau_{rot}} \Big( e_{rot}(T_{rot}) - e_{rot}(T_{tr}) \Big)
            +
            \frac{1}{\tau_{vib}} \Big( e_{vib}(T_{vib}) - e_{vib}(T_{tr}) \Big)
        \bigg)
        \bigg(
            \frac{C^2}{2RT_{tr}}
            -
            \frac{3}{2}
        \bigg)
        \\[0.5em] & + 
        \frac{1}{RT_{rot}}
        \bigg(
            \frac{1}{\tau_{rot}} \Big( e_{rot}(T_{tr}) - e_{rot}(T_{rot}) \Big)
        \bigg)
        \bigg(
            \frac{\varepsilon_{rot}}{RT_{rot}}
            -
            1
        \bigg)
        \\[0.5em] & + 
        \frac{1}{c_{v,\,vib}(T_{vib})\, T_{vib}}
        \bigg(
            \frac{1}{\tau_{vib}} \Big( e_{vib}(T_{tr}) - e_{vib}(T_{vib}) \Big)
        \bigg(
            \varepsilon_{vib} \frac{\Theta_{vib}}{T_{vib}}
            - 
            \frac{\delta_{vib}}{2}
        \bigg)
    \Bigg]
    ,
\end{aligned}
\end{equation}
where $\mathcal{F}_M$ is the Maxwellian distribution, defined as follows:
\begin{equation}
    \mathcal{F}_M
    =
    \rho\,
    \mathcal{M}_{tr}(\boldsymbol{c})\,
    \mathcal{M}_{rot}(\varepsilon_{rot})\,
    \mathcal{M}_{vib}(\varepsilon_{vib})
    ,
\end{equation}
with
\begin{equation}
    \mathcal{M}_{tr}(\boldsymbol{c})
    =
    \frac{1}{ (2\pi RT_{tr})^{3/2} }
    \exp{ \Big( -\frac{C^2}{2RT_{tr}} \Big) }
    ,
\end{equation}
\begin{equation}
    \mathcal{M}_{rot}(\varepsilon_{rot})
    =
    \frac{1}{ RT_{rot} }
    \exp{ \Big( -\frac{\varepsilon_{rot}}{RT_{rot}} \Big) }
    ,
\end{equation}
\begin{equation}
    \mathcal{M}_{vib}(\varepsilon_{vib})
    =
    \bigg[1 - \exp{\Big( -\frac{\Theta_{vib}}{T_{vib}} \Big)} \bigg]
    \exp{ \Big( -\frac{\varepsilon_{vib}}{R T_{vib}}  \Big) }
    .
\end{equation}
The USP-BGK method uses two auxiliary PDFs, whereas the USP-FPM method uses only the real PDF. To address this difference, the rescaled moments $\check{\phi}$ are introduced, following the same normalization strategy employed in the USP-FP formulation \cite{Kim24_USP-FP}:
\begin{equation}
    e_{tr}({T}_{tr})
    =
    e_{tr}(\check{T}_{tr})
    -
    \frac{\Delta t}{2 \tau_{rot}}
    \Big( 
        e_{rot}(\check{T}_{rot}) - e_{rot}(\check{T}_{tr})
    \Big)
    -
    \frac{\Delta t}{2 \tau_{vib}}
    \Big( 
        e_{vib}(\check{T}_{vib}) - e_{vib}(\check{T}_{tr})
    \Big)
    ,
\label{eq:T-tr-check}
\end{equation}
\begin{equation}
    e_{rot}({T}_{rot})
    =
    e_{rot}(\check{T}_{rot})
    -
    \frac{\Delta t}{2 \tau_{rot}}
    \Big( 
        e_{rot}(\check{T}_{tr}) - e_{rot}(\check{T}_{rot})
    \Big)
    ,
\label{eq:T-rot-check}
\end{equation}
\begin{equation}
    e_{vib}({T}_{vib})
    =
    e_{vib}(\check{T}_{vib})
    -
    \frac{\Delta t}{2 \tau_{vib}}
    \Big( 
        e_{vib}(\check{T}_{tr}) - e_{vib}(\check{T}_{vib})
    \Big)
    ,
\label{eq:T-vib-check}
\end{equation}
\begin{equation}
    \sigma_{ij}
    =
    \Big( 1 - \frac{\Delta t}{2\, \mathrm{Pr} \, \tau'} \Big)
    \check{\sigma}_{ij}
    ,
\label{eq:sigma_check}
\end{equation}
\begin{equation}
    q_{tr,\,i}
    =
    \Big( 1 - \frac{\Delta t}{2 \, \tau'} \Big)
    \check{q}_{tr,\,i}
    ,
\label{eq:q_tr_check}
\end{equation}
\begin{equation}
    q_{rot,\,i}
    =
    \Big( 1 - \frac{\Delta t}{2 \, \tau'} \Big)
    \check{q}_{rot,\,i}
    ,
\label{eq:q_rot_check}
\end{equation}
\begin{equation}
    q_{vib,\,i}
    =
    \Big( 1 - \frac{\Delta t}{2 \, \tau'} \Big)
    \check{q}_{vib,\,i}
    .
\label{eq:q_vib_check}
\end{equation}
To express the rescaled temperatures $\check{T}$ in terms of real temperature $T$, Equations (\ref{eq:T-tr-check}) to (\ref{eq:T-vib-check}) are first combined and rearranged:
\begin{multline}
    e_{tr}(\check{T}_{tr})
    +
    \bigg[ 
        \frac{\Delta t}{\Delta t + 2\tau_{vib}}
    \bigg] 
    e_{rot}(\check{T}_{tr})
    +
    \bigg[ 
         \frac{\Delta t}{\Delta t + 2\tau_{vib}}
    \bigg] 
    e_{vib}(\check{T}_{tr})
    \\ = 
    e_{tr}({T}_{tr})
    +
    \bigg[ 
        \frac{\Delta t}{\Delta t + 2\tau_{rot}}
    \bigg] 
    e_{rot}({T}_{rot})
    +
    \bigg[ 
        \frac{\Delta t}{\Delta t + 2\tau_{vib}}
    \bigg] 
    e_{vib}({T}_{vib})
    .
\end{multline}
The mean value theorem is applied to the function $e_{vib}$ to get the equation:
\begin{multline}
    c_{v,\, tr}\, \check{T}_{tr}
    +
    \bigg[ 
        \frac{\Delta t}{\Delta t + 2\tau_{rot}}
    \bigg] 
    c_{v,\,rot}\,  \check{T}_{tr}
    +
    \bigg[ 
         \frac{\Delta t}{\Delta t + 2\tau_{vib}}
    \bigg] 
    c_{v,\,vib}(T_1)\, \check{T}_{tr}
    \\ = 
    c_{v,\, tr}\, {T}_{tr}
    +
    \bigg[ 
        \frac{\Delta t}{\Delta t + 2\tau_{rot}}
    \bigg] 
    c_{v,\,rot}\, {T}_{rot}
    +
    \bigg[ 
        \frac{\Delta t}{\Delta t + 2\tau_{vib}}
    \bigg] 
    c_{v,\,vib}(T_1)\, {T}_{vib}
    ,
\end{multline}
where $T_1$ is a temperature between $\check{T}_{tr}$ and $T_{vib}$ satisfying the following equation:
\begin{equation}
    c_{v,\,vib}(T_1)
    =
    \frac{e_{vib}(\check{T}_{tr})-e_{vib}(T_{vib})}{\check{T}_{tr}-T_{vib}}
    . 
\label{eq:cv_vib}
\end{equation}
The rescaled temperatures are then expressed as follows:
\begin{equation}
\begin{aligned}
    e_{tr}(\check{T}_{tr})
    = 
    e_{tr}({T}_{tr})
    & +
    \frac{
            \frac{\Delta t}{\Delta t + 2\tau_{rot}}
            \,c_{v,\, tr}
            \,c_{v,\,rot}
          }
          {
            c_{v,\, tr}
            +
            \frac{\Delta t}{\Delta t + 2\tau_{rot}}
            c_{v,\,rot}
            +
            \frac{\Delta t}{\Delta t + 2\tau_{vib}}
            c_{v,\,vib}(T_1)
          }
    \Big( {T}_{rot} - T_{tr} \Big)
    \\[0.5em] & +
    \frac{
            \frac{\Delta t}{\Delta t + 2\tau_{vib}}
            \,c_{v,\, tr}
            \,c_{v,\,vib}(T_1)
          }
          {
            c_{v,\, tr}
            +
            \frac{\Delta t}{\Delta t + 2\tau_{rot}}
            c_{v,\,rot}
            +
            \frac{\Delta t}{\Delta t + 2\tau_{vib}}
            c_{v,\,vib}(T_1)
          }
    \Big( {T}_{vib} - T_{tr} \Big)
    ,
\label{eq:T_tr_check-T}
\end{aligned}
\end{equation}
\begin{equation}
\begin{aligned}
    e_{rot}(\check{T}_{rot})
    = 
    e_{rot}({T}_{rot})
    & +
    \frac{
            \frac{\Delta t}{\Delta t + 2\tau_{rot}}
            \,c_{v,\, tr}
            \,c_{v,\,rot}
          }
          {
            c_{v,\, tr}
            +
            \frac{\Delta t}{\Delta t + 2\tau_{rot}}
            c_{v,\,rot}
            +
            \frac{\Delta t}{\Delta t + 2\tau_{vib}}
            c_{v,\,vib}(T_1)
          }
    \Big( {T}_{tr} - T_{rot} \Big)
    \\[0.5em] &+
    \frac{
            \frac{\Delta t}{\Delta t + 2\tau_{rot}}
            \frac{\Delta t}{\Delta t + 2\tau_{vib}}
            \,c_{v,\,rot}
            \,c_{v,\,vib}(T_1)
          }
          {
            c_{v,\, tr}
            +
            \frac{\Delta t}{\Delta t + 2\tau_{vib}}
            c_{v,\,rot}
            +
            \frac{\Delta t}{\Delta t + 2\tau_{vib}}
            c_{v,\,vib}(T_1)
          }
    \Big( {T}_{vib} - T_{rot} \Big)
    ,
\label{eq:T_rot_check-T}
\end{aligned}
\end{equation}
\begin{equation}
\begin{aligned}
    e_{vib}(\check{T}_{vib})
    = 
    e_{vib}({T}_{vib})
    & +
    \frac{
            \frac{\Delta t}{\Delta t + 2\tau_{vib}}
            \,c_{v,\, tr}
            \,c_{v,\,rot}
          }
          {
            c_{v,\, tr}
            +
            \frac{\Delta t}{\Delta t + 2\tau_{rot}}
            c_{v,\,rot}
            +
            \frac{\Delta t}{\Delta t + 2\tau_{vib}}
            c_{v,\,vib}(T_1)
          }
    \Big( {T}_{tr} - T_{vib} \Big)
    \\[0.5em] & +
    \frac{
            \frac{\Delta t}{\Delta t + 2\tau_{rot}}
            \frac{\Delta t}{\Delta t + 2\tau_{vib}}
            \,c_{v,\,rot}
            \,c_{v,\,vib}(T_1)
          }
          {
            c_{v,\, tr}
            +
            \frac{\Delta t}{\Delta t + 2\tau_{vib}}
            c_{v,\,rot}
            +
            \frac{\Delta t}{\Delta t + 2\tau_{vib}}
            c_{v,\,vib}(T_1)
          }
    \Big( {T}_{rot} - T_{vib} \Big)
    .
\label{eq:T_vib_check-T}
\end{aligned}
\end{equation}


To derive translational energy relaxation in the USP-BGK method, Equation (\ref{eq:USP-BGK-collision}) is multiplied by $C^2/2$ and the ensemble average is taken:
\begin{equation}
    \langle \frac{1}{2} C^2 \mathcal{F}^{n+1} \rangle
    =
    \langle \frac{1}{2} C^2 \mathcal{F}^{n}  \rangle
    \exp{(-\frac{\Delta t}{ \tau' })}
    +     
    \Big( 1 - \exp{(-\frac{\Delta t}{ \tau' })} \Big)
    \langle \frac{1}{2} C^2 \check{{\mathcal{F}}}_{U}^{n}  \rangle
    .
\end{equation}
This equation is rearranged as follows:
\begin{equation}
\begin{aligned}
    e_{tr}& (T_{tr}^{n+1})
    = 
    e_{tr}(T_{tr}^{n})
    \exp{(-\frac{\Delta t}{ \tau' })}
    \\ & \quad \quad \quad \quad +      
    \Bigg( 1 - \exp{(-\frac{\Delta t}{ \tau' })} \Bigg)
    \Bigg[ 
    \langle \frac{1}{2} C^2 \check{{\mathcal{F}}}_{G}^{n}  \rangle
    +
    \Bigg(
        \frac{\Delta t}{2}
        \frac{1 + \exp{(-\frac{\Delta t}{\tau'})}}
             {1 - \exp{(-\frac{\Delta t}{\tau'})}}
        -
        \tau'
    \Bigg)
    \langle \frac{1}{2} C^2 \check{\mathcal{Q}}_C^n  \rangle
    \Bigg]
    \\[1.0em] & \quad  = 
    e_{tr}(T_{tr}^{n})
    \exp{(-\frac{\Delta t}{ \tau' })}
    \\ & \quad \quad  +
    \Bigg( 1 - \exp{(-\frac{\Delta t}{ \tau' })} \Bigg)
    \Bigg[ 
    e_{tr}(\check{T}_{tr}^{rel,\,n})
    +
    \Bigg(
        \frac{\Delta t}{2}
        \frac{1 + \exp{(-\frac{\Delta t}{\tau'})}}
             {1 - \exp{(-\frac{\Delta t}{\tau'})}}
        -
        \tau'
    \Bigg)
    \frac{e_{tr}(\check{T}_{tr}^{rel,\,n}) - e_{tr}(\check{T}_{tr}^{n})}{\tau'}
    \Bigg]
    \\[1.0em] & \quad  =
    e_{tr}(T_{tr}^{n})
    \exp{(-\frac{\Delta t}{ \tau' })}
    +
    \Bigg( 1 - \exp{(-\frac{\Delta t}{ \tau' })} \Bigg)
    e_{tr}(\check{T}_{tr}^{n})
    \\[0.5em]  &  \quad \quad   -
    \Bigg( 1 + \exp{(-\frac{\Delta t}{ \tau' })} \Bigg)
    \Bigg(
        \frac{\Delta t}{2 \tau_{rot}}
        \Big( 
            e_{rot}(\check{T}_{rot}) - e_{rot}(\check{T}_{tr})
        \Big)
        +
        \frac{\Delta t}{2 \tau_{vib}}
        \Big( 
            e_{vib}(\check{T}_{vib}) - e_{vib}(\check{T}_{tr})
        \Big)
    \Bigg)
    .
\end{aligned}
\end{equation}
By substituting Equations (\ref{eq:T_tr_check-T})-(\ref{eq:T_vib_check-T}), this equation is now rearranged as follows:
\begin{equation}
\begin{aligned}
    e_{tr} (T_{tr}^{n+1}) 
    = 
    e_{tr}(T_{tr}^n)
    \;+ & \;
    \frac
    {
        \frac{2\Delta t}{\Delta t + 2\tau_{rot}}
        c_{v,\, tr} \, c_{v,\,rot}
    }
    {
        c_{v,\, tr} 
        + \frac{\Delta t}{\Delta t + 2\tau_{rot}} c_{v,\,rot} 
        + \frac{\Delta t}{\Delta t + 2\tau_{vib}} c_{v,\,vib}(T_1)
    }
    (T_{rot}^n - T_{tr}^n) 
    \\[0.5em]
    + & \;
    \frac
        {
            \frac{2\Delta t}{\Delta t + 2\tau_{vib}} 
            c_{v,\, tr} \, c_{v,\,vib}(T_1)
        }
        {
            c_{v,\, tr} 
            + \frac{\Delta t}{\Delta t + 2\tau_{rot}} c_{v,\,rot} 
            + \frac{\Delta t}{\Delta t + 2\tau_{vib}} c_{v,\,vib}(T_1)
        }
    (T_{vib}^n - T_{tr}^n)
    \\[1.0em]= 
    e_{tr}(T_{tr}^n)
    \; -& \;
    \gamma_{tr-rot} 
    (T_{tr}^n - T_{rot}^n) 
    - 
    \gamma_{tr-vib} 
    (T_{tr}^n - T_{vib}^n) 
    ,
\end{aligned}
\end{equation}
where coefficients $\gamma_{tr-rot}$ and $\gamma_{tr-vib}$ are defined as follows:
\begin{equation}
    \gamma_{tr-rot} = 
    \frac
    {
        \frac{2\Delta t}{\Delta t + 2\tau_{rot}}
        c_{v,\, tr} \, c_{v,\,rot}
    }
    {
        c_{v,\, tr} 
        + \frac{\Delta t}{\Delta t + 2\tau_{rot}} c_{v,\,rot} 
        + \frac{\Delta t}{\Delta t + 2\tau_{vib}} c_{v,\,vib}(T_1)
    }
    ,
\label{eq:gamma_tr_rot}
\end{equation}
\begin{equation}
    \gamma_{tr-vib} = 
    \frac
        {
            \frac{2\Delta t}{\Delta t + 2\tau_{vib}} 
            c_{v,\, tr} \, c_{v,\,vib}(T_1)
        }
        {
            c_{v,\, tr} 
            + \frac{\Delta t}{\Delta t + 2\tau_{rot}} c_{v,\,rot} 
            + \frac{\Delta t}{\Delta t + 2\tau_{vib}} c_{v,\,vib}(T_1)
        }
    .
\label{eq:gamma_tr_vib}
\end{equation}
In the same way, the rotational and vibrational energy relaxations in the USP-BGK method are derived by multiplying Equation (\ref{eq:USP-BGK-collision}) by $\varepsilon_{rot}$ and $\varepsilon_{vib}$ and taking the ensemble average, respectively:
\begin{equation}
    e_{rot} (T_{rot}^{n+1}) = e_{rot}(T_{rot}^n)
    + \gamma_{\mathrm{tr-rot}} (T_{tr}^n - T_{rot}^n) 
    - \gamma_{\mathrm{rot-vib}} (T_{rot}^n - T_{vib}^n) 
    ,
\end{equation}
\begin{equation}
    e_{vib}(T_{vib}^{n+1}) = e_{vib}(T_{vib}^n)
    + \gamma_{tr-vib} (T_{tr}^n - T_{vib}^n) 
    + \gamma_{rot-vib} (T_{rot}^n - T_{vib}^n) 
    .
\end{equation}
where the coefficient $\gamma_{rot-vib}$ is defined as:
\begin{equation}
    \gamma_{rot-vib} = 
    \frac
    {
        \frac{2\Delta t}{\Delta t + 2\tau_{rot}}
        \frac{\Delta t}{\Delta t + 2\tau_{vib}}
        \, c_{v,\,rot} 
        \, c_{v,\,vib}(T_1)
    }
    {
        c_{v,\, tr} 
            + \frac{\Delta t}{\Delta t + 2\tau_{rot}} c_{v,\,rot} 
            + \frac{\Delta t}{\Delta t + 2\tau_{vib}} c_{v,\,vib}(T_1)
    }
    .
\label{eq:gamma_rot_vib}
\end{equation}


The viscous stress relaxation in the USP-BGK method is derived by multiplying Equation (\ref{eq:USP-BGK-collision}) by $(C_iC_j-C^2/3\,\delta_{ij})$ and taking the ensemble average:
\begin{equation}
\begin{aligned}
    \langle (C_iC_j-C^2/3\,\delta_{ij}) \mathcal{F}^{n+1} \rangle
    \;= &\;
    \langle (C_iC_j-C^2/3\,\delta_{ij}) \mathcal{F}^{n}  \rangle
    \exp{(-\frac{\Delta t}{ \tau' })}
    \\[0.5em] & +      
    \Big( 1 - \exp{(-\frac{\Delta t}{ \tau' })} \Big)
    \langle (C_iC_j-C^2/3\,\delta_{ij}) \check{{\mathcal{F}}}_{U}^{n}  \rangle
    .
\end{aligned}
\end{equation}
This equation is rearranged as follows:
\begin{equation}
\begin{aligned}
    \sigma_{ij}^{n+1}
    \;=\; &
    \sigma_{ij}^{n}
    \exp{(-\frac{\Delta t}{ \tau' })}
    \\[0.5em] & + 
    \Big( 1 - \exp{(-\frac{\Delta t}{ \tau' })} \Big)
    \Bigg[ 
    \langle (C_iC_j-C^2/3\,\delta_{ij}) \mathcal{F}_{G}^{n}  \rangle
    \\[0.5em] & \quad +
    \Bigg(
        \frac{\Delta t}{2}
        \frac{1 + \exp{(-\frac{\Delta t}{\tau'})}}
             {1 - \exp{(-\frac{\Delta t}{\tau'})}}
        -
        \tau'
    \Bigg)
    \langle (C_iC_j-C^2/3\,\delta_{ij}) \mathcal{Q}_{C}^{n}  \rangle
    \Bigg]
    \\[1.0em] \;=\; & 
    \sigma_{ij}^{n}
    \exp{(-\frac{\Delta t}{ \tau' })}
    \\[0.5em] & + 
    \Big( 1 - \exp{(-\frac{\Delta t}{ \tau' })} \Big)
    \Bigg[ 
    (1 - \frac{1}{\mathrm{Pr}}) \check{\sigma}_{ij}^{n}
    +
    \Bigg(
        \frac{\Delta t}{2}
        \frac{1 + \exp{(-\frac{\Delta t}{\tau'})}}
             {1 - \exp{(-\frac{\Delta t}{\tau'})}}
        -
        \tau'
    \Bigg)
    \frac{1}{\tau'}
    (-\frac{1}{\mathrm{Pr}} \check{\sigma}_{ij}^{n})
    \Bigg]
    \\[1.0em] \;=\; & 
    \sigma_{ij}^{n}
    \exp{(-\frac{\Delta t}{ \tau' })}
    + 
    \Big( 1 - \exp{(-\frac{\Delta t}{ \tau' })} \Big)
    \Bigg(
        1 - 
        \frac{1}{\mathrm{Pr}}
        \frac{\Delta t}{2}
        \frac{1 + \exp{(-\frac{\Delta t}{\tau'})}}
             {1 - \exp{(-\frac{\Delta t}{\tau'})}}
    \Bigg)
    \check{\sigma}_{ij}^{n}
    .
\end{aligned}
\end{equation}
By substituting Equation (\ref{eq:sigma_check}), this equation is now rearranged as follows:
\begin{equation}
    \sigma_{ij}^{n+1}
    =
    \sigma_{ij}^{n}
    \Bigg(
    \frac{2(\mu/p) - \Delta t}
         {2(\mu/p) + \Delta t}
    \Bigg)
    .
\end{equation}


The translational heat flux in the USP-BGK method is derived by multiplying Equation (\ref{eq:USP-BGK-collision}) by $C^2C_i/2$ and taking the ensemble average:
\begin{equation}
\begin{aligned}
    q_{tr,\,i}^{n+1}
    \;=\; &
    \langle \frac{1}{2}C^2C_i\, \mathcal{F}^{n+1} \rangle
    \\[0.5em] \;=\; &
    \langle \frac{1}{2}C^2C_i\, \mathcal{F}^{n}  \rangle
    \exp{(-\frac{\Delta t}{ \tau' })}
     +     
    \Big( 1 - \exp{(-\frac{\Delta t}{ \tau' })} \Big)
    \langle \frac{1}{2}C^2C_i\, \check{{\mathcal{F}}}_{U}^{n}  \rangle
    \\[0.5em] \;=\; &
    \langle \frac{1}{2}C^2C_i\, \mathcal{F}^{n}  \rangle
    \exp{(-\frac{\Delta t}{ \tau' })}
    \\[0.5em] & + 
    \Big( 1 - \exp{(-\frac{\Delta t}{ \tau' })} \Big)
    \Bigg[ 
    \langle \frac{1}{2}C^2C_i\, \mathcal{F}_{G}^{n}  \rangle
     +
    \Bigg(
        \frac{\Delta t}{2}
        \frac{1 + \exp{(-\frac{\Delta t}{\tau'})}}
             {1 - \exp{(-\frac{\Delta t}{\tau'})}}
        -
        \tau'
    \Bigg)
    \langle \frac{1}{2}C^2C_i\, \mathcal{Q}_{C}^{n}  \rangle
    \Bigg]
    \\[0.5em] \;=\;  &
    q_{tr,\,i}^{n}
    \exp{(-\frac{\Delta t}{ \tau' })}
     + 
    \Big( 1 - \exp{(-\frac{\Delta t}{ \tau' })} \Big)
    \Bigg(
        1 - 
        \frac{\Delta t}{2\tau'}
        \frac{1 + \exp{(-\frac{\Delta t}{\tau'})}}
             {1 - \exp{(-\frac{\Delta t}{\tau'})}}
    \Bigg)
    \check{q}_{tr,\,i}^n
    .
\end{aligned}
\end{equation}
By substituting Equation (\ref{eq:q_tr_check}), this equation is now rearranged as follows:
\begin{equation}
    q_{tr,\,i}^{n+1}
    =
    q_{tr,\,i}^{n}
    \Bigg(
    \frac{2(\mu/p) - \mathrm{Pr} \, \Delta t}
         {2(\mu/p) + \mathrm{Pr} \, \Delta t}
    \Bigg)
    .
\end{equation}
In the same way, the rotational and vibrational heat flux relaxations in the USP-BGK method are derived by multiplying Equation (\ref{eq:USP-BGK-collision}) by $C_i\varepsilon_{rot}$ and $C_i\varepsilon_{vib}$ and taking the ensemble average, respectively:
\begin{equation}
    q_{rot,\,i}^{n+1}
    =
    q_{rot,\,i}^{n}
    \Bigg(
    \frac{2(\mu/p) - \mathrm{Pr} \, \Delta t}
         {2(\mu/p) + \mathrm{Pr} \, \Delta t}
    \Bigg)
    ,
\end{equation}
\begin{equation}
    q_{vib,\,i}^{n+1}
    =
    q_{vib,\,i}^{n}
    \Bigg(
    \frac{2(\mu/p) - \mathrm{Pr} \, \Delta t}
         {2(\mu/p) + \mathrm{Pr} \, \Delta t}
    \Bigg)
    .
\end{equation}


\bibliography{USP-FPM_bibliography}

\end{document}